\documentclass[a4paper]{article}

\usepackage{amsthm}
\usepackage{amssymb}
\usepackage{amsmath}
\usepackage{graphicx}
\include{now_macro}
\makeatletter

\def\fixNumberingInArticle{
\@addtoreset{figure}{section}
\@addtoreset{equation}{section}
\renewcommand{\thefigure}{\thesection.\arabic{figure}}
\renewcommand{\theequation}{\thesection.\arabic{equation}}
}

\def\fixNumberingInAppendix{
\setcounter{section}{0}
\renewcommand{\theequation}{\Alph{section}.\arabic{equation}}
\renewcommand{\thesection}{\Alph{section}}
\setcounter{equation}{0}
}

\makeatother

\newcommand{\ie}{{\it i.e.\/}, }
\newcommand{\eg}{{\it e.g.\/}, }
\newcommand{\cf}{{\it cf.\/}\ }
\newcommand{\Eq}{{Eq.\/}\ }

\newcommand{\RR}{\mathbb{R}}
\newcommand{\CC}{\mathbb{C}}
\newcommand{\Ltwo}{\set{L}_2}

\newcommand{\intd}[2][ ]{\mathrm{d}^{#1}#2\ }
\newcommand{\ldotsk}{\ldots\hspace{0pt}}  %to remove additional space
\newcommand{\lexp}[1]{\mathrm{e}^{#1}}

\newcommand{\iu}{\mathrm{i}}
\newcommand{\sgn}[1]{\mathop{\mathrm{sgn}(#1)}}

\newcommand{\set}[1]{\mathrm{#1}}
\newcommand{\domain}[1]{\set{D}( #1 )}
\newcommand{\range}[1]{\set{R}\left( #1 \right)}

\newcommand{\closure}[1]{\left\{ #1 \right\}_{\mathrm{cl}} }
\newcommand{\adjoint}[1]{#1^{*}}
\newcommand{\commute}[2]{[#1,#2]}
\newcommand{\Res}[2]{\mathrm{Res}\left\{ #1; #2 \right\}}
\newcommand{\bin}[2]{\begin{pmatrix} #1 \\ #2 \end{pmatrix}}

\newcommand{\od}{\mathrm{d}}

\newcommand{\setSobolev}[1]{\set{H}^{#1}}

\newcommand{\mSnorm}[2]{\| #2\|_{#1}}
\newcommand{\mSdot}[3]{( #2 , #3 )_{#1}}
\newcommand{\mSob}[1]{\mathbf{H}^{#1}}

\newcommand{\mLs}{\mathbf{L}^{2}}  %the set
\newcommand{\mLnorm}[1]{\mSnorm{0}{#1}}
\newcommand{\mLdot}[2]{ \mSdot{0}{#1}{#2}}
\newcommand{\msLs}{\{\mathbf{L}^{2}, \mLdot{\cdot}{\cdot}\}} %weighted

\newcommand{\snorm}[1]{|#1|_{\rm s}}
\newcommand{\sdot}[2]{\langle #1, #2 \rangle_{\rm s}}

\newcommand{\RE}[1]{\set{Re}\left\{#1\right\}}
\newcommand{\IM}[1]{\set{Im}\left\{#1\right\}}
\newcommand{\RL}{\lambda_{\mathrm{R}}}
\newcommand{\IL}{\lambda_{\mathrm{I}}}
\newcommand{\rS}{s_{\mathrm{r}}}

\newcommand{\as}{\acute{s}}

\newcommand{\OP}[1]{\mathcal{#1}}
\newcommand{\symb}[1]{\mbox{\boldmath $#1$}}

\newcommand{\sr}{\symb{r}}
\newcommand{\ro}{\sr_{;-1}}
\newcommand{\rt}{\sr_{;-2}}
\newcommand{\rd}{\sr_{;-3}}
\newcommand{\rn}{\sr_{;-m-1}}
\newcommand{\rno}{\sr_{;-m}}
\newcommand{\rnt}{\sr_{;-m+1}}

\newcommand{\hp}{\set{hp}}
\newcommand{\hq}{\set{hq}}
\newcommand{\hP}{\set{hP}}
\newcommand{\OO}{\set{W}}
\newcommand{\diag}{\mathop{\mathrm{diag}}}
\newcommand{\diagT}{\mathop{\mathrm{diag}_2}}

\newcommand{\EL}{\OP{E}_{s,\lambda}}

\newcommand{\el}{\symb{e}}
\newcommand{\eu}{\el_{;1}}

\newcommand{\ac}{\symb{\alpha}_{;1}}
\newcommand{\aS}{\symb{a}}
\newcommand{\auu}{\aS_{11;1}}
\newcommand{\aud}{\aS_{12;1}}
\newcommand{\adu}{\aS_{21;1}}
\newcommand{\add}{\aS_{22;1}}

\newcommand{\inv}[1]{\hat{#1}}
\newcommand{\iauu}{\inv{\aS}_{11;1}}
\newcommand{\iaud}{\inv{\aS}_{12;1}}
\newcommand{\iadu}{\inv{\aS}_{21;1}}
\newcommand{\iadd}{\inv{\aS}_{22;1}}
\newcommand{\ieu}{\inv{\el}_{;1}}

\newcommand{\Ac}{\OP{A}}
\newcommand{\Aco}{\Ac_{s,\lambda}}
\newcommand{\oL}{\OP{L}}
\newcommand{\oV}{\OP{V}}

\newcommand{\qL}[1]{\left. \oL\right|_{#1}}

\newcommand{\pmN}{\OP{N}^{\pm}}
\newcommand{\pmL}{\oL^{\pm}}
\newcommand{\qpmL}[1]{\left. \pmL \right|_{#1}}
\newcommand{\sbb}{\symb{b}}
\newcommand{\psb}{\sbb_{;0}}
\newcommand{\oB}{\OP{B}}
\newcommand{\qB}[1]{\left. \OP{B} \right|_{#1}}

\newcommand{\qN}[1]{\left. \OP{N} \right|_{#1}}
\newcommand{\qpN}[1]{\left. \OP{N}^{+} \right|_{#1}}
\newcommand{\qmN}[1]{\left. \OP{N}^{-} \right|_{#1}}
\newcommand{\qV}[1]{\left. \OP{V} \right|_{#1}}

\newcommand{\DtN}{\OP{Z}}

\newcommand{\Strut}{\mbox{\rule{0pt}{2.4ex}}}
\newcommand{\smooth}[2][0]{\smash{\overset{\mbox{\raisebox{-.4 ex}{\hspace{#1 em}$\scriptscriptstyle{\circ}$}}}{#2}}}
\newcommand{\sB}{\smooth{\OP{B}}\Strut}
\newcommand{\sV}{\smooth{\OP{V}}\Strut}
\newcommand{\sL}{\smooth[.2]{\OP{L}}\Strut}

\newcommand{\spmL}{\smooth[.2]{\OP{L}}^{\pm}\Strut}
\newcommand{\sA}{\smooth[.5]{\Ac}\Strut}
\newcommand{\spmN}{\smooth[.2]{\OP{N}}^{\pm}\Strut}
\newcommand{\spN}{\smooth[.2]{\OP{N}}^{+}\Strut}
\newcommand{\smN}{\smooth[.2]{\OP{N}}^{-}\Strut}
\newcommand{\spmDtN}{\smooth[.2]{\DtN}^{\pm}\Strut}
\newcommand{\sN}{\smooth[.2]{\OP{N}}\Strut}
\newcommand{\swpm}{\smooth{\OP{S}}^{\pm}\Strut}
\newcommand{\swp}{\smooth{\OP{S}}^{+}\Strut}
\newcommand{\swm}{\smooth{\OP{S}}^{-}\Strut}
\newcommand{\setC}[1][4]{\set{C}^{\infty}(\RR^2,\CC^{#1})}

\newcommand{\intKn}{\lambda \in \set{K}_{n}}

\newcommand{\sKn}{\set{K}_{n}}
\newcommand{\lLb}{\tau}
\newcommand{\lowS}{S_{\mathrm{R}}}
\newcommand{\lowEu}{\hat{\epsilon}_{1}}
\newcommand{\lowMu}{\hat{\mu}_{1}}
\newcommand{\lowEz}{\hat{\epsilon}_{0}}
\newcommand{\lowMz}{\hat{\mu}_{0}}

\newcommand{\rpm}{\mu}
\newcommand{\rpe}{\epsilon}
\newcommand{\rM}{\nu}
\newcommand{\rE}{\varepsilon}

\newcommand{\w}{\OP{S}}
\newcommand{\WP}{\w^{+}}
\newcommand{\WM}{\w^{-}}
\newcommand{\wpm}{\w^{\pm}}

\newtheorem{prop}{Proposition}
\newtheorem{lemm2}{Lemma}[prop]
\newtheorem{remark}{Remark}[prop]
\newtheorem{cor}{Corollary}[prop]

\fixNumberingInArticle

\title{Wave splitting of Maxwell's equations with anisotropic heterogeneous constitutive relations}
\author{B. L. G. Jonsson}

\begin{document}
\maketitle \pagestyle{myheadings}

\markboth{B. L. G. Jonsson}{Wave splitting of Maxwell's
equations\ldots}

\NowFoot

%%%%%%%%%%%%%%%%%%%%%%%%%%%%%%%%%%%%%%%%%%%%%%%%%%%%%%%%%%%%%%%%%%%%%%%%
% Document                                                             %
%%%%%%%%%%%%%%%%%%%%%%%%%%%%%%%%%%%%%%%%%%%%%%%%%%%%%%%%%%%%%%%%%%%%%%%%

% abstract
%
\begin{abstract}
 The equations for the electromagnetic field in an anisotropic media
are written in a form containing only the transverse field
components relative to a half plane boundary. The operator
corresponding to this formulation is the electromagnetic system's
matrix. A constructive proof of the existence of directional
wave-field decomposition with respect to the normal of the boundary
is presented.

In the process of defining the wave-field decomposition
(wave-splitting), the resolvent set of the time-Laplace
representation of the system's matrix is analyzed. This set is shown
to contain a strip around the imaginary axis. We construct a
splitting matrix as a Dunford-Taylor type integral over the
resolvent of the unbounded operator defined by the electromagnetic
system's matrix. The splitting matrix commutes with the system's
matrix and the decomposition is obtained via a generalized
eigenvalue-eigenvector procedure. The decomposition is expressed in
terms of components of the splitting matrix. The constructive
solution to the question on the existence of a decomposition also
generates an impedance mapping solution to an algebraic Riccati
operator equation. This solution is the electromagnetic
generalization in an anisotropic media of a Dirichlet-to-Neumann
map.


%\noindent {\bf Mathematics Subject Classifications (2000):} 35P25,
%78A25, 78A40, 78A46, 35Q60, 46N20, 47N20.
\end{abstract}

%
%
% main text
\section{Introduction}

Wave field decomposition is a tool for analyzing and computing waves
in a configuration characterized by a certain directionality. The
wave-field decomposition, or wave splitting, has been used to
separate the wave field constituents which are of importance for the
analysis on a boundary, both for direct and inverse scattering
problems~\cite{Kristensson+Kruger85, Fishman94, MdeHoop9606,
He+Strom+Weston98, Gustafsson00, Stolk+deHoop2006, Wu2007} and for
the analysis of boundary conditions (see for example
\cite{Engquist+Majda77, Mittra+etal89, Cao98, Jonsson2004}).

A remaining challenge in seismic prospecting methods is to incorporate
anisotropy into the analysis. The enormous data sets used for studying
such inverse problems are on the border or beyond today's
computers~\cite{Commer2008}. A common method to access such problems
is to use wave field approximations. Such approximations have been
developed for and applied to a wide range of hyperbolic equations
describing wave propopagation in isotropic media. One class of such
approximations is based on a decomposition of the wave-field into
up-/down-going components. Such a decomposition is usually denoted a
wave splitting or a wave-field decomposition. There are essentially
two types of limitations to the present theory of wave splitting: 
The traditional operator based approach of wave-splitting has been
limited to heterogeneous isotropic materials see
\eg~\cite{Weston86,MdeHoop9606,Gustafsson00,Morro2004} or up-/down
symmetric media~\cite{deHoopSquare}. This method is based essentially
on constructing a certain square root operator and it fails, once the
media becomes inherently anisotropic. Whereas wave-splitting by
spectral decomposition of a certain matrix is restricted to
homogeneous or depth-independent material. Here both anisotropic and
bi-anisotropic materials have been
considered~\cite{deHon96,Rikte2001}.

The present paper removes both of these limitations. We present a
derivation of a three dimensional wave splitting for electromagnetic
fields in the presence of inherently anisotropic loss-less
heterogeneous constitutive relations. We show the existence of a
decomposition by a constructive argument. The decomposition is given
for media with anisotropic permittivity and permeability that is
described by self-adjoint, heterogeneous, positive definite
matrices. These conditions are sufficient but not necessary material
conditions for the resolvent set of a certain operator to contain
the strip around the imaginary axis. The requirements of the material
parameters, in time-Laplace domain, are corresponding to a medium
with only instantaneous lossless response. The analysis which use
pseudodifferential calculus is straightforward if one assumes that
the material coefficients depend smoothly on the spatial variables.
In order to simplify the analysis this assumption is made.
Physically this should not be regarded as a restriction since the
smooth functions densely approximate the square integrable ones.

The decomposition is constructed through a generalized
eigenvalue-eigenvector procedure and a certain commutation of two
operators. The construction of the commuting operator, the splitting
matrix, is made by means of a functional analysis approach using the
resolvent of the electromagnetic system's matrix and is analyzed
with pseudodifferential calculus with parameters, to prove its
existence and to study its behavior. The method is a generalization
of the stratified-media case first presented in~\cite{deHon96} and
extends the theory form the linear acoustic~\cite{Jonsson+deHoop01}
case to the considerably more complex electromagnetic case. The
challenge in going from the anisotropic-acoustic case to the
electromagnetic case includes a more complex differential operator
with a non-trivial null-space, as well as the analysis of a
resolvent operator which here is the inverse of a 4x4-matrix of
operators.

There is a wealth of literature on wave splitting and their
applications. A few references are mentioned below. For time-domain
wave-splitting see~\cite{He+Strom+Weston98}, where both the wave
equation and the Maxwell's equations are considered with both
applications and theory. Wave-splitting in connection with Bremmer
series for linear acoustics~\cite{MdeHoop9606,Gustafsson2000,
  Gustafsson00} and uniform asymptotics and normal
modes~\cite{Fishman+deHoop+vanStralen00, deHoop+Gautesen} has been
used to analyze the wave-field constituents. An extension to include
dispersion is presented in~\cite{Morro2004} and wave-splitting on
structural elements in~\cite{Johansson2006}. The square-root of a
certain operator is a key step in isotropic wave-splitting this
operator has been carefully studied in~\cite{HRB2003}. A reciprocity
theorem approach to decomposition is used by~\cite{vandenBerg2008}.
The results for anisotropic media
includes~\cite{deHon96,deHoopSquare,Jonsson+deHoop01}.

Applications of the existing wave-splitting techniques include several
successful analyzing tools of the wave-field including Bremmer series,
normal modes and uniform asymptotics. Another spin off is the
development of fast numerical codes to calculate the wave
fields. Among their implementations we have `rational approximations'
and `generalized screens' and
`multiple-forescattering-single-backscattering approximation'
\cite{vanStralen+deHoop+Blok98,deHoop+Jerome00,LeRousseau2003,Wu2007}.
In the active field of time-reversal mirrors see \eg
\cite{deRosnyFink2007,Yavuz2008}, the wave-splitting techniques have
been used see \eg \cite{Jonsson2004}. It is our hope that the
extension of the wave-splitting techniques to the inherently
anisotropic case will provide a base for generalization of the above
mentioned applications to analysis and fast numerical codes to general
anisotropic media.

The present paper is organized in a set of three propositions that
step by step introduce and prove the necessary properties and tools
to obtain the decomposition. The analysis is preformed in the
time-Laplace domain and the procedures impose limitations on the
Laplace parameter.  In \S \ref{sec2:motion} the problem is
formulated after a rewriting of Maxwell's equations to a suitable
form. In \S \ref{sec2:prop} the properties, mostly the spectral
properties, of the electromagnetic system's matrix are stated and
proved using functional analysis. The propositions impose only the
natural condition that the Laplace parameter has to belong to the
right-hand half plane of the complex space.  In \S \ref{sec2:P} the
splitting matrix is constructed and several of its properties are
shown. The analysis utilize that the material parameters are
self-adjoint, positive, and furthermore, require a mild constraint
on the Laplace parameter, in order to obtain a certain ellipticity
condition that is needed in the subsequent analysis. The most
important property shown in this section is that the generalized
eigenvectors of the splitting matrix can be obtained explicitly in
terms of the elements of the splitting matrix. In \S \ref{sec:5} the
decomposition is derived in terms of the generalized eigenvectors of
the splitting matrix.  The last section concludes with a discussion
and some observations.

Some lengthy intermediate derivations of the electromagnetic system's
matrix in \S \ref{sec2:motion} are detailed in
Appendix~\ref{app:calcA}. In Appendix~\ref{app:iso} the special case
of an isotropic homogeneous medium is treated using the approach
developed in the present paper and the results are compared with
traditional methods. In Appendix~\ref{app:det} the determinant of
the symbol of the electromagnetic systems matrix is given. Furthermore,
explicit integrations of the resolvent in symbol representation are
presented in terms of residue calculus and the integrals are stated
in terms of the roots of the determinant of the principal symbol of
the electromagnetic system's matrix.

\section{Directional wave-field decomposition}
\label{sec2:motion}

\subsection{The two-way equations for Maxwell equations}
\label{ssec2:reduced}
% Eeq:mr

We consider electromagnetic wave motion in heterogeneous anisotropic
media with instantaneous response. Let $x\in \RR^3$ be a point in
space and $t\in \RR$ is time. The media is assumed to be independent
of time. The following initial conditions of the fields ensures
causality,
\begin{eqnarray}\label{init}
    E(x,t)=0\; , \ D(x,t)=0 & \mbox{for}\ t<0 & \mbox{and all} \ x\in \RR^3 \; ,
  \\
    H(x,t)=0\; , \ B(x,t)=0 & \mbox{for}\ t<0 & \mbox{and all} \ x\in \RR^3 \; .
    \nonumber
\end{eqnarray}
where $B =$ magnetic flux density [T], $E =$ electric field strength
[V/m], $D=$ electric flux density [$\mbox{C/m}^{2}$] and $H=$
magnetic field strength [A/m]. The quantities $E,D,H,B$ are all
functions of space, $x\in \RR^3$ and time, $t\in \RR$, with values
in $\RR^3$, we return to which function spaces that they belong at a
latter point in the present paper. Above we have used standard units
and notation see \cite{ISOstandard93}.

The electromagnetic field satisfies the first-order hyperbolic
system of partial differential equations in time domain, Maxwell's
equations see \eg~\cite{ATdeHoop95,Jackson99}.
%\eg~\cite[pp.610-612]{ATdeHoop95},\cite[p.248]{Jackson99}.
We consider Maxwell's equations in time-Laplace domain. That is,
\begin{equation}
\begin{split}
        s B + \nabla \times E & =  K^{\rm e} \; , \\
        -s D + \nabla \times H & = J^{\rm e} \; .
\end{split}
    \label{Eeq:mr2}
\end{equation}
where $J^{\rm e}=$ external electric current density
[$\mbox{A/m}^{2}$] and $K^{\rm e}=$ external magnetic current
density [$\mbox{V/m}^{2}$]. The external currents are applied,
prescribed, sources. The causality of the field is taken into
account by requiring that all field quantities are bounded functions
of the time-Laplace parameter $s$, that is in general complex valued
and lies in the right-hand plane $\RE{s}>0$. With the specified
initial condition \eqref{init}, we have $\partial_{t} \rightarrow
s$. In this paper $x=\{x_{1},x_{2},x_{3}\}$ are right-handed
orthogonal Cartesian coordinates. All the subsequent analysis is
carried out in the domain of $(x,s)$ hence there is no need to
distinguish between the time dependent field, $E(x,t)$, and the
Laplace-parameter dependent field, $E(x,s)$.

To explicitly introduce the material parameters into the equations we
assume the following constitutive relations
\begin{equation}
\begin{split}
        B(x,s) & =  \rpm(x) \mu_{0} H(x,s)  \\
        D(x,s) & =  \rpe(x) \epsilon_{0} E(x,s)
\end{split}
    \label{Eeq:mr3}
\end{equation}
where $\rpm=$ relative anisotropic permeability tensor, $\rpe=$
relative anisotropic permittivity tensor, $\mu_{0}=$ empty space
permeability [H/m] and $\epsilon_{0}=$ empty space permittivity
[F/m]. The relative permeability and permittivity are assumed to be
self adjoint and positive definite $3 \times 3$ tensors of second
rank, that is, the media under consideration has only instantaneous
response. Inserting the constitutive relations into the Maxwell
equations gives
\begin{equation}
\begin{split}
        s \rpm \mu_{0} H + \nabla \times E & =  K^{\rm e} \; , \\
        -s \rpe \epsilon_{0} E + \nabla \times H & =  J^{\rm e} \; .
\end{split}
    \label{Eeq:mr4}
\end{equation}

Before proceeding we re-scale and change dimension of the equations
analogous to \eg \cite[p.10]{Gustafsson00} to simplify the
subsequent analysis:
\begin{eqnarray}
    c_{0}^{-2}:=\mu_{0}\epsilon_{0}  \;, \ \
    \as := \frac{s}{c_{0}} \; , \ \
    \acute{H} := \sqrt{\mu_{0}} H \; , \ \
    \acute{E} := \sqrt{\epsilon_{0}} E \; , \nonumber \\
    \acute{J}^{\rm e} := \sqrt{\mu_{0}}J^{\rm e} \; , \ \
    \acute{K}^{\rm e} := \sqrt{\epsilon_{0}}K^{\rm e} \; ,
    \label{Eeq:mr4a}
\end{eqnarray}
where $c_{0}$ is the speed of light in vacuum.
Upon substituting \eqref{Eeq:mr4a} into \eqref{Eeq:mr4},
\begin{equation}
\begin{split}
    \as \rpm \acute{H} + \nabla \times \acute{E} & =  \acute{K}^{\rm e} \; , \\
    -\as \rpe \acute{E} + \nabla \times \acute{H} & =  \acute{J}^{\rm e} \; .
\end{split}
    \label{Eeq:mr4b}
\end{equation}
All the following considerations refer to this transformed space and
for notational simplicity we remove the $\acute{\cdot}$, but
remember the change in dimension, in particular that $\as$ has
dimension m$^{-1}$, $\acute{H}$ and $\acute{E}$ have dimension
$(\mbox{J/(m}^3))^{1/2}$, $\acute{J}^{\rm e}$ and $\acute{K}^{\rm
e}$ has dimension $(\mbox{J/(m}^5))^{1/2}$. The transformation above
is for dimensional convenience, in particular in the calculus of
pseudodifferential operators see \S \ref{sec2:P}.

The `evolution' of the wave field in space, along a direction of
preference, can be expressed in terms of the change of the wave
field in the directions perpendicular to it. The direction of
preference  is taken to be along the $x_{3}$-axis (or `vertical'
axis) and the remaining (`horizontal') coordinates are denoted by
$x_{\alpha},x_{\beta}, \ \alpha,\beta \in \{1,2\}$ or
$x'=\{x_{1},x_{2}\}$ when convenient. The procedure requires a
separate treatment of the vertical components of $E$ and $H$. From
\eqref{Eeq:mr4b} we find the vertical field components to be
\begin{equation}
\begin{split}
        s \rpm_{33} H_{3} &= - s \rpm_{3\beta} H_{\beta} - (\nabla \times
        E)_{3} + K_{3}^{\rm e}  \; ,\\
        s \rpe_{33} E_{3} &= - s \rpe_{3\beta} E_{\beta} + (\nabla \times
        H)_{3} - J_{3}^{\rm e} \; ,
\end{split}
    \label{Eeq:mr5}
\end{equation}
where Einstein's summation convention for Cartesian tensors are
employed for repeated indices $\alpha,\beta\in \{1,2\}$, \eg
$\rpm_{3\beta} H_{\beta}= \sum_{\beta=1}^2 \rpm_{3\beta} H_{\beta}$.
To project out the third component of a vector we have used the
subscript 3, to explicitly show the notation consider
\begin{equation}
    (\nabla \times E)_{3} := \partial_{1}E_{2} - \partial_{2}E_{1} \; .
    \label{Eeq:m5b}
\end{equation}
Thus, \Eq \eqref{Eeq:mr5} relates the vertical components of the electric
and magnetic field strength to the horizontal components.
The remaining equations contained in \eqref{Eeq:mr4b} are
\begin{equation}
\begin{split}
        s \rpm_{\alpha \beta} H_{\beta} + s\rpm_{\alpha 3} H_{3} + (\nabla \times
        E)_{\alpha} & = K_{\alpha}^{\rm e}  \; ,\\
        -s \rpe_{\alpha \beta} E_{\beta} - s\rpe_{\alpha 3}E_{3} + (\nabla
        \times H)_{\alpha} & = J_{\alpha}^{\rm e} \; .
\end{split}
    \label{Eeq:mr6}
\end{equation}
By replacing $E_3$, $H_3$ in \eqref{Eeq:mr6} with \eqref{Eeq:mr5} we
arrive (the derivation is detailed in Appendix \ref{app:calcA}) to
\begin{equation}
    (I\partial_{3} + \Ac)F = N \; ,
    \label{Eeq:m7}
\end{equation}
where $N$ is a linear combinations of the sources and their
derivatives \cf \eqref{Eeq:mr14} and where the elements of the
electromagnetic field matrix, $F$, are given by
\begin{equation}
    F_{1} := E_{1} \; , \  F_{2} := -E_{2} \ \ \mbox{and} \ \ F_{3} := H_{2}
    \; , \ F_{4}:=H_{1} \; .
    \label{Eeq:m10}
\end{equation}
To simplify some of the following calculations we introduce the
notation
\begin{equation}
    \tilde{E}:=(E_{1},-E_{2})^{T} \ \ \mbox{and}\ \
    \tilde{H}:=(H_{2},H_{1})^{T} \; .
    \label{Eeq:m11}
\end{equation}
There are several possible orderings of the transverse components of
$E$, $H$ in $F$. The particular choice of combinations given in
\eqref{Eeq:m10} has two advantages. First, the given choice ensures
that both the matrix operators $\Ac_{12}$ and $\Ac_{21}$ are
invertible. Secondly, we have that the third component of the
Poynting vector, equals $(E \times \bar{H})_{3} =
\tilde{E}^{T}\overline{\tilde{H}}=
(F_{1},F_{2})(\bar{F}_{3},\bar{F}_{4})^T$, where $\bar{\cdot}$
denotes the complex conjugate. The 4x4 electromagnetic system's
matrix, $\Ac$, is here represented by four 2x2 blocks
\begin{equation}
    \Ac=\begin{pmatrix}
       \Ac_{11} & \Ac_{12} \\        \Ac_{21} & \Ac_{22}
    \end{pmatrix}
\end{equation}
where each block-matrix is given by
\begin{equation}
\begin{split}
    \Ac_{11} & =  \rpm_{33}^{-1}\left( \begin{array}{rr}
            \rpm_{23}\partial_{2} & \rpm_{23} \partial_{1} \\
            \rpm_{13}\partial_{2} & \rpm_{13}\partial_{1}
        \end{array} \right)  +
        \left( \begin{array}{rr}
            \partial_{1}\rpe_{31} & -\partial_{1}\rpe_{32} \\
            -\partial_{2}\rpe_{31} & \partial_{2}\rpe_{32}
        \end{array} \right) \rpe_{33}^{-1}  \; , \\ \\
    \Ac_{12} & =  s \left(\begin{array}{rr}
        \nu_{22} & \nu_{21} \\  \nu_{12} & \nu_{11} \end{array}\right) -
        s^{-1} \left(\begin{array}{rr}
             \partial_{1}\rpe_{33}^{-1} \partial_{1} &
            -\partial_{1}\rpe_{33}^{-1} \partial_{2} \\
            -\partial_{2}\rpe_{33}^{-1} \partial_{1} &
             \partial_{2}\rpe_{33}^{-1} \partial_{2}
        \end{array}\right)  \; , \\ \\
    \Ac_{21} & =  s\left(\begin{array}{rr}
            \varepsilon_{11} & -\varepsilon_{12} \\
            -\varepsilon_{21} & \varepsilon_{22}
        \end{array}\right) - s^{-1}\left(\begin{array}{rr}
            \partial_{2}\rpm_{33}^{-1}\partial_{2} &
            \partial_{2}\rpm_{33}^{-1}\partial_{1} \\
            \partial_{1}\rpm_{33}^{-1}\partial_{2} &
            \partial_{1}\rpm_{33}^{-1}\partial_{1}
        \end{array}\right)  \; , \\ \\
    \Ac_{22} & =  \left(\begin{array}{rr}
            \partial_{2}\rpm_{32} & \partial_{2}\rpm_{31} \\
            \partial_{1}\rpm_{32} & \partial_{1}\rpm_{31}
        \end{array}\right) \rpm_{33}^{-1} +
        \rpe_{33}^{-1}\left(\begin{array}{rr}
            \rpe_{13}\partial_{1} & -\rpe_{13}\partial_{2} \\
            -\rpe_{23}\partial_{1} & \rpe_{23}\partial_{2}
        \end{array}\right) \; ,
\end{split}
    \label{Eeq:m13}
\end{equation}
in which
\begin{equation}
\begin{split}
        \varepsilon_{\alpha\beta} &= \rpe_{\alpha\beta} -
        \rpe_{\alpha 3}\rpe_{33}^{-1}\rpe_{3\beta} \; , \\
        \nu_{\alpha\beta} & =  \rpm_{\alpha\beta} - \rpm_{\alpha 3} \rpm_{33}^{-1}
        \rpm_{3 \beta} \; .
\end{split}
    \label{Eeq:m14}
\end{equation}
The permeability and permittivity are symmetric, $3 \times 3$
tensors of rank (tensor order) 2 that are bounded from below and
from above. Hence the upper-left $2 \times 2$ matrices of $\rpe$ and
$\rpm$ are bounded below by the constants $\lowEz$ and $\lowMz$
respective. To show the notation we have
\begin{equation}
    u_{\alpha}\rpe_{\alpha\beta}\bar{u}_{\beta} \geq
    \lowEz u_{\alpha}\bar{u}_{\alpha} \; ,
    \label{Eeq:m16}
\end{equation}
for any complex field $u=(u_{1},u_{2})$. From the definitions of
$\rE_{\alpha\beta}$ and $\rM_{\alpha\beta}$ it is clear that they
are symmetric matrices (since $\rE$ and $\rM$ are symmetric and all
elements are real valued. Furthermore, each is bounded below by the
constants $\lowEu$ and $\lowMu$ respectively. This follows from the
identity (summation over repeated index $\alpha,\beta\in \{1,2\}$
and $j,k\in \{1,2,3\}$)
\begin{equation}
    v_j \rpe_{jk} \bar{v}_k =
    u_{\alpha} \varepsilon_{\alpha\beta}\bar{u}_{\beta} \; ,
    \label{Eeq:m17}
\end{equation}
where
\begin{equation}
    v_{i} = u_{\alpha}(\delta_{i \alpha} -
    \rpe_{33}^{-1}\rpe_{3\alpha}\delta_{3 i}) \; ,\ \mbox{for} \
    i = \{1,2,3\} \; ,
    \label{Eeq:m18}
\end{equation}
for any complex field $u$. Since $\rpe$ is positive definite
$\varepsilon$ must also be positive definite and analogously for $\rpm$ and $\nu$.

\subsection{Preliminaries}

We consider the electromagnetic system's matrix and other operators
on Sobolev spaces. Let $\setSobolev{r}(\RR^2;\CC)$ be the set of
functions belonging to the Sobolev space of order $r\in \mathbb{N}$
with domain in $\mathbb{R}^{2}$ and values in $\CC$, with a weighted
inner product to compensate for the dimension of the derivative. To
extend this scalar space to vectors we introduce the notation
\begin{equation}
    \mSob{r} :=  \setSobolev{r}(\RR^2;\CC^4)
\end{equation}
for a $4 \times 1$ matrix, with each element in the set of functions
belonging to the Sobolev space of order $r$. Let $F =
(\tilde{E}^{\rm a},\tilde{H}^{\rm a})$ and $G = (\tilde{E}^{\rm
b},\tilde{H}^{\rm b})$. Then we define the inner product to be
%\begin{equation}
\begin{multline}
    \mSdot{r}{F}{G} = \int_{\mathbb{R}^{2}} \intd[2]{x'}
    \sum_{|k|\leq r} y_{0}^{2|k|}\left(
    \partial_{x'}^k \overline{\tilde{E}_{1}^{\rm a}} \
    \partial_{x'}^k
    \tilde{E}_{1}^{\rm b} +
    \partial_{x'}^k \overline{\tilde{E}_{2}^{\rm a}} \
    \partial_{x'}^k
    \tilde{E}_{2}^{\rm b} \right. \nonumber
    \\
    + \left.
    \partial_{x'}^k \overline{\tilde{H}_{1}^{\rm a}} \
    \partial_{x'}^k
    \tilde{H}_{1}^{\rm b} +
    \partial_{x'}^k \overline{\tilde{H}_{2}^{\rm a}} \
    \partial_{x'}^k
    \tilde{H}_{2}^{\rm b}
    \right) \; ,
%   \label{Eeq:sd1}
\end{multline}
%\end{equation}
where $y_{0}$ is a constant of dimension length and it is used to
normalize the change of dimension from the derivatives. All
components of the field depend on $x$, but inner products and norms
refer to $x'$ and we treat $x_{3}$ as a parameter. We have adopted
the multi-index notation of pseudodifferential
calculus~\cite{Shubin87} above and use $k\in \mathbb{N}^{2}$
together with
\begin{equation}
    |k|=k_{1}+k_{2} \; .
    \label{Eeq:sd2}
\end{equation}
The norm corresponding to the inner product is
\begin{equation}
    \mSnorm{r}{F} = \sqrt{\mSdot{r}{F}{F}} \; .
    \label{Eeq:sd3}
\end{equation}
The set $\mSob{r}$ with the inner product $\mSdot{r}{\cdot}{\cdot}$
is a Hilbert space. For the case $r=0$ we recover the Lebesgue space
of square integrable functions, $\msLs$.

The analysis of unbounded operators --- such as the electromagnetic
system's matrix --- requires that one specifies the domain of the
operator and its embedding space. Below we consider operators on the
space $\msLs$, that is $\Ac: \msLs \rightarrow \msLs$. Since the
operator is an unbounded operator, we also need to specify its
domain, which is $\mSob{2}$. The domain of the operator on a space
is fundamental for the analysis. Here all operators have dense
domains and when necessary, with restrictions to dense subsets of
the their domains for the operation under consideration to be
defined. In the case of such a restriction we use the notation
$\left. \Ac\right|_{q}$ for the operator restricted to this dense
subset of its domain and indicate by $q$ what dense subset is
understood to be the restricted domain.

One can also consider $\Aco$ as an operator on
$\{\mSob{r},\mSdot{r}{\cdot}{\cdot}\}$, (if $r>0$ this is a
restriction of the operator defined above) with domain $\mSob{r-2}$,
and the analysis extends trivially to this case. An alternative
method was detailed in~\cite{Jonsson+deHoop01}, where one instead of
$\Aco$ consider the operator $\gamma^{-r}\Aco\gamma^r$, where
$\gamma^r:=(1-y_0^2\partial_\alpha\partial_\alpha)^{r/2}$. The
results in this paper hold also for this class of operators.

\subsection{Formulation of the problem}

To be able to solve the scattering process along the vertical
direction separately from the scattering process in the horizontal
directions, we diagonalize the operator on the left-hand side of
\eqref{Eeq:m7}. This procedure will possibly lead to an additional
source term on the right-hand side that accounts for the coupling.
To achieve this, we construct a linear operator $\OP{L}$ which
convert two-way fields $F$ to one-way field constituents $W$, by
\begin{equation}
    F = \OP{L} W \; .
    \label{Eeq:4.17}
\end{equation}
We require that $\OP{L}$ when introduced into \eqref{Eeq:m7} gives,
\begin{equation}
   \OP{L} \, (I \partial_3  +
   \OP{V}) W =
   - (\partial_3 \OP{L}) \, W + N
\label{Eeq:4.18}
\end{equation}
so as to make $\OP{V}$, defined by
\begin{equation}
   \Ac \OP{L} = \OP{L} \OP{V} \; ,
\label{Eeq:4.19}
\end{equation}
a block diagonal matrix of operators. We call $\OP{L}$ the
composition operator, and $W$ the wave matrix. The elements of the
wave matrix represent locally the down- and up-going constituents.
The expression in parentheses on the left-hand side of
\eqref{Eeq:4.18} represents the two so-called {\em one-way} wave
operators. The first term on the right-hand side of \eqref{Eeq:4.18}
represents the scattering due to variations of the medium properties
in the vertical direction. The scattering due to variations of the
medium properties in the horizontal directions is contained in
$\OP{V}$ and, implicitly, in $\OP{L}$ also.

To investigate whether solutions $\{\OP{L},\OP{V}\}$ of
\eqref{Eeq:4.19} exist, we introduce the column matrices, or
generalized eigenvectors, $\OP{L}^{\pm}$, according to
\begin{equation}
   \OP{L} = \left(\begin{array}{cc} \OP{L}^{+} &  \OP{L}^{-}
                    \end{array} \right) \; .
\label{Eeq:4.20}
\end{equation}
Upon writing the block diagonal elements of $\OP{V}$ (generalized
eigenvalues)  as
\begin{equation}
   \OP{V} = \left(\begin{array}{cc} \WP & 0 \\ 0 & \WM \end{array}
   \right) \; .
\label{Eeq:4.23}
\end{equation}
Eqn.~\eqref{Eeq:4.19} decomposes into the two systems of equations
\begin{equation}
   \Ac \OP{L}^{\pm} = \OP{L}^{\pm} \wpm \; ,
\label{Eeq:4.24}
\end{equation}
where $\wpm$ are $2 \times 2$ matrices.
The central problem that we consider in the present paper is
to show that there exists an operator pair, $\{\OP{L},\OP{V}\}$, such that the
above operator equation, \eqref{Eeq:4.19}, is satisfied. Since the
operators are unbounded we need to modify \eqref{Eeq:4.19} and
\eqref{Eeq:4.24} with respect
to the domain of the respective operator.

Note that the upper $2 \times 2$ matrix of the operators $\OP{L}^{\pm}$
combines the transverse electric field strength, and the
lower, the magnetic field
strength, whereas the elements of $W$ may be physically `non-observable'.

We now focus on the fundamental question: Does there exist a
composition operator $\oL$ that decomposes $\Ac$ in the above sense?
To begin to show that there exists such a decomposition of $\Ac$, we
derive properties of the resolvent set of $\Ac$ which enable us to
define a certain operator which commute with $\Ac$.

\section{Properties of the $\Aco$ operator}
\label{sec2:prop}

In this section we show that the directional decomposition of the
electromagnetic field is closely related to the spectral properties
of the operator $\Ac$. The definition of the splitting matrix
requires that there exists a region around the imaginary axis which
is free from the spectrum. We therefore state the definition of the
spectra explicitly. Consider first the operator $\Aco$ defined on
$\msLs$ with domain $\mSob{2}$:
\begin{equation}
   \Aco = \Ac - \lambda I \ :\ \mLs \to \mLs \; .
   \label{Eeq:pao1}
\end{equation}
Following \cite[\S 5, p.253]{Taylor58}, \cite[\S 6.5,
p.412]{Naylor+Sell82} and \cite[\S VIII.1, p.209]{Yosida80}, we
define the spectrum of $\Ac$, for fixed $s$ as follows: if the
scalar $\lambda\in \mathbb{C}$ is such that the range of $\Aco$ is
dense in $\msLs$ and $\Aco$ has a bounded inverse, $\lambda$ is in
the \textit{resolvent set}, $\set{P}(\Ac)$, of $\Ac$, and we denote
this inverse by $\Aco^{-1}$ and call it the \textit{resolvent} (at
$s,\lambda$) of $\Ac$. All complex numbers not in the resolvent set
form a set $\set{\Sigma}(\Ac)$ called the \textit{spectrum} of
$\Ac$.

To simplify some of the upcoming calculations we use the notations
\begin{equation}
    s = \rS \lexp{\iu \sigma} = \rS \cos{\sigma} + \iu \rS \sin{\sigma} \ \
    \mbox{and} \ \ \lambda = \RL + \iu \IL \; .
    \label{Eeq:c7a}
\end{equation}
The following
proposition gives as a corollary that there exists a
strip that belongs to the resolvent set of $\Ac$.
\begin{prop} \label{prop2:spec}
    Let $\Aco = \Ac - \lambda$, be defined through \eqref{Eeq:m13} and
    \eqref{Eeq:pao1}.
    Let
    \begin{displaymath}
        \set{Q} := \left\{ \{s,\lambda\}\in \mathbb{C}^{2} : \RE{s}>0\
        \mbox{and}\ (\RE{\lambda})^{2}<(\RE{s})^{2} \lowEu\lowMu
        \right\} \; ,
    \end{displaymath}
    where $\lowEu$ and $\lowMu$ are defined in \S \ref{ssec2:reduced}.
    Then for $\{s,\lambda\}\in \set{Q}$, $\Aco$
    \begin{enumerate}
        \item  is bounded from below;

        \item  is one-to-one;

        \item  has dense range;

        \item  is closable;

        \item  has an inverse:
    \begin{displaymath}
    \Ac^{-1}=\left(\begin{array}{cc}
        -\Ac_{21}^{-1}\left(\Ac_{22}-\lambda\right)\EL^{-1} &
        \Ac_{21}^{-1}+\Ac_{21}^{-1}\left( \Ac_{22}-\lambda \right)
        \EL^{-1}\left(\Ac_{11}-\lambda\right)\Ac_{21}^{-1} \\
        \EL^{-1} &
        -\EL^{-1}\left(\Ac_{11}-\lambda\right)\Ac_{21}^{-1}
    \end{array}\right) \; ,
    \end{displaymath}
        where
        \begin{displaymath}
           \EL := \Ac_{12} - (\Ac_{11} - \lambda) \Ac_{21}^{-1} (\Ac_{22} -
           \lambda) \; .
        \end{displaymath}
    \end{enumerate}
\end{prop}
\begin{remark}\label{rem2:closed}
    The Hilbert identity or the (first) resolvent equation,
    \begin{displaymath}
        \Aco^{-1} - \Ac_{s,\lambda'}^{-1} = (\lambda - \lambda')
        \Aco^{-1} \Ac_{s,\lambda'}^{-1} \; ,
%       \label{Eeq:def12-sss}
    \end{displaymath}
    for $\{s,\lambda\},\{s,\lambda'\}\in \set{Q}$. If the operator is closed
    and if $\{s,\lambda\} \in \set{Q}$ then the resolvent is well
    defined and it is an analytic function of $\lambda$
    (\cf~\cite[\S III.6.1, pp.172-174]{Kato80},
    \cite[\S VIII.2, pp.211-212]{Yosida80} and
    \cite[p.84 \S 3.7.5]{Birman+Solomjak87}).
\end{remark}
\begin{remark}
    The underlying requirement of self-adjoint material parameters
    can be replaced by positivity of the real part of the
    eigenvalues of the two matrices $s\mu$, $s\epsilon$ in the case
    of up/down symmetric materials, i.e. when $\mu_{3\alpha}=\mu_{\alpha 3} =0$ and
    $\epsilon_{3\alpha}=\epsilon_{\alpha 3}=0$, for $\alpha=1,2$.
    This follows directly from the proof of part 1. It is not clear that
    this extension is valid for the general anisotropic case, we do not pursue this since
    our proof of Proposition \ref{prop2:split2} makes use of
    the self adjoint property of $\epsilon,\mu$.
\end{remark}
From Proposition \ref{prop2:spec} it directly follows that:
\begin{cor} \label{cor2:spec}
    For any fixed $s\in\mathbb{C}$ such that $\RE{s}>0$
    the resolvent set of the electromagnetic system's matrix contains
    the strip of all
    $\lambda\in\mathbb{C}$ such that
    \begin{equation}
        (\RE{\lambda})^{2}<(\RE{s})^{2} \lowEu\lowMu \; .
    \label{Eeq:ccc1}
    \end{equation}
    That is, the strip belongs to the resolvent set, $\set{P}(\Ac)$.
\end{cor}
\begin{remark}\label{spec:sym}
    Let $\Ac^*$ be the adjoint of $\Ac$ with respect to the standard
    inner product in $\mLs$, denote the spectrum of $\Ac$
    by $\Sigma(\Ac)$, and recall the relation
    $\Sigma(\Ac)=\overline{\Sigma(\Ac^*)}$. Let
    \begin{equation}
        J = \begin{pmatrix} 0 & I \\ -I & 0 \end{pmatrix},
    \end{equation}
    where $I$ is $2\times 2$ unit matrices. From \eqref{ssym} we note that $\Ac$
    satisfy the identity
    \begin{equation}
        \Ac(s)-\lambda = J^{-1}(\Ac(\bar{s}))^*J^* - \lambda =
            J(-(\Ac(\bar{s}))^*-\lambda)J^* \; ,
    \end{equation}
    and consequently when $\lambda\in\Sigma(\Ac(s))$ then $-\bar{\lambda}\in
    \Sigma(\Ac(\bar{s}))$.
\end{remark}

\begin{proof}[Proof of Corollary \eqref{cor2:spec}]
Given $\RE{s}>0$, and the strip \eqref{Eeq:ccc1} we find such
$(s,\lambda)\in \set{Q}$, and hence $\Ac$ satisfy the properties
1--3 of Proposition \ref{prop2:spec}. These properties are the
conditions needed for a point $\lambda$ to be in the resolvent set.
\end{proof}

\subsection{Properties of a quadratic form}
To prove Proposition~\ref{prop2:spec} we introduce some properties
on an auxiliary quantity, a quadratic form, $\OP{C}_{s,\lambda}[F]$,
defined in Lemma~\ref{lem2:sesq}. Using the notation introduced in
Proposition~\ref{prop2:spec} and in the definition of the norm
in~\eqref{Eeq:sd3}, we have:
\begin{lemm2}\label{lem2:sesq}
    Let $\{s,\lambda\} \in \set{Q}$ and define the quadratic form
    \begin{multline*}
        \OP{C}_{s,\lambda}[F] := \int_{\mathbb{R}^2}
    s \varepsilon_{\alpha\beta}
    \bar{E}_{\alpha} E_{\beta} +
    s^{-1}\mu_{33}^{-1}\left| (\nabla
    \times E)_{3} \right|^2  \\
+   s
    \nu_{\alpha\beta}\bar{H}_{\alpha}H_{\beta} + s^{-1}
    \epsilon_{33}^{-1} \left|(\nabla \times H )_{3}\right|^2  -
    2\lambda \RE{E_{1}\bar{H}_{2}-E_{2}\bar{H}_{1}}
+ \\
    2 \iu \mu_{33}^{-1}
    \IM{(\nabla \times E)_{3} \mu_{3\alpha}\bar{H}_{\alpha}}  +
    2 \iu \epsilon_{33}^{-1} \IM{\epsilon_{\alpha 3}\bar{E}_{\alpha}
    (\nabla \times H)_{3}}  \, \intd[2]{x'} \; .
    \end{multline*}
    where $F=(\tilde{E},\tilde{H})^{T}$.
    Then $\OP{C}_{s,\lambda}$ is well defined for $F \in \mSob{1}$ and
    \begin{displaymath}
        |\OP{C}_{s,\lambda}\adjoint{\tilde{H}}[F]| \geq C_{0}(s,\lambda )\mLnorm{F}^{2} \; ,
    \end{displaymath}
    where $C_{0}(s,\lambda)>0$ for $\{s,\lambda\}\in \set{Q}$.
\end{lemm2}

\begin{proof}
That $\OP{C}_{s,\lambda}$ is well defined for $F\in \mSob{1}$ is
clear as the quadratic form contain at most one derivative for each
field component in each term. To see that $\OP{C}_{s,\lambda}$ is
bounded from below we
take the real part of the integrand and
using the notation introduced in \eqref{Eeq:c7a} we obtain
\begin{multline}
(\rS \cos \sigma) \varepsilon_{\alpha\beta}
    \bar{E}_{\alpha} E_{\beta} + (\rS^{-1} \cos \sigma) \mu_{33}^{-1}\left| (\nabla
    \times E)_{3} \right|^2
+   (\rS \cos \sigma)
    \nu_{\alpha\beta}\bar{H}_{\alpha}H_{\beta}  \\ + (\rS^{-1} \cos
    \sigma)
    \epsilon_{33}^{-1} \left|(\nabla \times H )_{3}\right|^2
      -
    2\RL \RE{E_{1}\bar{H}_{2}-E_{2}\bar{H}_{1}}  \; ,
    \label{Eeq:qd2}
\end{multline}
due to that $\varepsilon$ and $\nu$ are self-adjoint. Since
$\{s,\lambda\} \in \set{Q}$, it follows that $\cos \sigma >0$ and we
have
\begin{equation}
    \mbox{\eqref{Eeq:qd2}} \geq
\rS \cos \sigma \left(\varepsilon_{\alpha\beta}
    \bar{E}_{\alpha} E_{\beta}
+   \nu_{\alpha\beta}\bar{H}_{\alpha}H_{\beta} \right) -   2\RL
\RE{E_{1}\bar{H}_{2}-E_{2}\bar{H}_{1}}  \; .
     \label{Eeq:qd3}
\end{equation}
For all $\eta>0$ we have the inequality
\begin{equation}
    2 \RL \RE{E_{1}\bar{H}_{2}-E_{2}\bar{H}_{1}}  =  2 \RL
    \RE{\tilde{E}^{T}  \overline{\tilde{H}}} \nonumber  \leq  \eta
    \lowEu |\tilde{E}|^2  \rS \cos \sigma +
    \frac{\RL^{2}}{\eta \lowEu \rS \cos \sigma} |\tilde{H}|^{2} \; ,
\end{equation}
which implies
\begin{equation}
    \mbox{\eqref{Eeq:qd3}} \geq
    \lowEu\rS \cos \sigma (1-\eta)|\tilde{E}|^2 +
    \left(\lowMu \rS \cos \sigma - \frac{\RL^{2}}{\eta \lowEu
    \rS \cos{\sigma}}\right)|\tilde{H}|^2 \; .
    \label{Eeq:qd4}
\end{equation}
Thus we require that $\eta \in (0,1)$. The largest $|\RL|$-strip is
obtained
in the limit $\eta \rightarrow 1$, thus
\begin{equation}
    \RL^2 < (\RE{s})^{2} \lowEu\lowMu \; .
    \label{Eeq:qd5}
\end{equation}
Hence for given fixed $s$ such that $\RE{s}>0$ and for a fixed $\lambda$ that
fulfils \eqref{Eeq:qd5}, there exists an optimal $\eta$ such
that the bound from below in \eqref{Eeq:qd4} is maximal. Thus the best
choice of bound from below with the given estimates is
\begin{equation}
    C_{0}(s,\lambda) = \max_{0<\eta<1} \min
    \left( \RE{s}\lowEu(1-\eta),
    \left(\lowMu \RE{s} - \frac{\RL^{2}}{\eta \lowEu
    \RE{s}}\right)\right) >0 \; .
\end{equation}
We note that the argument above is a continuous function of $\eta$
on $[0,1]$, and the maximal-value is attained (no need for $\sup$)
in the interior of the interval. Furthermore, the optimal value
$\eta$ for which the $C_0(s,\lambda)$ is obtained as a solution of a
second order equation in $\eta$, but here it suffices to know that
it exists and is positive. Upon integration we find that
\begin{equation*}
    |\OP{C}_{s,\lambda}[F]| \geq C_{0}(s,\lambda) (\mLnorm{\tilde{E}}^2 +
    \mLnorm{\tilde{H}}^2) = C_{0}(s,\lambda)\mLnorm{F}^2 \; .
    \qedhere
\end{equation*}
\end{proof}

\subsection{Proof of Proposition \ref{prop2:spec}, part 1}

To start the proof that the operator $\Aco$ is bounded from below, we
employ Schwartz' inequality
\begin{equation}
    \mLnorm{\Aco F} \mLnorm{F} = \mLnorm{K\Aco F}\mLnorm{F} \geq
    \left| \mLdot{F}{K\Aco F} \right| \; ,
    \label{Eeq:bb0}
\end{equation}
with $F \in \domain{\Ac} = \mSob{2}$ and
where
\begin{equation}
    K = \left(  \begin{array}{cc} 0 & I \\ I & 0 \end{array}\right)  \; ,
    \label{Eeq:bb0b}
\end{equation}
and $I$ is a $2\times 2$-matrix. $K$ is a unitary for
$\mLnorm{\cdot}$. On the block element level where $F =
(\tilde{E},\tilde{H})$ (\cf~\eqref{Eeq:m11} we have
\begin{equation}
    \bar{F}_{q}(K(\Ac-\lambda) F)_{q}  =
%    \label{Eeq:bb1}
    \nonumber
    \adjoint{\tilde{E}} \Ac_{21} \tilde{E} +
     \adjoint{\tilde{H}} \Ac_{12} \tilde{H} -
     2\lambda \RE{\adjoint{\tilde{E}}\tilde{H}}
     + \adjoint{\tilde{E}} \Ac_{22} \tilde{H} +
     \adjoint{\tilde{H}} \Ac_{11}\tilde{E} \; ,
\end{equation}
where the repeated index $q$ indicates summation over the four
components and $\adjoint{\tilde{E}}$ = $(\overline{\tilde{E}})^{T}$.
The first term becomes after integration over $x'$ and integration
by parts,
\begin{equation}
    \int_{\mathbb{R}^2} \adjoint{\tilde{E}}\Ac_{21}\tilde{E} \, \intd[2]{x'}=
    \int_{\mathbb{R}^2} s \varepsilon_{\alpha\beta}
    \bar{E}_{\alpha} E_{\beta} + s^{-1}\mu_{33}^{-1}\left| (\nabla
    \times E)_{3} \right|^2 \, \intd[2]{x'}
\end{equation}
and the second becomes
\begin{equation}
    \int_{\mathbb{R}^2} \adjoint{\tilde{H}}\Ac_{12}\tilde{H} \, \intd[2]{x'}=
    \int_{\mathbb{R}^2} s
    \nu_{\alpha\beta}\bar{H}_{\alpha}H_{\beta} + s^{-1}
    \epsilon_{33}^{-1} \left|(\nabla \times H )_{3}\right|^2 \, \intd[2]{x'} \; .
\end{equation}
The fourth term becomes after simplification and integration by parts
\begin{equation}
    \int_{\mathbb{R}^2} \adjoint{\tilde{E}}\Ac_{22}\tilde{H} \, \intd[2]{x'}
    = \int_{\mathbb{R}^2}
    \epsilon_{33}^{-1} \epsilon_{\alpha 3}\bar{E}_{\alpha}
    (\nabla \times H)_{3}
    -(\nabla \times
    \bar{E})_{3} \mu_{33}^{-1}\mu_{3\alpha}H_{\alpha}
    \, \intd[2]{x'} \; ,
    \label{Eeq:bb2a}
\end{equation}
and the last term
\begin{equation}
    \int_{\mathbb{R}^2} \adjoint{\tilde{H}} \Ac_{11} \tilde{E} \, \intd[2]{x'} =
    \int_{\mathbb{R}^2} (\nabla \times E)_{3}
    \mu_{33}^{-1} \mu_{\alpha 3}\bar{H}_{\alpha}
    - \epsilon_{33}^{-1}\epsilon_{3\alpha} E_{\alpha} (\nabla
    \times \bar{H})_{3} \, \intd[2]{x'} \; .
    \label{Eeq:bb3a}
\end{equation}
Since $\epsilon$ and $\mu$ are self-adjoint matrices the two terms
in \eqref{Eeq:bb2a} and \eqref{Eeq:bb3a} combine to
\begin{multline}
    \int_{\mathbb{R}^2} 2 \iu \IM{\adjoint{\tilde{E}}\Ac_{22}\tilde{H}} \, \intd[2]{x'} =
    \int_{\mathbb{R}^2} 2 \iu \mu_{33}^{-1}
    \IM{(\nabla \times \bar{E})_{3} \mu_{\alpha 3}H_{\alpha}} \\ +
    2 \iu \epsilon_{33}^{-1} \IM{\epsilon_{\alpha 3}\bar{E}_{\alpha}
    (\nabla \times H)_{3}} \, \intd[2]{x'} \; .
\end{multline}
Thus,
\begin{equation}
    \left| \mLdot{F}{K\Ac F} \right| = |\OP{C}_{s,\lambda}[F]| \; ,
    \label{Eeq:bb2}
\end{equation}
for $F\in \domain{\Aco}\subset \mSob{1}$. By Lemma \ref{lem2:sesq}
and \eqref{Eeq:bb0} we obtain
\begin{equation}
    \mLnorm{\Ac F} \geq C_{0}(s,\lambda) \mLnorm{F} \; ,
    \label{Eeq:bb3}
\end{equation}
where $C_{0}(s,\lambda)$ is defined in the lemma, and
$C_{0}(s,\lambda)>0$ for $\{s,\lambda\}\in \set{Q}$. \qed%\endproof

\subsection{Proof of Proposition \ref{prop2:spec}, part 2}
The inequality
\begin{equation}
    \mLnorm{\Aco F}\geq C_{0}(s,\lambda)\mLnorm{F}\; ,
\end{equation}
with $C_0>0$ from part 1, implies that the null space only contains
the zero element. By \cite[p.171, theorem 4.4.1]{Naylor+Sell82} an
operator with trivial null space is one-to-one (injective). Hence,
the operator $\Aco$ is one-to-one for $\{s,\lambda\}\in \set{Q}$.
\qed

\subsection{Proof of Proposition \ref{prop2:spec}, part 3}
Let $\Aco^*$ denote the adjoint of $\Aco$ on $\mLs$. To show that
the operator has dense range it is sufficient to show that the
kernel of $\Aco^{*}$ is trivial. That is,
\begin{equation}
    \Aco^{*}G = 0 \ , \ G \in \domain{\adjoint{\Ac}}\ \Rightarrow\ G=0
    \; ,
\end{equation}
thus if $\Aco^{*}$ is bounded from below then the desired result
follows directly, \cf Proposition~\ref{prop2:spec}, part~2.

The adjoint of $\Aco$ with respect to the inner product
$\mLdot{\cdot}{\cdot}$ is
\begin{equation}
    \Aco^{*} = \left( \begin{array}{cc}
        \Ac_{11}^{*}-\bar{\lambda}I & \Ac_{21}^{*} \\
        \Ac_{12}^{*} & \Ac_{22}^{*}-\bar{\lambda} I
    \end{array} \right) \; ,
\end{equation}
where
\begin{align*}
    \Ac_{11}^{*} & =  - \left( \begin{array}{rr}
            \partial_{2}\bar{\mu}_{23} & \partial_{2}\bar{\mu}_{13}  \\
            \partial_{1}\bar{\mu}_{23} & \partial_{1}\bar{\mu}_{13}
        \end{array} \right)\mu_{33}^{-1}  -
         \epsilon_{33}^{-1} \left( \begin{array}{rr}
            \bar{\epsilon}_{31}\partial_{1} & -\bar{\epsilon}_{31}\partial_{2} \\
            -\bar{\epsilon}_{32}\partial_{1} & \bar{\epsilon}_{32}\partial_{2}
        \end{array} \right) \; , \\
    \Ac_{12}^{*} & =  \bar{s} \left(\begin{array}{rr}
        \bar{\nu}_{22} & \bar{\nu}_{12} \\  \bar{\nu}_{21} & \bar{\nu}_{11} \end{array}\right) -
        \bar{s}^{-1} \left(\begin{array}{rr}
             \partial_{1}\bar{\epsilon}_{33}^{-1} \partial_{1}  &
            -\partial_{1}\bar{\epsilon}_{33}^{-1} \partial_{2}  \\
            -\partial_{2}\bar{\epsilon}_{33}^{-1} \partial_{1}  &
             \partial_{2}\bar{\epsilon}_{33}^{-1} \partial_{2}
        \end{array}\right)  \; , \\
    \Ac_{21}^{*} & =  \bar{s}\left(\begin{array}{rr}
            \bar{\varepsilon}_{11} & -\bar{\varepsilon}_{21} \\
            -\bar{\varepsilon}_{12} & \bar{\varepsilon}_{22}
        \end{array}\right) - \bar{s}^{-1}\left(\begin{array}{rr}
            \partial_{2}\bar{\mu}_{33}^{-1}\partial_{2} &
            \partial_{2}\bar{\mu}_{33}^{-1}\partial_{1} \\
            \partial_{1}\bar{\mu}_{33}^{-1}\partial_{2} &
            \partial_{1}\bar{\mu}_{33}^{-1}\partial_{1}
        \end{array}\right)  \; , \\
    \Ac_{22}^{*} & =  -\mu_{33}^{-1} \left(\begin{array}{rr}
            \bar{\mu}_{32}\partial_{2} & \bar{\mu}_{32}\partial_{1}  \\
            \bar{\mu}_{31}\partial_{2} & \bar{\mu}_{31}\partial_{1}
        \end{array}\right)  -
        \left(\begin{array}{rr}
            \partial_{1}\bar{\epsilon}_{13} & -\partial_{1}\bar{\epsilon}_{23} \\
            -\partial_{2}\bar{\epsilon}_{13} & \partial_{2}\bar{\epsilon}_{23}
        \end{array}\right)\epsilon_{33}^{-1}  \; ,
\end{align*}
where we have used that $\epsilon$, $\mu$ are self adjoint, and
hence that their diagonals are real-valued.  The domain of the
adjoint is the set
\begin{multline}
    \domain{\adjoint{\Aco}}  = \{ \ G \in \mLs :\\  \exists H \in \mLs \ \mbox{such that} \
     \mLdot{H}{F} = \mLdot{G}{\Aco F}\ \ \forall F \in
    \domain{\Aco}\  \} \; .
\end{multline}
To show that $\adjoint{\Aco}$ is bounded from below we will use the
same method as in Proposition~\ref{prop2:spec}, part~1. First we
need a small enough set that contains the domain; from the form of
$\adjoint{\Aco}$ it is clear that if $G=\adjoint{\Aco}F \in \mLs$,
then at least $F\in \mSob{1}$. Thus the domain is contained in
$\mSob{1}$, that is
\begin{equation}
    \domain{\adjoint{\Aco}}\subset \mSob{1} \; .
    \label{Eeq:den1}
\end{equation}
To obtain the quadratic form needed to use Lemma \ref{lem2:sesq}, we
once again use the Schwartz estimate. Let $G\in
\domain{\adjoint{\Aco}}$, and $F:=KG$, where the matrix $K$ was
introduced in \eqref{Eeq:bb0b} and is unitary on
$\mLdot{\cdot}{\cdot}$ and has the properties $K=\adjoint{K}=K^{-1}$
and $\mLnorm{KG}=\mLnorm{G}$. Similarly to~\eqref{Eeq:bb0} we find
\begin{multline}
    \mLnorm{\Aco^{*}G}\mLnorm{G} = \mLnorm{\Aco^{*}G}\mLnorm{F}
    \geq |\mLdot{F}{\adjoint{\Aco}G}|
    =  |\mLdot{F}{\adjoint{\Aco} K F}|  = \\
    |\mLdot{K\Aco F}{F}| =  |\overline{\OP{C}_{s,\lambda}[F]}| \; ,
    \label{Eeq:den1a}
\end{multline}
and with $F=\{\tilde{E},\tilde{H}\}$. Let $\{s,\lambda\}\in
\set{Q}$, from the properties of $\OP{C}_{s,\lambda}$, we find
\begin{equation}
    |\mLdot{F}{\adjoint{\Aco} K F}|= |\overline{\OP{C}_{s,\lambda}[F]}|
    \geq C_{0}(s,\lambda)\mLnorm{F}^{2} =
    C_{0}(s,\lambda)\mLnorm{G}^{2} \; ,
\end{equation}
where $\OP{C}_{s,\lambda}$ is defined for $F\in \mSob{1}$. From
\eqref{Eeq:den1a} and \eqref{Eeq:den1} it follows that
$\adjoint{\Aco}$ is bounded from below for all $G\in
\domain{\adjoint{\Aco}}\subset \mSob{1}$. For the bound from below
of $\adjoint{\Aco}$ it directly follows that $\adjoint{\Aco}$ has
trivial kernel and thus $\Aco$ has dense range for the condition of
$\{s,\lambda\}\in \set{Q}$. \qed

\subsection{Proof of Proposition \ref{prop2:spec}, part 4}

In Section~\ref{sec2:P} below we consider a Dunford-Taylor integral
over the resolvent and the analysis simplifies if $\Ac$ is closed.
From the form of $\Aco^{*}$ we note that
$\set{C}^{\infty}(\RR^2,\CC^4) \subset \domain{\adjoint{\Aco}}$ and
hence it is densely defined for $\{s,\lambda\}\in \set{Q}$ and thus,
by \cite[p.168, \S III.5.5]{Kato80}, the operator $\Aco$ is
closable. The closure is denoted by $\closure{\Aco}$.

The range for the closed operator is still dense in $\msLs$ since
\begin{equation}
    \range{\Aco} \subset \range{\closure{\Aco}} \subset \mLs \; .
\end{equation}
for $\{s,\lambda\}\in \set{Q}$.

To show that the closed operator is bounded from below, we rely on
Corollary VI.1.19 \cite{Kato80}. This corollary applies to
sesquilinear forms in Hilbert spaces, but due to example 1.23 and
example 1.3 in \cite{Kato80} we draw the conclusion that we can
construct the sesquilinear form $\mLdot{\Aco F}{\Aco G}$ and that it
is only closable when $\Aco$ is closable. Thus, by the above
mentioned corollary, we obtain that the closed form is bounded from
below with the same constant and thus, the closed operator is
bounded from below for $\{s,\lambda\}\in \set{Q}$. \qed

\subsection{Proof of Proposition \ref{prop2:spec}, part 5}

Given $(s,\lambda)\in \set{Q}$ we have shown in part 1-3 that $\Aco$
is one-to-one, has dense range and it is bounded from below.
Consequently we know that the inverse exists and is unique. The
operator $\Aco$ can be explicitly inverted in terms of the inverse
of two $2 \times 2$ matrix operators through a
quasi-diagonalization. Once again introduce the matrix $K$ (\cf
\eqref{Eeq:bb0b}), with $K^{-1} = K$. We find
\[
   K\Aco =
   \left(\begin{array}{cc} \Ac_{21} & \Ac_{22} - \lambda \\
                 \Ac_{11} - \lambda & \Ac_{12}
         \end{array}\right) \; ,
\]
and from the form of $\Ac_{21}$, \eqref{Eeq:m13} we find
\begin{align}
    |\mLdot{\tilde{E}}{\Ac_{21} \tilde{E}}| & =
    \left| \int_{\mathbb{R}^2} s\varepsilon_{\alpha\beta}\bar{E}_{\alpha}E_{\beta} +
    s^{-1}\mu_{33}^{-1}|(\nabla\times E)_{3}|^{2} \, \intd[2]{x'}
    \right|
    \nonumber \\
    & \geq \RE{s}
    \lowEu \mLnorm{\tilde{E}}^{2} \; ,
\end{align}
hence the inverse exists and is bounded for $\RE{s}>0$. Thus
$\Ac_{21}^{-1}$ is well defined.
The operator $K\Aco$ is then diagonalized as follows
\begin{equation}
   K\Aco \OP{T}_{2;s,\lambda} = \OP{T}_{1;s,\lambda} \OP{D}_{s,\lambda} \; ,
\label{Eeq:qd.3}
\end{equation}
where
\begin{equation}
   \OP{T}_{1;s,\lambda}
   = \left(\begin{array}{cc} 1  & 0 \\
           (\Ac_{11} - \lambda) \, \Ac_{21}^{-1} & 1
           \end{array}\right)
\label{Eeq:dq.4a}
\end{equation}
and
\begin{equation}
   \OP{T}_{2;s,\lambda}
   = \left(\begin{array}{cc} 1 &
           -\Ac_{21}^{-1} (\Ac_{22} - \lambda) \\
                   0 & 1
           \end{array}\right) \; ,
\label{Eeq:dq.5}
\end{equation}
while
\begin{equation}
   \OP{D}_{s,\lambda}
   = \left(\begin{array}{cc} \Ac_{21} & 0 \\
                   0      & \EL
           \end{array}\right) \; ,
\label{Eeq:qd.4b}
\end{equation}
with
\begin{equation}
   \EL
    = \Ac_{12} - (\Ac_{11} - \lambda) \Ac_{21}^{-1} (\Ac_{22} -
     \lambda)  \; .
\label{Eeq:qd.5}
\end{equation}
The characteristic operator, $\EL$, a matrix extension of the of the
`transverse Helm\-holtz' operator \cite{MdeHoop9606}. For each fixed
$\lambda$, the operators $\OP{T}_{1;s,\lambda},
\OP{T}_{2;s,\lambda}$ have the inverses:
\begin{equation}
   \OP{T}_{1;s,\lambda}^{-1}
   = \left(\begin{array}{cc} 1  & 0 \\
           -(\Ac_{11} - \lambda) \, \Ac_{21}^{-1} & 1
           \end{array}\right)
\label{Eeq:dq.6}
\end{equation}
and
\begin{equation}
   \OP{T}_{2;s,\lambda}^{-1}
   = \left(\begin{array}{cc} 1 &
           \Ac_{21}^{-1} (\Ac_{22} - \lambda) \\
                   0 & 1
           \end{array}\right) \; ,
\label{Eeq:dq.7}
\end{equation}
respectively. From the quasi-diag\-onal\-iza\-tion \eqref{Eeq:qd.3}
we obtain an explicit expression for $\Aco^{-1}$ in terms of
$\EL^{-1}$. That $\EL^{-1}$ is well defined follows by considering
\begin{equation}
|\mLdot{\tilde{E}}{\EL \tilde{E}}| =
|\mLdot{\tilde{E}}{\Ac_{12}\tilde{E}} + \mLdot{(\Ac_{22}-\iu
\lambda_I)\tilde{E}}{\Ac_{21}^{-1}(\Ac_{22}-\iu \lambda_I)\tilde{E}}
- \lambda_R^2 \mLdot{\tilde{E}}{\Ac_{21}^{-1} \tilde{E}}|
\end{equation}
where we have used $\Ac_{11}^*=-\Ac_{22}$. The bounds from below
$\Ac_{12}\geq\RE{s}\lowMu$ and $\Ac_{21}\geq\RE{s}\lowEu$, gives
\begin{equation}
|\mLdot{\tilde{E}}{\EL \tilde{E}}| \geq
\frac{1}{\RE{s}\hat{\epsilon}_1}
((\RE{s})^2\lowEu\lowMu-\lambda_R^2) \mLnorm{\tilde{E}}^{2} \; ,
\end{equation}
and since $(s,\lambda)\in \set{Q}$ the bound from below is positive
and $\EL$ is invertible. Starting from \eqref{Eeq:qd.3} and
inverting term by term, gives
\begin{eqnarray}
    \lefteqn{\Aco^{-1}  =
    \OP{T}_{2;s,\lambda}\OP{D}_{s,\lambda}^{-1}\OP{T}_{1;s,\lambda}^{-1}K =}
%   \label{Eeq:dq.8}
    \\
    && \left(\begin{array}{cc}
        -\Ac_{21}^{-1}\left(\Ac_{22}-\lambda\right)\EL^{-1} &
        \Ac_{21}^{-1}+\Ac_{21}^{-1}\left( \Ac_{22}-\lambda \right)
        \EL^{-1}\left(\Ac_{11}-\lambda\right)\Ac_{21}^{-1} \\
        \EL^{-1} &
        -\EL^{-1}\left(\Ac_{11}-\lambda\right)\Ac_{21}^{-1}
    \end{array}\right) \; .
    \nonumber
%    \hfil{\qed}
    \qedhere
\end{eqnarray}
\qed\endproof

\section{The Splitting Matrix}
\label{sec2:P}

We proceed with the decomposition of the electromagnetic system's
matrix. As the spectrum is absent from the strip (see Corollary
\ref{cor2:spec}), we define a certain commuting operator through a
resolvent integral, this operator will satisfy a number of
properties, and will be called the splitting matrix. We note that if
an operator has a spectral resolution, or even a part of the
spectrum which is bounded, then one can define a projector with help
of a Cauchy type integral, also called Dunford's integral, over the
resolvent with integration path around the bounded spectral region,
see \cite[III.6.4]{Kato80}, and also \cite{Yosida80, Shubin87}. For
the electromagnetic system's matrix such information about the
spectrum is not known, we do know however that the spectrum is
separated into two parts by a strip around the imaginary axis. The
idea here is to accomplish a decomposition by introducing an
operator defined by an integral over the resolvent of $\Ac$, as to
try to split the two parts we know exist, similarly to the case of
bounded spectral regions. We use the Dunford-Taylor integral applied
to a closed, unbounded, operator as in \cite{Dunford+Schwartz+1+64},
\cf \cite{Kato80} for accretive operators. This theory is given only
for closed paths, or absolutely bounded integrals, hence the
extension needed here to non-closed paths require that we prove that
the operator is well defined.

In this section we prove a number of properties of the splitting
matrix, among them that it is well defined as a pseudodifferential
operator with a parameter, that it is an involution, and that it
commutes with the electromagnetic system's matrix. Once the
splitting matrix is shown to be well defined we derive its
generalized eigenvalues and eigenvectors; the generalized
eigenvectors are the key components for the decomposition detailed
in the next section.

\subsection{Definition of the Splitting Matrix}

Given a fixed positive constant $\lowS>0$, let
\begin{equation}
    \set{Q}_{1} = \left\{ \{s,\lambda\}\in \mathbb{C}^2:
%   \right. \\ \left.
    \RE{s}>\lowS, \ |\arg s|< \pi/2\ \ \mbox{and}\ \ |\RE{\lambda}| <
    \lowS \sqrt{\lowEu \lowMu} \right\} \; .
    \label{Eeq:p0}
\end{equation}
From Proposition \ref{prop2:spec} we note that
$\set{Q}_{1} \subset \set{Q}$, hence by Corollary \ref{cor2:spec} the
strip $|\RE{\lambda}| < \lowS \sqrt{\lowEu \lowMu}$ belongs to the
resolvent set of the electromagnetic system's matrix.
Thus we can consider the operator defined through
\begin{equation}
    \oB =  \lim_{n\rightarrow \infty}
    \frac{1}{\pi\iu} \int_{\intKn} \intd{\lambda}
    \left( \closure{\Ac}-I\lambda \right)^{-1} \; .
    \label{Eeq:p1}
\end{equation}
The spatial and time-Laplace dependence is present but
not explicit in the notation. The integration path is:
\begin{equation}
    \sKn = \left\{\lambda \in \mathbb{C}: \RE{\lambda}= \lLb/2 \ \
    \mbox{and} \ \ |\IM{\lambda}| \leq n \right\} \; ,
    \label{Eeq:p2}
\end{equation}
where
\begin{equation}
    \lLb = \lowS \sqrt{\lowEu \lowMu} \; ,
    \label{Eeq:p3}
\end{equation}
and hence the integral path is in the resolvent set. Some of the
considerations that follow become simpler if we consider the
operators restriction to $\set{C}^{\infty}(\RR^2,\CC^4)$, and
similarly for any operator with the notation
$\smooth{\phantom{\OP{S}}}\Strut$. With the above introduction, we
have the following proposition.
\begin{prop}\label{prop2:split2}
    Let $\oB$ be defined as
    in \eqref{Eeq:p1} with
    $\{s,\sKn\}\in \set{Q}_{1}$
    then $\oB$
    \begin{enumerate}
        \item  is a pseudodifferential operator with parameters of order 0;

        \item  has a restriction $\qB{q}$, which maps $\mSob{q}$ into
        $\mSob{q-1}$;

        \item  `commutes' with $\Ac$ in the sense that on the
        set $\mSob{3}$ we have
        \begin{displaymath}
            \qB{1}\Ac = \Ac \qB{3} \; ;
        \end{displaymath}

        \item  has a restriction, $\sB$, that is an involution;

        \item has a restriction $\sB$ that has a generalized eigenvector
        $\spmL$; unique up to a normalization, and
        with a corresponding scalar `eigenvalue' $\gamma=\pm 1$, satisfying the equation
            \begin{displaymath}
                \sB\spmL = \gamma \spmL \; .
            \end{displaymath}
        The explicit form of $\spmL$ is
            \begin{displaymath}
            \spmL = \begin{pmatrix} (\pm I +
            \sB_{11})\spmN \\ (\sB_{21})\spmN \end{pmatrix}   \; ,
        \end{displaymath}
        where $\spmN$ is a normalization in the form of
        invertible $2 \times 2$ operator matrices;

        \item has a restriction $\sB$ that
        is one-to-one on a core and thus its element $\sB_{21}$ is invertible on
        its range;

        \item has `generalized eigenvectors', \ie that is the extension of
        $\spmL$ exists. With proper choice of normalization,
        $\qpmL{q}:\setSobolev{q}(\RR^2,\CC^2) \rightarrow \mSob{q-1}$.
    \end{enumerate}
\end{prop}

\begin{remark}
  With the choice of $\spmN$ as $\sB_{21}^{-1}$, we identify
  $\spmL_{1}$ as a mapping between $\tilde{H}$ and $\tilde{E}$, \ie an
  impedance mapping. The corresponding map in linear acoustic is a map
  between the pressure and the vertical particle velocity
  \cf~\cite{Jonsson+deHoop01}.  Both these mappings are the acoustic,
  and electromagnetic respective equivalent maps to a
  Dirichlet-to-Neumann map for the wave
  equation~\cite{He+Strom+Weston98}.
\end{remark}
\begin{remark}
  To ensure that the above defined splitting matrix is
  non-trivial we have to exclude two cases: that the integral
  \eqref{Eeq:p1} collapses to the identity or to the zero operator.
  Whether this happens depends on the non-triviality of the
  spectrum. To ensure that this is not always the case we consider
  the homogeneous-isotropic case, see Appendix \ref{app:iso} where we
  obtain the explicit form of $\oB$. We note that if the medium in the
  neighborhood of a point is isotropic, then microlocally, at that
  point, the operator $\oB$ reduces to the isotropic case, which is
  clearly different from the identity and the zero operator, and hence
  $\oB$ can not be the unity or the zero operator for such media.
\end{remark}

\subsection{Proof of Proposition~\ref{prop2:split2}, part~1}

The operator $\oB$ is defined through an improper integral over the
resolvent. To prove that $\oB$ is well defined as a
pseudodifferential operator with parameter we consider first the
parametrix of $\Aco$ and then integrate each term of the asymptotic
expansions with respect to $\lambda$ and prove that this step is
well defined. Hence we obtain an asymptotic expansion for the symbol
of $\oB$, via the usual calculus of pseudodifferential operators we
thus construct a well defined operator $\oB$.

\subsubsection{Pseudodifferential preliminaries}

The calculus of pseudodifferential operators can be introduced by
means of a Fourier transform, thus defining signs and symbols. For
simplicity we use standard notation for the symbols and their
compositions. Throughout this paper we use the left symbol (in the
notation of \cite{Shubin87}). The Fourier transform, $\OP{F}$, in
the plane with respect to the first two variables
$x'=\{x_{1},x_{2}\}$ has an inverse given by
\begin{equation}
    \tilde{E}(x,s) = \frac{1}{(2\pi)^2} \int_{\mathbb{R}^2} \intd[2]{\xi'}
    \lexp{\iu \xi'\cdot x'}  (\OP{F}\tilde{E})(\xi',x_{3};s)
    \label{Eeq:ep1}
\end{equation}
for the complex field $\tilde{E}\in \Ltwo(\RR^2,\CC^2)$. Here
$\xi'\cdot x' = \xi_{1}x_{1} + \xi_{2}x_{2}$. To obtain the left
symbol of $\Ac_{21}$ we let it act upon \eqref{Eeq:ep1} and obtain
that the integrand expression in front of $\lexp{\iu\xi'\cdot x'}
(\OP{F}\tilde{E})(\xi',x_{3},s)$ is
\begin{multline}
    \aS_{21}(x,\xi';s) = s
    \begin{pmatrix}
%   \left( \begin{array}{rr}
        \varepsilon_{11} & -\varepsilon_{12} \\
        -\varepsilon_{21} & \varepsilon_{22}
    \end{pmatrix}
%   \end{array} \right)
    + (s \mu_{33})^{-1}
    \left( \begin{array}{rr}
        \xi_{2}^2 & \xi_{1}\xi_{2} \\ \xi_{1}\xi_{2} & \xi_{1}^2
    \end{array} \right)
%   \end{array} \right)
    \\
     -  \iu s^{-1}
    \begin{pmatrix}
%   \left( \begin{array}{cc}
        (\partial_{2} \mu_{33}^{-1})\xi_{2} & (\partial_{2} \mu_{33}^{-1})\xi_{1} \\
        (\partial_{1} \mu_{33}^{-1})\xi_{2} & (\partial_{1} \mu_{33}^{-1})\xi_{1}
    \end{pmatrix} \; .
%   \end{array} \right)
    \nonumber
\end{multline}
This is the left symbol of $\Ac_{21}$. To find the appropriate
behavior of the symbols we have to consider symbols with parameters.
We consider $s$ to be a parameter of the same order as $\xi_i$. We
hence find the principal symbol of $\aS_{21}$ to be
\begin{equation}
    \adu(x,\xi'; s)= s
    \left( \begin{array}{rr}
        \varepsilon_{11} & -\varepsilon_{12} \\
        -\varepsilon_{21} & \varepsilon_{22}
    \end{array} \right) + (s \mu_{33})^{-1}
    \left( \begin{array}{rr}
        \xi_{2}^2 & \xi_{1}\xi_{2} \\ \xi_{1}\xi_{2} & \xi_{1}^2
    \end{array} \right) \; ,
    \label{Eeq:ep3}
\end{equation}
and $\adu$ is homogeneous of order one in $(\xi',s)$.

\subsubsection{Ellipticity of $\Aco$}
\label{sec:ellips}
For matrix valued operators it is the determinant
of the symbol that controls the regularity and existence of its
parametrix. We require that the coefficients to $(\xi',s,\lambda)$
are arbitrarily smooth for each term in $\Aco$, and thus we can use
the criteria in Definition 5.1 together with Proposition $5.1'$ of
\cite[pp.38,39]{Shubin87} to define ellipticity of $\Aco$. To
construct the principal symbol of the operator $\Ac$,
$\aS_{;1}(x,\xi';s)$, we proceed as above and obtain for the
remaining elements
\begin{align}
    \auu &= \iu \mu_{33}^{-1} \left( \begin{array}{rr}
        \mu_{23}\xi_{2} & \mu_{23}\xi_{1} \\ \mu_{13} \xi_{2} & \mu_{13} \xi_{1}
    \end{array}\right) + \iu \epsilon_{33}^{-1} \left( \begin{array}{rr}
        \epsilon_{31}\xi_{1} & - \epsilon_{32}\xi_{1} \\
        -\epsilon_{31}\xi_{2} & \epsilon_{32}\xi_{2}
    \end{array}\right) \; , \\
    \aud &=
    s\left( \begin{array}{rr}
        \nu_{22} & \nu_{21} \\ \nu_{12} & \nu_{11}
    \end{array}\right) + s^{-1} \epsilon_{33}^{-1}
    \left( \begin{array}{rr}
            \xi_{1}^{2} & -\xi_{1}\xi_{2} \\ -\xi_{1}\xi_{2} & \xi_{2}^{2}
        \end{array} \right) \; , \\
    \add &= \iu \mu_{33}^{-1} \left( \begin{array}{rr}
            \mu_{32} \xi_{2} & \mu_{31} \xi_{2} \\ \mu_{32}\xi_{1} & \mu_{31}\xi_{1}
        \end{array}\right) + \iu \epsilon_{33}^{-1} \left(\begin{array}{rr}
            \epsilon_{13}\xi_{1} & -\epsilon_{13} \xi_{2} \\
            -\epsilon_{23}\xi_{1} & \epsilon_{23} \xi_{2}
        \end{array}\right) \; .
\end{align}
Let
\begin{equation}
   \ac := \aS_{;1} - \lambda I.
\end{equation}
Then $\ac$ have homogeneity degree 1 in $(\xi',s,\lambda)$. The
remaining part of the symbol of $\Aco$ has a lower degree of
homogeneity in $(\xi',s,\lambda)$. To ensure the ellipticity of an
operator in the parameters $(\xi',s,\lambda)$ the following estimate
is needed
\begin{equation}
    C_{1} (|\xi'|^2 + |s|^2 + |\lambda|^2)^2 \leq |
    \det \ac | \leq C_{2} (|\xi'|^2 + |s|^2 + |\lambda|^2)^2
    \label{Eeq:ep8}
\end{equation}
for some $R$ such that $|\xi'|^2 + |s|^2+ |\lambda|^2> R^2$ and with
proper restrictions on the parameters $\{s,\lambda\}$. The properly
supported requirement for ellipticity follows from the fact that
$\Ac$ is a {\em classical pseudodifferential operator} with smooth
coefficients \cite{Shubin87}. The upper limit of \eqref{Eeq:ep8}
follows directly from the fact that $\det \ac$ is a polynomial,
homogeneous of order four in $(\xi',s,\lambda)$ (see Appendix
\ref{app:det}), together with the fact that we can dominate this
polynomial by $(|\xi'|^2+|s|^2+|\lambda|^2)^2$ and a constant, for
some constant $R_{0}$ such that
$|\xi'|^2+|s|^2+|\lambda|^2>R_{0}^2$. To prove the lower limit we
need a more subtle method.

\paragraph{The lower limit of \eqref{Eeq:ep8}:}

\begin{figure}
   \centering
   \centerline{\includegraphics{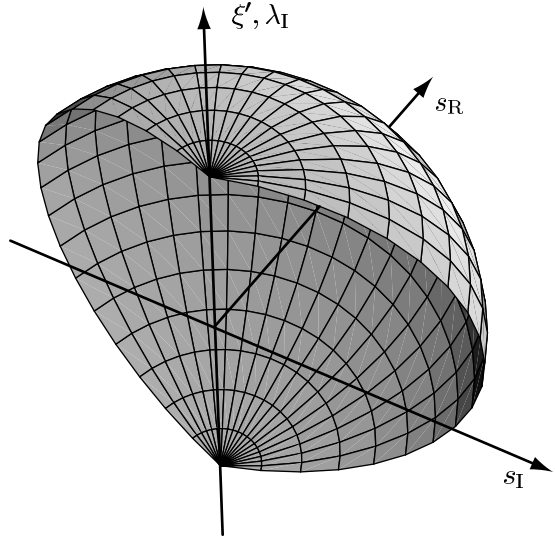}}
   \parbox{\textwidth}{
   \caption{A schematic picture of the surface $|\xi'|^2 + |s|^2 +
   |\IL|^2=R_{e}^2$, $|\arg s|<\pi/2$.
   In particular note that the condition $|\arg s|<\pi/2$ shrinks the
   surface away from a half sphere.}
    \label{fig:con} }
\end{figure}

This is a multi-step proof, and a complication arises since the
region in $\xi',s,\lambda$ is not conical. First we prove that the
determinant is non-zero on a surface (see Figure \ref{fig:con}),
then we use a scaling argument to extend this to a bound from below
of the form \eqref{Eeq:ep8} for a conical region with the surface in
figure \ref{fig:con} as `bottom surface'. In the last step we extend
the obtained result to the non-conical domain, so as to include
$\RL$.

\subparagraph{Non-zero determinant of $\ac$}

Let $\RL=0$, to explicitly show that the determinant of $\ac$ is
non-zero is difficult due to the large number of terms that it
contains (see Appendix \ref{app:det}). Our scaling argument needs
only that the determinant is non-zero on a surface, here part of a
sphere, see figure \ref{fig:con}. Thus let
$|s|^2+|\xi|^2+|\IL|^2=R_{e}^2$, for some positive constant $R_e$.
We consider two cases; $s=0$ and $s\neq 0$. For the first case with
$s = 0$, we find from Appendix~\ref{app:det} that
\begin{multline}
    \label{Eeq:ap1}
    \left. \det \ac \right|_{s=0} = \\ \left( \IL^2 -
    \IL \rpe_{33}^{-1}(\rpe_{\alpha
    3} + \rpe_{3 \alpha} ) \xi_{\alpha} + \rpe_{33}^{-1}\rpe_{\alpha\beta}
    \xi_{\alpha}\xi_{\beta} \right)
    \left(\IL^2 - \IL\rpm_{33}^{-1}(\rpm_{\gamma 3} + \rpm_{3 \gamma})\xi_{\gamma} +
    \rpm_{33}^{-1}\rpm_{\gamma\delta}\xi_{\gamma}\xi_{\delta}  \right) \; .
\end{multline}
Using the restriction that $\rpm$ and $\rpe$ are self adjoint, together
with the estimate
\begin{equation}
    2 \IL \xi_{\alpha} \rpe_{33}^{-1}\RE{\rpe_{3\alpha}} \leq
    \eta_{1}^{-1}\IL^2 + \eta_{1} \rpe_{33}^{-2} \rpe_{\alpha 3}\rpe_{3 \beta}
    \xi_{\alpha}\xi_{\beta}
    \label{Eeq:ap4}
\end{equation}
for $\eta_{1}>0$ and similarly for
$2\IL\rpm_{33}^{-1}\RE{\rpm_{\gamma 3}}\xi_{\gamma}$ gives
\begin{align}
    \nonumber
    |\det \ac|_{s=0} & \geq \left((1-\eta_{2}^{-1})|\IL|^2 +
    \rpm_{33}^{-1}\rM^{(\eta_{2})}_{\alpha\beta} \xi_{\alpha}\xi_{\beta}
    \right)\left((1-\eta_{1}^{-1})|\IL|^2 +
    \rpe_{33}^{-1}\rE^{(\eta_{1})}_{\gamma\delta}
    \xi_{\delta}\xi_{\gamma}\right) \\ &
    \geq C_{a} \left(|\IL|^2 + |\xi'|^2\right)^2 = \left. C_{a}R_{e}^4\right|_{s=0, \lambda_R=0}\; ,
    \label{Eeq:ap2}
\end{align}
where $\left. R_{e}^2\right|_{s=0,\lambda_R=0}=|\IL|^2+|\xi'|^2$ and
\begin{equation}
    \rE^{(\eta_{1})}_{\alpha\beta} = \rpe_{\alpha\beta} -
    \eta_{1} \rpe_{\alpha 3} \rpe_{33}^{-1} \rpe_{3 \beta} \ \
    \mbox{and} \ \
    \rM^{(\eta_{2})}_{\alpha\beta}= \rpm_{\alpha\beta} -
    \eta_{2} \rpm_{\alpha 3} \rpm_{33}^{-1} \rpm_{3 \beta} \; .
    \label{Eeq:ap3}
\end{equation}
To find an explicit expression for $C_{a}$, let us choose $\eta_{1}>1$ so that
\begin{equation}
    0< 1-\eta_{1}^{-1} = \inf_{x'} \rpe_{33}^{-1} \inf_{|\xi'|=1}
    \rE^{(\eta_{1})}_{\gamma\delta} \xi_{\gamma}\xi_{\delta} \; .
    \label{Eeq:ap5}
\end{equation}
The right-hand side of this equation gives the minimum of the lower
eigenvalue of the matrix $\rE^{(\eta_{1})}_{\gamma\delta}$
normalized with $\rpe_{33}$. There exists an $\eta_{1}>1$ that
fulfils this equation since $\lowEu>0$, hence the left-hand side of
the above expression is positive. The same way we find an
$\eta_{2}>1$ such that
\begin{equation}
    0 < 1-\eta_{2}^{-1} = \inf_{x'} \rpm_{33}^{-1} \inf_{|\xi'|=1}
    \rM^{(\eta_{2})}_{\gamma\delta} \xi_{\gamma}\xi_{\delta}   \; .
    \label{Eeq:ap6}
\end{equation}
The constant $C_{a}$ becomes
\begin{equation}
    C_{a} = (1-\eta_{1}^{-1})(1-\eta_{2}^{-1}) \; .
    \label{Eeq:ap7}
\end{equation}

For the case where $s \neq 0$ we use Schwartz' inequality on an inner
product. Thus we introduce the `matrix' norm
\begin{equation}
    \snorm{F}^2 = \sum_{i=1}^{4}|F_{i}|^2 \ \
    \mbox{and} \ \ \snorm{\ac} = \sup_{\snorm{F}=1} \snorm{\ac
    F} \; ,
    \label{Eeq:ep9b2}
\end{equation}
with a corresponding inner product defined
analogously and denoted by $\sdot{\cdot}{\cdot}$. Both the norm and
inner product depend on $(x,\xi',s,\IL,0)$.
Consider the normal $4 \times 4$
matrix $\adjoint{\ac} \ac$, for $\RL=0$, that have
eigenvalues $\kappa_{1},\ldotsk,\kappa_{4}$ each with a variable
dependence $(x,\xi';s,\IL,0)$.
From the definition of eigenvalues it follows that
$0\leq\kappa_{i}\in \mathbb{R}$, where $i=1,\ldotsk,4$, and we use the
convention
$\kappa_{4}\geq \cdots\geq \kappa_{1}$.
From the relation
\begin{equation}
    \kappa_{1}^4 \leq \kappa_{4}\kappa_{3}\kappa_{2}\kappa_{1}
    = \left. \det{\adjoint{\ac}\ac}  \right|_{\RL=0}
    = \left|\det{\ac}\right|_{\RL=0}^2  \; ,
    \label{Eeq:ep9c}
\end{equation}
we find that it is enough to prove that $\kappa_{1}\neq 0$.
Schwartz' inequality (\cf
\eqref{Eeq:bb0}) gives
\begin{equation}
    \snorm{\ac F} \snorm{F} \geq |{\sdot{F}{ K \ac F}}| \; .
    \label{Eeq:ep11a}
\end{equation}
Thus if for $\RL=0$ and $s\neq 0$ we can obtain an estimate of the form
\begin{equation}
    | \sdot{F}{K \ac F}| \geq
    C_{b} \snorm{F}^{2} \; ,
    \label{Eeq:ep11}
\end{equation}
where $C_{b}>0$, then from \eqref{Eeq:ep9b2} and \eqref{Eeq:ep11} it follows that
\begin{equation}
    \kappa_{1}^{-1/2} = \snorm{\ac^{-1}} \leq C_{b}^{-1} \ \
    \mbox{hence}\ \
    \kappa_{1} \geq C_{b}^{2}  \; .
    \label{Eeq:ep10}
\end{equation}
By \eqref{Eeq:ep9c} we obtain,
\begin{equation}
    \left| \det \ac \right|_{\RL=0} \geq C_{b}^{4} \; ,
    \label{Eeq:ep13}
\end{equation}
under some restrictions on $s$, to be derived. Now with the explicit
form of $\ac$ we obtain
\begin{multline}
    \left. \sdot{F}{K\ac F}\right|_{\RL=0} = s \left( \varepsilon_{\alpha\beta}
    \overline{E}_{\alpha} E_{\beta}
    +
    \nu_{\alpha\beta}\overline{H}_{\alpha}H_{\beta} \right)
    \\ +s^{-1} \left( \mu_{33}^{-1}\left| (\xi
    \times E)_{3} \right|^2
    +
    \epsilon_{33}^{-1} \left|(\xi \times H )_{3}\right|^2 \right) - 2
    \iu \IL \RE{E_{1}\overline{H}_{2}-E_{2}\overline{H}_{1}}
    \\ +
    2 \iu \left( \mu_{33}^{-1}
    \IM{(\xi \times \overline{E})_{3} \mu_{\alpha 3}H_{\alpha}} +
     \epsilon_{33}^{-1} \IM{\epsilon_{\alpha 3}\overline{E}_{\alpha}
    (\xi \times H)_{3}} \right) \; ,
    \label{Eeq:ep15a}
\end{multline}
where we have used the notation of Proof of
Proposition~\ref{prop2:spec}, part~1. Since $s\neq 0$ we take the
real part and obtain
\begin{equation}
    |\sdot{F}{K\ac F}|_{\RL=0} \geq \RE{s} \left( \lowEu |\tilde{E}|^2 + \lowMu
    |\tilde{H}|^2\right) > C_{b} \snorm{F}^2 \; ,
    \label{Eeq:ep15b}
\end{equation}
if $\RE{s}>0$ and here
\begin{equation}
    C_{b} = |s|\min\{\lowEu,\lowMu\} \cos \sigma
    \label{Eeq:ap10}
\end{equation}
is positive if $|\sigma|=|\arg s| < \pi/2$ and $s \neq 0$. By the
above argument, \eqref{Eeq:ap2} and \eqref{Eeq:ep13} the determinant
is non-zero on the surface
\begin{equation}
    R_{e}^2 = |\xi'| + |s|^2 + |\IL|^2\; , \ |\sigma|<\pi/2
    \label{Eeq:ap11}
\end{equation}
if $R_{e} \neq 0$. Hence there exists a lower constant
$C_{e} = \min(C_{b},C_{a}R_{e}^4)$ such that
\begin{equation}
    \left|\det \ac\right|_{\RL=0} \geq C_{e}>0
    \label{Eeq:ap12}
\end{equation}
on this surface.

\subparagraph{A scaling argument:}
\label{eub:cone}

\begin{figure}
   \centering
   \centerline{\includegraphics{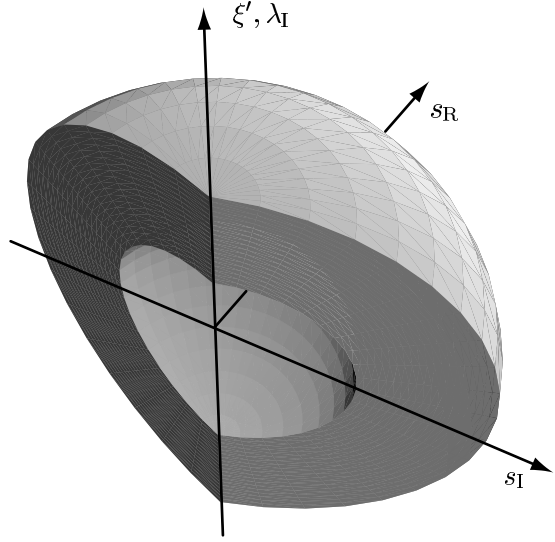}}
   \parbox{\textwidth}{
   \caption{A schematic picture of the conical region in $\xi',\ s,\
   \IL$.
   Note that the outer surface can have an arbitrary radius larger
   than the inner radius $R_{e}$.}
    \label{fig:conR} }
\end{figure}

To extend the result
\begin{equation}
    \left|\left. \det \ac(x,\xi';\rS,\sigma,\IL,\RL)
    \right|_{\RL=0}\right| \geq
    C_{e} > 0
    \label{Eeq:ep9}
\end{equation}
for $|\xi'| + |s|^2 + |\IL|^2 = R_{e}^2$ and $|\sigma|<\pi/2$
to a proper bound
from below, we use a scaling argument.
The homogeneity of $\det \ac$ allow us to scale
$\xi',s,\IL$, to an arbitrary radius greater then $R_{e}$ and
\begin{equation}
    \det \ac(x,\xi';\rS,\sigma,\IL,0)  =
    R_{e}^{-4}(|\xi'|^2+|s|^2+|\IL|^2
    )^2 \det
    \ac(x,\tilde{\xi'};\tilde{\rS},\sigma,\tilde{\IL},0) \; ,
    \label{Eeq:ep9a}
\end{equation}
where the $\tilde{\cdot}$ variables are normalized to lie on the
surface $|\xi'|^2 + |s|^2 + |\IL|^2 = R_{e}^2$.
Thus from \eqref{Eeq:ep9} we have obtained
\begin{equation}
    \left| \det \ac(x,\xi';\rS,\sigma,\IL,0) \right|
    \geq C_{e} R_{e}^{-4}(|\xi'|^2 + |s|^2 + |\IL|^2)^2  \; ,
    \label{Eeq:ep9b}
\end{equation}
in the conical domain,
\begin{equation}
    \xi' \in \mathbb{R}^2\; , \ \IL \in \mathbb{R}\; , \ s\in
    \mathbb{C}\ \
    \mbox{and} \ \ |\arg s|<\pi/2 \; ,
\end{equation}
see Figure~\ref{fig:conR}.

\subparagraph{The case $\RL \neq 0$:} To extend the argument above
to include the case $\RL\neq 0$ we impose the condition
$\RE{s}>\lowS=\frac{1}{2}S_R\sqrt{\lowEu\lowMu}$. Let $\RL \leq
\tau$ and $R_{e}>\lowS$. For large enough $R_{e}$, where
$|\xi'|^2+|s|^2+|\IL|^2\geq R_{e}^2$, the worst case for the bound
from below of the determinant is
\begin{equation}
    |\det \ac|  \geq  C_{e} R_{e}^{-4}\left(|\xi'|^4 + |s|^4 + |\IL|^4
    \right) - \tilde{C}_{\tau} \lLb \left(|\xi'|^3 + |s|^3 + |\IL|^3
    \right)
\end{equation}
where $\tilde{C}_{\tau}$ is chosen to include the sum of the maximal
material parameters in front of $\RL$. Using the H\"{o}lder and the
Jensen inequalities \cite[p.28, Theorem 19]{Hardy52} gives
\begin{align}
    |\det \ac| &\geq C_{e} R_{e}^{-4}\left(|\xi'|^4 + |s|^4 + |\IL|^4
    \right) - \tilde{C}_{\tau} \lLb \left(|\xi'|^3 + |s|^3 + |\IL|^3
    \right) \nonumber \\
    &\geq  C_{e}R_{e}^{-4}\left(|\xi'|^4 + |s|^4 + |\IL|^4 +|\RL|^4
    \right)
     - C_{\tau} \lLb \left(|\xi'|^3 + |s|^3 + |\IL|^3 +
    |\RL|^3 \right) \nonumber \\
    &\geq  C_{e}R_{e}^{-4} 4^{-1} |z|^4 - C_{\tau}\tau |z|^3 =
    4^{-1}C_{e}R_e^{-4}|z|^3 (|z|-4
    \frac{C_{\tau}R_{e}^{4}}{C_{e}}\tau) \; ,
    \label{Eeq:ep14}
\end{align}
where
\begin{equation}
    C_{\tau} = \max\{\tilde{C}_{\tau},C_{e}R_{e}^{-4}\} \;
    \label{Eeq:ep15}
\end{equation}
and
\begin{equation}
    |z|^2 = |\xi'|^2 + |s|^2 + |\lambda|^2 \; .
    \label{Eeq:ep16}
\end{equation}
Comparing with \eqref{Eeq:ep8} we obtain the condition
\begin{equation}
    C_{1}|z|^4 \leq 4^{-1} C_{e}R_{e}^{-4}|z|^4 - C_{\tau}|z|^3 \tau \; .
    \label{Eeq:ep16b}
\end{equation}
Thus for some given, arbitrary $R_e>0$, there exists a large enough
$R$ such that $|\det \ac|> C_1 |z|^4$, for $|z|^2>R^2$ when
$(C_1,R)$ satisfy the following constraints:
\begin{equation}
    0< C_{1} < 4^{-1}C_{e}R_{e}^{-4}
    \label{Eeq:ep16c}
\end{equation}
and
\begin{equation}
    |z|^2 \geq R^2 \geq \max\{ R_{0}^2 , R_{e}^2 + \tau^2 , \left(\frac{C_{\tau}
    \tau}{4^{-1}C_{e}R_{e}^{-4} - C_{1}}\right)^2 \} \; .
    \label{Eeq:ep16d}
\end{equation}
Hence for $\xi' \in \mathbb{R}^2$ and
$\{s,\lambda\}\in \set{Q}_{1}$, the operator is elliptic.

Note that the quadratic form argument of Lemma~\ref{lem2:sesq} can
be applied to the symbol $\ac$ to yield a positive lower bound.
Consequently \eqref{Eeq:ep10} and \eqref{Eeq:ep9c} imply
\begin{equation}
    |\det \ac| \neq 0  \; , \label{Eeq:nzd}
\end{equation}
for $\{s,\lambda\}\in \set{Q}$ and $|s|^2+|\xi'|^2+|\lambda|^2 \neq
0$. However the desired increase in $|s|^2+|\xi'|^2+|\lambda|^2$
does not follow directly, since the domain is not conical. Observe
also that \eqref{Eeq:nzd} is true in the region $\{s,\lambda\} \in
\set{Q}_{1}$. This result will be used in the end of the proof of
part~1.

\subsubsection{The parametrix of $\Aco$}

We have above shown that $\Aco$ is elliptic in pseudodifferential
sense, hence the corresponding parametric is well defined. In the
subsequent analysis we are interesting only in the principal part.
From Proposition~\ref{prop2:spec}, part~5, we know that the inverse
can be efficiently expressed in terms of $\EL^{-1}$ for
$s,\lambda\in Q$ and different combinations of $2\times 2$ matrices.
Therefore we introduce the notation of $\hat{\cdot}$ on $2\times 2$
matrices defined by
\begin{equation}
    \iadu: = \left(\begin{matrix} (\adu)_{22} & -(\adu)_{12} \\
            -(\adu)_{21} & (\adu)_{11} \end{matrix}
            \right) \; ,
    \label{Eeq:pa0}
\end{equation}
where $(\cdot)_{ij}$ is the $(ij)$-element of the 2x2 matrix. From
the definition it follows directly that
\begin{gather}
%\begin{split}
    \inv{\inv{\aS}}_{21;1} = \adu \; ,\   \ ( \auu \adu)\inv{\phantom{a}} = \iadu
    \iauu \; , \label{Eeq:pa0b}
    \\
    \ (\auu +\add ) \inv{\phantom{a}} = \iauu +
    \iadd
    \; ,
%\end{split}
    \label{Eeq:pa0a}
\end{gather}
and
\begin{equation}
    \iadu \adu = \det \adu I \; .
\end{equation}
The principal symbol of the characteristic operator, $\EL$, is
\begin{equation}
    \eu = \aud - (\auu - I\lambda)\iadu(\add-I\lambda) (\det\adu)^{-1} \; ,
    \label{Eeq:ap0b}
\end{equation}
and using \eqref{Eeq:pa0b}--\eqref{Eeq:pa0a} we find
\begin{equation}
    \ieu = \iaud - (\iadd-I\lambda)\adu(\iauu-I\lambda)(\det\adu)^{-1} \; ,
    \label{Eeq:ap0c}
\end{equation}
thus
\begin{equation}
    (\det \eu) I  = \ieu \eu\; .
\end{equation}
From writing out all terms we find that
\begin{equation}
    \det \ac = (\det \eu)(\det \adu) \; ,
    \label{Eeq:ap0d}
\end{equation}
which is a polynomial homogeneous of order 4 in $\xi',s,\lambda$.
That $|\det \eu|\neq 0$ follows directly from that $|\det \ac|\neq
0$ (see \eqref{Eeq:nzd}) together with the observation that $|\det
\adu|\neq 0$ and that $|\det \adu|$ is bounded above and below by
constants times $|\xi'|^2+|s|^2$. Hence it follows that the
parametrix of $\EL$ is well defined. This is to be expected since
the inverse of $\EL$ was shown to be well defined in Proposition
\ref{prop2:spec}, part~5. With the above consideration we find that
the components of the principal symbol of the resolvent,
$\symb{r}_{;-1}:=\ac^{-1}=(\aS-\lambda I)^{-1}$, are (\cf
Proposition \ref{prop2:spec} part~5)
\begin{align}
      (\symb{r}_{;-1})_{11} &:= -(\det \ac)^{-1}\iadu(\add - I\lambda)\ieu
    \; , \nonumber \\
    \begin{split}
      (\symb{r}_{;-1})_{12} &:=
        (\det \adu)^{-1}\iadu \big(I
%        & \phantom{=}
        +(\det \ac)^{-1}(\add - I\lambda)\ieu(\auu - I
        \lambda)\iadu \big)
    \; , \nonumber
\end{split}
    \\
    (\symb{r}_{;-1})_{21} &:= (\det \ac)^{-1}\ieu (\det \adu)
    \; ,    \label{Eeq:inver1}
    \\
    (\symb{r}_{;-1})_{22}  &:=-(\det \ac)^{-1}
         \ieu(\auu-I\lambda)\iadu
    \; . \nonumber
%   \label{Eeq:inver4}
\end{align}
Upon integration of the parametrix with respect to $\lambda$, the
element $(\symb{r}_{;-1})_{12}$ has an unsuitable form, therefore
we use the identity $\ieu \eu = I \det \eu$ and rewrite $(\symb{r}_{;-1})_{12}$
into
\begin{align}
    (\det \ac)(\symb{r}_{;-1})_{12} &=
    \iadu (\ieu \eu  + (\det \adu)^{-1}(\add - I\lambda)\ieu(\auu - I
        \lambda)\iadu) \nonumber \\
     &=     \iadu (\ieu \aud
    + (\det \adu)^{-1}
     \commute{\add}{\ieu (\auu - I\lambda)\iadu}) \; ,
    \label{Eeq:pa6}
\end{align}
where $\commute{\cdot}{\cdot}$ is the standard commutator
$\commute{A}{B}=AB-BA$. By inserting the explicit expression of
$\ieu$ in the commutator together with the two relations
$(\iauu-I\lambda)(\auu-I\lambda)=I\det(\auu-I\lambda)$ and
$\iadu\adu=\det \adu I$ we find that \eqref{Eeq:pa6} reduce to
\begin{equation}
    (\det \ac)(\symb{r}_{;-1})_{12} = % \\
    \iadu (\ieu \aud
    + (\det \adu)^{-1}
     \commute{\add}{\iaud (\auu - I\lambda)\iadu}) \; .
    \label{Eeq:pa6b}
\end{equation}
An alternative form of \eqref{Eeq:pa6b} is obtain by inserting the
explicit form of $\ieu$ and simplifying:
\begin{multline}
    (\det \ac)(\symb{r}_{;-1})_{12} = % \\
    \iadu (I\det\aud-(\det \adu)^{-1}(\iadd-I\lambda)\adu(\iauu-I\lambda)\aud
    \\ + (\det \adu)^{-1}
     \commute{\add}{\iaud (\auu - I\lambda)\iadu}) \; .
\label{Eeq:pa6bb}
\end{multline}

With the above expression we have obtained the principal part of the
symbol of $\Aco^{-1}$ and each term in the matrix has the form
\begin{equation}
    \frac{\lambda^n}{\det \ac} \; ,
    \nonumber
%   \label{Eeq:pa8}
\end{equation}
times a constant homogeneous in $\xi,s$ of order $3-n$, here
$n=0,\ldots,3$. For the lower order terms we use the recursive
construction formula of \eg \cite[pp.44,45 \S I.5.5]{Shubin87} to
deduct that symbols of lower orders have the $\lambda$ dependence
\begin{equation}
    \frac{\lambda^{n'}}{(\det \ac)^m} \; ,
    \label{Eeq:pa9}
\end{equation}
where $4 m - n' > 1$.  We have above deduced the $\lambda$
dependence for all terms in the symbolic expansion of the symbol of
$\Aco^{-1}$, and furthermore, we have the principal part explicitly.

\subsubsection{$\oB$ is a pseudodifferential with parameter of order 0}
\label{sec:Bpsdo}

Given the parametrix of the resolvent, we integrate each term of the
asymptotic series with respect to $\lambda$. To validate this
procedure we show below that each of the terms is finite and
that the integration does not rearrange the terms with respect to
order, \ie the principal term remains the principal term. We also show
that $\oB$ is an operator corresponding to a symbol that is
homogeneous of order 0 in $(\xi,s)$.

As shown in the previous section, each element of the resolvent has
the form
\begin{equation}\label{mn}
    \frac{\lambda^n}{(\det \ac)^m} \; ,
    \nonumber
\end{equation}
with $4m-n\geq 1$. The principal symbol has $m=1$ and $n=0,1,2,3$,
all other terms have homogeneity 0 or lower. In the evaluation of the
integral we distinguish between two
different cases, the principal valued integral corresponding to
$n=3, m=1$ and the other cases. We observe that due to the homogeneity
of the parametrix terms the case $n=3$, $m=1$ is the only principal
integral.

The case $m=1$, $n=3$ gives a finite result, which we find by
evaluation the integral over the integrand \eqref{mn}. In order to
do this claim that we can use the representation
\begin{equation}
    \det \ac = (\lambda-\lambda_{1}^+)(\lambda-\lambda_{2}^+)
               (\lambda-\lambda_{1}^-)(\lambda-\lambda_{2}^-) \; ,
    \label{Eeq:pp2}
\end{equation}
where the eigenvalues, \ie the roots of the fourth order polynomial
$\det \ac$, are denoted by $\lambda^\pm_{1},\lambda^\pm_{2}$ where
the $+(-)$ indicates that they have positive (negative) real part.
Indeed, to show that two of the eigenvalues of $\aS_{;1}$ have
positive(negative) real part we consider the isotropic case. The
isotropic $\det \ac$ have the $\lambda$-roots (\cf Appendix
\ref{app:iso})
\begin{equation}
    \lambda = \pm \sqrt{s^2 \epsilon^{\mathrm{iso}}\mu^{\mathrm{iso}} +
    |\xi|^2} \; ,
    \label{Eeq:pp3}
\end{equation}
\ie two double roots on each side of the strip
$|\RE{\lambda}|<S_{R}\inf_{x'}\sqrt{\epsilon^{\mathrm{iso}}\mu^{\mathrm{iso}}
}$. The lower bound on $\Aco$ in Proposition~\ref{prop2:spec},
part~1 shows that for each anisotropic material with instantaneous
response, the area around the imaginary axis is free from
eigenvalues. We introduce a parameter $\gamma$ in the material
coefficients by
\begin{equation}
    \epsilon^{(\gamma)} := \epsilon^{\rm{iso}} + \gamma(\epsilon-\epsilon^{\rm
    iso})
    \label{Eeq:pp4}
\end{equation}
and analogously for $\mu^{(\gamma)}$. The lower bound on $\ac$, that
is obtained by applying Lemma~\ref{lem2:sesq} and
Proposition~\ref{prop2:spec}, part~1-3, to $\ac$, and apparent in
\eqref{Eeq:nzd}, ensures us that for all $\gamma\in[0,1]$ the $|\det
\ac^{(\gamma)}|$ is bounded from below and that there are no
eigenvalues on the strip around the imaginary axis, and since the
eigenvalues of a matrix depend point-wise continuous on its
coefficient \cite[pp.107-108, \S 2.5.1]{Kato80}, the eigenvalues
vary continuously, but not discontinuously on each side of the
imaginary axis. In the case $\gamma=0$ there are two eigenvalues on
each side, by counting multiplicity and hence, by the continuity of
the eigenvalues, this has to be the case for all $\gamma\in[0,1]$.
Thus we have shown the claim. The symmetry of $\Ac$ discussed in
Remark~\ref{spec:sym} can be used to show the same result for the
special case $s=\RE{s}$. With the representation \eqref{Eeq:pp2} we
evaluate the integral
\begin{equation}
    \lim_{n\rightarrow \infty}\int_{-n}^{n}
    \intd{\IL} \frac{\lambda^3}{\det \ac} \; ,
    \label{Eeq:pp1}
\end{equation}
by partial fraction decomposition. Assume initially that
there are no equal roots then
\begin{equation}
    \frac{\lambda^3}{\det \ac} =
    \frac{D_{1,+}}{\lambda-\lambda_{1}^+} + \frac{D_{2,+}}{\lambda-\lambda_{2}^+} +
    \frac{D_{1,-}}{\lambda-\lambda_{1}^-} +
    \frac{D_{2,-}}{\lambda-\lambda_{2}^-} \; ,
    \label{Eeq:pp5}
\end{equation}
where all $D$ depended only on
$\lambda_{1,2}^{\pm}$. Each such fraction is integrated over the
imaginary axis to become
\begin{equation}
\begin{split}
    \lim_{n\rightarrow \infty} \int_{-n}^{n} \intd{\IL}
    \frac{1}{\lambda - \lambda_{1,2}^{\pm}} & =
    \lim_{n\rightarrow \infty} \left.  \ln (\lambda -
    \lambda_{1,2}^{\pm}) \right|_{\IL = -n}^{\IL = n}
    \\ &=
    \iu \pi \sgn{\RE{\RL -\lambda_{1,2}^{\pm}}} =
    \mp \iu \pi \; ,
\end{split}
    \label{Eeq:pp6}
\end{equation}
where the branch cut is along the negative imaginary axis
and where $|\RL|< |\RE{\lambda_{1,2}^{\pm}}|$. Concerning the choice of
branch cut, observe that the integral above should be summed over each
eigenvalue, thus the branch cut of the logarithm has to be chosen
such that it agrees for all eigenvalues, hence the negative imaginary
axis. Thus
\begin{equation}
    \lim_{n\rightarrow \infty} \int_{-n}^{n} \intd{\IL}
        \frac{\lambda^3}{\det \ac} = \iu \pi (D_{1,+} + D_{2,+}
        -D_{1,-}-D_{2,-})
    \label{Eeq:pp7}
\end{equation}
for $n=0,\ldotsk,3$.
Here
\begin{align}
    D_{1,+} &:= \frac{(\lambda_{1}^{+})^{3}}{(\lambda_{1}^+ -
    \lambda_{2}^+)(\lambda_{1}^+-\lambda_{1}^-)(\lambda_{1}^+ -\lambda_{2}^-)}
    \; , \nonumber \\
    D_{2,+} & := \frac{-(\lambda_{2}^{+})^{3}}
    {(\lambda_{1}^{+}-\lambda_{2}^{+})(\lambda_{2}^{+}-\lambda_{1}^{-})
    (\lambda_{2}^{+}-\lambda_{2}^{-})} \; , \nonumber \\
    D_{1,-} &:= \frac{(\lambda_{1}^-)^{3}}
    {(\lambda_{1}^+-\lambda_{1}^-)(\lambda_{2}^+-\lambda_{1}^-)
    (\lambda_{1}^--\lambda_{2}^-)} \; , \nonumber \\
    D_{2,-} &:= \frac{-(\lambda_{2}^-)^{3}}
    {(\lambda_{1}^+-\lambda_{2}^-)(\lambda_{2}^+-\lambda_{2}^-)
    (\lambda_{1}^--\lambda_{2}^-)} \nonumber \; ,
\end{align}
and each term is homogeneous of order $0$. To show that the sum is bounded
from above we have to eliminate $(\lambda_{1}^{+}-\lambda_{2}^{+})$
and $(\lambda_{1}^{-}-\lambda_{2}^{-})$ from the denominator. We find
that
\begin{equation}
    D_{1,+}+D_{2,+} = \frac{\lambda_{1}^{+}\lambda_{2}^{+}
    (\lambda_{1}^{+} + \lambda_{2}^{+}) - \lambda_{1}^{-} (
        (\lambda_{1}^{+})^2+\lambda_{1}^{+}\lambda_{2}^{+}+(\lambda_{2}^{+})^2) }
    {(\lambda_{1}^{+}-\lambda_{1}^{-})
      (\lambda_{1}^{+}-\lambda_{2}^{-})(\lambda_{2}^{+}-\lambda_{1}^{-})
    (\lambda_{2}^{+}-\lambda_{2}^{-})}
    \label{Eeq:pp13}
\end{equation}
and
\begin{equation}
    D_{1,-} + D_{1,-} = \frac{\lambda_{1}^{-}\lambda_{2}^{-}
    (\lambda_{1}^{-} + \lambda_{2}^{-}) - \lambda_{1}^{+} (
        (\lambda_{1}^{-})^2+\lambda_{1}^{-}\lambda_{2}^{-}+(\lambda_{2}^{-})^2) }
    {(\lambda_{1}^{+}-\lambda_{1}^{-})
      (\lambda_{1}^{+}-\lambda_{2}^{-})(\lambda_{2}^{+}-\lambda_{1}^{-})
    (\lambda_{2}^{+}-\lambda_{2}^{-})} \; ,
    \label{Eeq:pp14}
\end{equation}
hence the denominator is bounded away from zero by $(2\lLb)^4$,
since $|\RL^{\pm}|\geq\lLb=S_R\sqrt{\lowEu\lowMu}$, this follows
from the Corollary~\ref{cor2:spec} that shows that the strip
$|\RL|<\lLb$ is free from eigenvalues. Hence the integral is
bounded.

The case with equal eigenvalues follows similarly. Assume
$\lambda_{2}^+=\lambda_{1}^+$ and $\lambda_{1}^- \neq \lambda_{2}^-$ then
\begin{equation}
    \frac{\lambda^3}{\det \ac} =
    \frac{D_{1,+,2}}{(\lambda-\lambda_{1}^+)^2} + \frac{D_{1,+,1}}{\lambda-\lambda_{1}^+} +
    \frac{D_{1,-}}{\lambda-\lambda_{1}^-} +
    \frac{D_{2,-}}{\lambda-\lambda_{2}^-} \; ,
    \label{Eeq:pp15}
\end{equation}
where
\begin{align}
    D_{1,+,2} &:= \frac{(\lambda_{1}^{+})^{3}}{
    (\lambda_{1}^+-\lambda_{1}^-)(\lambda_{1}^+ -\lambda_{2}^-)} \; ,
    \nonumber \\
    D_{1,+,1} &:= \frac{\lambda_{1}^{+}\lambda_{2}^{+}
    (\lambda_{1}^{+}\lambda_{2}^{+}-\lambda_{1}^{-}\lambda_{2}^{-}) +
    (\lambda_{1}^{+} +\lambda_{2}^{+})(\lambda_{1}^{+}+\lambda_{2}^{+}
    \lambda_{1}^{-}\lambda_{2}^{-} - (\lambda_{1}^{-}-\lambda_{2}^{-})\lambda_{1}^{+}\lambda_{2}^{+})
    }
    {(\lambda_{1}^+-\lambda_{1}^-)^2(\lambda_{1}^+ -\lambda_{2}^-)^2}
    \nonumber
    \; .
\end{align}
With the integral
\begin{equation}
    \lim_{n\rightarrow \infty} \int_{-n}^{n} \intd{\IL}
    \frac{1}{(\lambda - \lambda_{1}^{+})^2}  =
    \lim_{n\rightarrow \infty} \left. \frac{1}{\lambda_{1}^+-\lambda} \right|_{\IL = -n}^{\IL =
    n} =0
    \label{Eeq:pp16}
\end{equation}
and \eqref{Eeq:pp6} we find that
\begin{equation}
    \lim_{n\rightarrow \infty} \int_{-n}^{n} \intd{\IL}
        \frac{\lambda^3}{\det \ac} = \iu \pi (D_{1,+,1}-D_{1,-}-D_{2,-}) \; ,
    \label{Eeq:pp17}
\end{equation}
and hence, it is bounded from above and
homogeneous of order $0$. The case where
$\lambda_{1}^{-} = \lambda_{2}^{-}$ and
$\lambda_{2}^+\neq\lambda_{1}^+$ is totally analogous.

For the case with two equal eigenvalues we have
\begin{equation}
    \frac{\lambda^3}{\det \ac} =
    \frac{D_{+,2}}{(\lambda-\lambda_{1}^+)^2} + \frac{D_{+,1}}{\lambda-\lambda_{1}^+} +
    \frac{D_{-,2}}{(\lambda-\lambda_{1}^-)^2} + \frac{D_{-,1}}{\lambda-\lambda_{2}^-}
    \label{Eeq:pp18}
\end{equation}
and from \eqref{Eeq:pp6} and \eqref{Eeq:pp16} we find that
\begin{equation}
    \lim_{n\rightarrow \infty} \int_{-n}^{n} \intd{\IL}
        \frac{\lambda^3}{\det \ac} = \iu \pi (D_{+,1}-D_{-,1}) \; .
    \label{Eeq:pp19}
\end{equation}
Hence we need only to find $D_{+,1}$ and $D_{-,1}$,
\begin{equation}
    D_{+,1} = \frac{(\lambda_{1}^+)^2(\lambda_{1}^+ - 3\lambda_{1}^-)}
                {(\lambda_{1}^+-\lambda_{1}^-)^3} \; , \ \
    D_{-,1}  = \frac{(\lambda_{1}^-)^2(3\lambda_{1}^+ -\lambda_{1}^-)}
    {(\lambda_{1}^+-\lambda_{1}^-)^3}
    \nonumber
%   \label{Eeq:pp20}
\end{equation}
and hence the integral is homogeneous of order 0 and bounded from
above since the denominator is bounded from below.  We have thus
shown that the integral~\eqref{Eeq:pp1} is well defined, and
homogeneous of order 0 for all possible combinations of
$\lambda$-roots in $\det \ac$.

Next we show that the remaining terms have homogeneous degree +1
compared to the corresponding term in the polyhomogeneous expansion
of $\symb{r}$ and consequently that the integral over $\lambda$ does
not rearrange the symbol expansion. From the construction of the
parametrix of $\Aco$ we know that its asymptotic symbol expansion
has a $\lambda$ dependence of the form \eqref{Eeq:pa9}. Using that
the determinant $\det \ac$ is homogeneous of degree 4 in
$(\xi',s,\lambda)$, we find that
\begin{equation}
    I_{n,m}(x,\xi';s) := \int \intd{\lambda} \frac{\lambda^n}{\left(\det
    \ac(x,\xi';s,\lambda)\right)^m}
    \label{Eeq:ei1}
\end{equation}
is homogeneous of degree $n-4m+1$ if $4m-n\geq 2$. Indeed,
\begin{align}
    \eta^{n-4m+1} I_{n,m}(x,s;\xi') &= \int \intd{\lambda} \eta \frac{(\eta\lambda)^n}
    {\left(\det \ac(x,\eta \xi';\eta s,\eta \lambda)\right)^m}
    \label{Eeq:ei2} \\
    &= \int \intd{\tilde{\lambda}} \frac{\tilde{\lambda}^n}
    {\left(\det \ac(x,\eta \xi';\eta s,\tilde{\lambda})\right)^m} =
    I_{n,m}(x,\eta\xi';\eta s) \; ,
    \nonumber
\end{align}
where we have used 1) that the limits of the integral goes to
infinity, 2) that the $\lambda$-roots scale with $\eta$, implying
that the strip $\tilde{\lambda}_{\mathrm{R}} \leq \eta \tau$ is free
from poles, 3) the scaled integration path is equivalent to the
integration path of $\tilde{\lambda}_{\mathrm{R}} = \tau/2$ since
$\Aco^{-1}$ is analytical in $\lambda$ in the resolvent set and 4)
that $\eta$ is such that $\eta \RE{s}> S_{\mathrm{R}}$. Thus
$I_{n,m}$ is homogeneous of degree $n-4m+1$ in $\xi',s$. Each term
in the polyhomogeneous expansion of $\symb{r}$ have a
$\lambda$-dependence in the form of $I_{n,m}$ with a $\lambda$
independent coefficient. It follows that each integrated term of the
expansion has a homogeneous degree that is one order higher than the
homogeneous degree of each term of $\symb{r}$.

Let
\begin{equation}
    |z|^2 := |\xi'|^2 + |s|^2 \; .
    \nonumber
%   \label{Eeq:ib0}
\end{equation}
To show that each of the integrals $I_{n,m}$ is bounded from above,
for fixed $\xi' , s$ such that $z \neq 0$ we use the estimate of the
lower bound of the determinant for $(s,\lambda)\in Q_1$ (\cf
\S\ref{sec:ellips} and \eqref{Eeq:nzd}),
\begin{equation}
    |\det \ac| \geq C_{1}(|\lambda|^2 + |z|^2)^2 \; .
    \label{Eeq:ib1}
\end{equation}
The principal case $4m-n=1$ is taken care of above see
\eqref{Eeq:pp1}. For $4m-n>1$ we have
\begin{multline}\label{1}
    \left|\int_{\mathbb{R}} \intd{\IL} \frac{\lambda^n}{|\det \ac|^m}\right| \leq
    2 C_{1}^{-m} \int_{0}^\infty \intd{\IL} \frac{|\lambda|^n}
    {(|\lambda|^2+|z|^2)^{2m}} \\
    = 2 C_{1}^{-m} \left(
    \int_{0}^{|z|}  \frac{|\lambda|^n}
    {(|\lambda|^2+|z|^2)^{2m}} \intd{\IL} + \int_{|z|}^{\infty}
    \frac{|\lambda|^n}{(|\lambda|^2+|z|^2)^{2m}} \intd{\IL} \right) \;
    ,
\end{multline}
where for $|\IL|<|z|$ we use
\begin{equation}
    \frac{|\lambda|^n}{(|\lambda|^2 + |z|^2)^{2m}} \leq
    \frac{1}{(|z|^2+|\RL|^2)^{2m-n/2}} \leq \frac{1}{|z|^{4m-n}}\; ,
    \label{Eeq:ib2}
\end{equation}
and for $|\IL|>|z|$ we use the estimate
\begin{equation}
    \frac{|\lambda|^n}{(|\lambda|^2 + |z|^2)^2} \leq
    \frac{1}{(|\RL|^2+|\IL|^{2})^{2m-n/2}} \leq \frac{1}{|\IL|^{4m-n}}
    \; .
    \label{Eeq:ib3}
\end{equation}
Inserting the above estimates into \eqref{1} yields
\begin{equation}
\begin{split}
    \left|\int_{\mathbb{R}} \intd{\IL} \frac{\lambda^n}{|\det \ac|^m}\right|
    & \leq 2 C_{1}^{-m} \left(
    \frac{|z|}{|z|^{4m-n}}  + \int_{|z|}^{\infty} |\IL|^{n-4m}
    \intd{\IL} \right) \\ & =
    \frac{2}{C_{1}^m |z|^{4m-n-1}} \left(
    \frac{4m - n}{(4m-n-1)} \right)
    \label{Eeq:ib4}
\end{split}
\end{equation}
and hence the integral is bounded from above since $|z|>0$ and
$4m-n-1>0$. Thus we find that the asymptotic series expansion of
$\symb{r}$ can be integrated, since each term is finite for $s>S_R$
and $\arg s<\pi/2$. Furthermore, the $\lambda$-integral of
$\symb{r}_-m$, which is homogeneous of order $-m$, results in
$\symb{b}_{-m+1}$ which is homogeneous of order $-m+1$. We have hence
a well defined polyhomogeneous asymptotic expansion of a
pseudodifferential operator with a parameter of homogeneous degree 0
in $\{\xi',s\}$, the corresponding operator is represented in the
usual way through an oscillatory integral.

One can use the residue theorem to evaluate the integrals in
terms of the roots of the equation $\det \ac=0$. This is done for
arbitrary $I_{m,n}$, $4m-n>1$ in Appendix \ref{app:det}.

We have above found an oscillatory integral representation of the
desired operator $\oB$ through the $\lambda$-integral of the symbol
expansion of the resolvent. Its principal symbol is $\int
\intd{\lambda} \symb{r}_{;-1}$, where as usual
$\symb{r}_{;-1}:=\ac^{-1}=(\aS-\lambda I)^{-1}$.  One question
remains in order to associate $\oB$ with $\int \intd{\lambda}
\Aco^{-1}$. It can be reduced to a question of the order of iterated
integrals. Towards this end we use an alternative representation of the
$\lambda$-integral. We note that
\begin{equation}
\int_{\lambda\in K} \intd{\lambda} \ac^{-1} = \int_{\IL\in
[0,\infty], \RL\leq\tau} (\aS_{;1} - I(\RL-\iu \IL ))^{-1}2(\aS_{;1}
- I\RL)(\aS_{;1} - I(\RL+\iu \IL))^{-1}
\end{equation}
where we used the following identity which similar to the (first)
resolvent equation:
\begin{multline}\label{ident}
\big(\aS_{;1} - I(\RL+\iu \IL)\big)^{-1} + \big(\aS_{;1} - I(\RL-\iu
\IL )\big)^{-1} \\ = \big(\aS_{;1} - I(\RL-\iu \IL
)\big)^{-1}2(\aS_{;1} - I\RL)\big(\aS_{;1} - I(\RL+\iu \IL
)\big)^{-1}.
\end{multline}
Denote the right-hand side of the above identity
$w_{;-1}(\lambda,\xi';s,x)$. Clearly this identity holds also if
$\aS_{;1}$ is replaced with the operator $\Ac$. By analyticity of the
resolvent we can choose to integrate along the positive imaginary
axis, \ie $\RL=0$.

Let $u$ be an arbitrary vector in $\mSob{1}$, and consider the two
integrals
\begin{multline}
V_1(x;s):=\int_{\RR^2} \intd[2]{\xi'} (\int_{\RL=0,\IL\geq 0} \intd{\lambda}
\lexp{\iu \xi'\cdot x'}w_{;-1}(\lambda,\xi';s,x) (\OP{F}u)(\xi',x_3))
\\ = \mathcal{F}^{-1}_{\xi'\rightarrow x'}[\symb{b}_{;0}(\xi',s;x)(\mathcal{F}u)(\xi';x_3)]
\end{multline}
and
\begin{multline}
  V_2(x;s):=\int_{\RL=0,\ \IL\geq 0} \intd{\lambda} (\int_{\RR^2} \intd[2]{\xi'}
  \lexp{\iu \xi'\cdot x'}w_{;-1}(\lambda,\xi';s,x)(\OP{F}u)(\xi',x_3) ) \\
=
  \int \intd{\lambda} \mathcal{F}^{-1}_{\xi'\rightarrow x'}[\aS_{;1}^{-1}(\xi',\lambda,s,;x)
  (\mathcal{F}u)(\xi',x_3)]
\end{multline}
Here we have once again used $\mathcal{F}$ to denote the Fourier
transform with respect to $x'$ and $\mathcal{F}^{-1}_{\xi'\rightarrow
  x'}$ to denote the inverse Fourier transform from $\xi'$ to $x'$
variables.  The first integral is the standard way of representing the
action of principal part of $\oB_{;0}$ on $u$. That is
$\oB_{;0}u=V_1$.  The second integral $V_2$ is the $\lambda$-integral
of the first term of the parametrix corresponding to $\Aco$. We thus
have two, possible different, representations of an operator. Below we
will show that the two representations are equal. For the principal
term the problem is reduced to showing that the two iterated integrals
exist and are equal, \eg that $V_1=V_2$. We have the following result
\begin{lemm2}\label{lem:vv}
Let $u\in \mSob{1}$ then for $\RE{s}>S_R>0$ and $\arg s<\pi/2$ it
follows that $V_1(\cdot;s)=V_2(\cdot;s)\in \mLs$.
\end{lemm2}
This result is shown after the proof of
Proposition~\ref{prop2:split2} part~2. We have defined the operator
$\oB$ as the oscillatory integral of the $\lambda$-integral of the
symbol representation of the resolvent expansion, and above shown
that such an operator exists. The desired splitting matrix is
however the $\lambda$-integral over the oscillatory integral over
the resolvent expansion. The above lemma shows that the principal
term of both these expressions are equal for functions on a dense
set in the domain. To continue and show that the remaining terms in
the respective symbol expansions are equal we can once again
construct two iterated integrals and apply the proof of
Lemma~\ref{lem:vv}, \eg the Fubini theorem on this term, and since
all the assumptions carry over the result remains the same.

Hence we have
shown that the two representations of the splitting matrix indeed are
equal and can be applied to the wave-splitting procedure below.\qed

\subsection{Proof of Proposition~\ref{prop2:split2}, part~2}

The symbol of $\oB$ for the isotropic homogeneous medium case is given
in \eqref{Eeq:c98} (see Appendix \ref{app:iso}) and by counting its
powers of $\xi'$ it follows that the corresponding operator $\oB$ can
be restricted to an unbounded operator on $\msLs$ with domain
$\mSob{1}$ and range in $\mSob{0}$.

To show that this result holds also in general we need to show that
the $\xi'$-growth in each of the $\sbb$-terms is at most linear. To
obtain such a result we need good control of the shape of the
parametrix $\sr$ of $\Aco$, which is an asymptotic series of
poly-homogeneous terms $\sr=\ro+\rt+\rd+\ldots$. Recall that (see
\eg~\cite{Shubin87}) $\ro:=\ac^{-1}$ and
\begin{align}
\rt &
=-\ac^{-1}\left(\aS_{;0}\ro+\sum_{|\eta|=1}\left[\partial^\eta_\xi
\aS_{;1}D_x^\eta \ro \right] \right) \\
\rn &= \ac^{-1}\left(\aS_{;0}\rno + \aS_{;-1}\rnt+
\sum_{|\eta|=1}\left[\partial^\eta_\xi \aS_{;0}D_x^\eta \rnt +
\partial^\eta_\xi \aS_{;1}D_x^\eta \ro\right] +\right. \nonumber \\ & \qquad \left. \sum_{|\eta|=2}\partial^\eta_\xi \aS_{;1}D_x^\eta \rnt
 \right), \qquad\ m\geq 2, \label{rn}
\end{align}
where $\eta\in \mathbb{N}^2$ \ie a multi-index,
$D_{x_j}=\frac{1}{\iu}\partial_{x_j}$ and if $\eta=(\eta_1,\eta_2)$,
then $D_x^\eta=D_{x_1}^{\eta_1}D_{x_2}^{\eta_2}$.

Let the linear space of homogeneous polynomials of order $n$ be
denoted by $\hp_n$. The explicit shape of $\det\ac$ in
Appendix~\ref{app:det} ensure that $\det\ac\in
\hp_4(s,\lambda,\xi';x)$. Here we use $\hp_n(s,\lambda,\xi';x)$ to
indicate that $\det\ac$ is a homogeneous polynomial in
$s,\lambda,\xi'$, and have $\set{C}^{\infty}$-coefficients depending on
$x$. Given the homogeneous polynomials $p_1\in\hp_n$, $p_2\in\hp_m$ we can
consider the space of $\hq_{n-m}$ of homogeneous rational functions,
as elements of the form $q=p_1/p_2$ and $q\in \hq_{n-m}$. We will
restrict $\hq_n$ even further and require that $p_2$ is a power of
$\det\ac$. We note two useful properties: Let $q\in
\hq_n(s,\lambda,\xi';x)$, then $D_x^\eta q\in \hq_n(s,\lambda,\xi';x)$
and if $q_1\in\hq_n$, $q_2\in \hq_m$ then $q_1q_2\in\hq_{n+m}$.

In addition to these two spaces we need two additional spaces. The
first is a space of block-diagonal matrices $\hP_n$, where the two 2x2
blocks have elements which are homogeneous polynomials of order $n$.
That is if $\symb{p}\in \hP_n$, then $\symb{p}=\diag_2 (P_1,P_2)$,
where $P_1$ and $P_2$ are 2x2-blocks with each element, $(P_m)_{ij}\in
\hp_n$, for $m=1,2$ and $i,j=1,2$. Clearly for
$\symb{g}\in\hP_n(\lambda,\xi';x)$ and $\symb{h}\in
\hP_m(\lambda,\xi';x)$ we have
$\symb{hg},\symb{gh}\in\hP_{m+n}(\lambda,\xi';x)$ and
$D_x^\eta\symb{g}\in \hP_n(\lambda,\xi',x)$. The second and final space
is $\OO_{-m}$ and an element $\symb{h}_{-m}$ is in $\OO_{-m}$ if it can be
written in the form
\begin{equation}\label{OO_org}
\symb{h}_{-m}(s,\lambda,\xi';x) = \sum_{k=0}^{4m} s^{-m+k}\sum_{n\in
N_k}\symb{g}_{4m-k}^{(m,k,n)}(\lambda,\xi';x)K^{m+k}q_{-4m}^{(m,k,n)}(s,\lambda,\xi';x)
\end{equation}
where $q_{j}^{(m,k,n)}\in \hq_{j}$ for a given $m$, $k=0,...,4m$,
and $n\in N_k$ and $\symb{g}_{j}^{(m,k,n)}\in \hP_{j}$ for a fixed
$m$, $k=0,...,4m$, and all $n$. The matrix $K$ is given in
\eqref{Eeq:bb0b}. Here we have used $\cdot^{(m,k,n)}$ as a way to
index the components of $\symb{h}_{-m}$. We require the number of
elements for each $(m,k)$-level to be finite, \ie $|N_k|<\infty$
where $N_k\subset \mathbb{N}$. We find here the nice properties that
if $\symb{h}\in \OO_{-m}$ then $D_x^\eta\symb{h}\in\OO_{-m}$ and if in addition
$\symb{g}\in \OO_{-n}$ then we have $\symb{gh}\in \OO_{-n-m}$. Note
that the representation~\eqref{OO_org} is not unique due to that
the numerator of $q_{-4m}^{m,\cdot,\cdot}$ may contain a power of $s$.
This non-uniqueness of the representation will be used
constructively in the proof of the lemma below.

We have the following technical lemma:
\begin{lemm2}\label{lem:OO}
  Given the above defined spaces $\OO_{-m}$, $\hp_n$,
  and let
\begin{equation}\label{uv}
  u:=(\xi_1,-\xi_2),\ \text{and}\ v:=(\xi_2,\xi_1).
\end{equation}
Then, for the homogeneous terms of order $-m$ of the symbol of the
parametrix of $\Aco$, $\rno$, we have that $\rno\in\OO_{-m}$, \ie
\begin{equation}\label{OO}
  \rno(s,\lambda,\xi';x) = \sum_{k=0}^{4m}
  s^{-m+k}\sum_{n\in N_k}\symb{g}_{4m-k}^{(m,k,n)}(\lambda,\xi';x)K^{m+k}
  q_{-4m}^{(m,k,n)}(s,\lambda,\xi';x).
\end{equation}
Furthermore, for a fixed $m$ denote the 2x2-block diagonal elements
of $\symb{g}_{4m}^{(m,0,n)}$ by $P_{1n},P_{2n}$, then $P_{1n}=u^T
w_{1n}$ and $P_{2n}=v^Tw_{2n}$, where each of the elements in the
(1,2)-vectors $w_{jn}$ are in $\hp_{4m-1}(\lambda,\xi';x)$ and
$|N_k|<\infty$ for each $k$.
\end{lemm2}
\begin{proof}
  A straightforward calculation shows that $\ac^{-1}=\ro\in \OO_{-1}$
  and $\rt\in\OO_{-2}$. Furthermore inspection of the leading terms
  $\symb{g}_{4}^{(1,0,0)}$ and $\symb{g}_{8}^{(2,0,0)}$ shows that
  their elements can be written as outer products. Indeed, let
  $P_{1},P_{2}$ denote the 2x2 blocks such that
  $\symb{g}_{4}^{(1,0,0)}=\diag_2 (P_{1},P_{2})$, then
  \begin{equation}\label{outer}
    P_{1}=u^Tu \gamma_{1}\ \text{and}\ P_2=v^Tv \gamma_{2},\ \text{where}\
    \gamma_{k}\in \hp_{2}(\lambda,\xi';x), \ k=1,2.
  \end{equation}
  Similarly let $P_{3},P_{4}$ denote the diagonal 2x2 blocks of
  $\symb{g}_{8}^{(2,0,0)}$ then each of these terms are of the form
  $u^Tw_{3}$ and $v^Tw_{4}$ respective, where $w_{k}$ are
  (1,2)-vectors with each element in $\hp_{7}(\lambda,\xi';x)$.

  The construction of $\rn$ in~\eqref{rn} is a product of finitely
  many terms, we consequently find that $\rn$ has at most a finite
  number of terms. This ensures that the $|N_k|<\infty$ for all $k$.

  To show that the lemma is valid for an arbitrary $m\geq 3$ we make
  the recursive assumption that $\rno\in\OO_{-m}$ and
  $\rnt\in\OO_{-m+1}$ with the desired outer-product structure on
  their respective leading matrices $\symb{g}^{(m,0,j)}_{4m}$,
  $\symb{g}_{4m-4}^{(m-1,0,k)}$ for the respective range of $j$ and
  $k$.

  We now calculate $\rn$ from~\eqref{rn} and show that it satisfies the
  lemma. Towards this end we consider the following three 4x4 matrices
  $\symb{c}=\diag_2(c_1,c_2)\in \hP_0(\xi';x)$,
  $\symb{f}=\diag_2(f_1,f_2)\in \hP_0(\xi';x)$, and $\symb{d}$.  The
  matrix $\symb{d}$ is defined by
  $\symb{d}:=\diag_2(u^Td_{11}+d_{12}^Tu,v^Td_{12}+d_{22}^Tv)$ where
  $d_{jk}$ are (1,2)-vectors with elements in $\hp_0(\xi';x)$, and
  $u,v$ are the (1,2)-vectors of $\xi'$-elements defined in~\eqref{uv}.
  Each term in~\eqref{rn} is of the form
\begin{equation}\label{cd}
\ac^{-1}(\symb{c}+\frac{1}{s}\symb{d}K)\symb{h_1},\ \text{where}\
\symb{h}_1\in \OO_{-m}
\end{equation}
or of the form
\begin{equation}\label{f}
\ac^{-1}\frac{1}{s}\symb{f}K\symb{h_2},\ \text{where}\ \symb{h}_2\in
\OO_{-m+1}
\end{equation}
Indeed, the terms containing $\aS_{;0}$ and
$\partial_\xi^\eta\aS_{;1}$ with $|\eta|=1$ belong to the kind
in~\eqref{cd}, and the terms containing $\aS_{;-1}$,
$\partial_\xi^\eta \aS_{;0}$, $|\eta|=1$ and $\partial_\xi^\eta
\aS_{;1}$, $|\eta|=2$ are of the kind~\eqref{f}.

The lemma follows if we can show that the resulting products of
\eqref{cd} and \eqref{f} are elements in $\OO_{-m-1}$ with the
desired outer product-structure on the leading order terms. The
three terms containing $\symb{c}$, $\symb{d}$ and $\symb{f}$
respectively are considered separately. For the first term we note
that $\symb{c}\symb{h}_1\in \OO_{-m}$. Since $\ac^{-1}\in \OO_{-1}$
we immediately find that $\ac^{-1}\symb{c}\symb{h}_1\in\OO_{-m-1}$.
The outer-product structure on the leading term survives since the
leading order term in $\ac^{-1}$ is of the form~\eqref{outer}.

Consider the second term containing $\symb{d}$. We explicitly write
out the leading order elements in
$\ac^{-1}\frac{1}{s}\symb{d}K\symb{h_1}$:
\begin{multline}
  \ac^{-1}\frac{1}{s}\symb{d}K\symb{h_1} =
  \left(\frac{1}{s}\symb{g}_{4}^{(1,0,0)}Kq_{-4}^{(1,0,0)}+s^0\sum_k
    \symb{g}_{3}^{(1,1,k)}q_{-4}^{(1,1,k)}+\cdots\right)\frac{1}{s}\symb{d}K
  \\ \left(\frac{1}{s^m}\sum_j
    \symb{g}_{4m}^{(m,0,j)}K^m q_{-4m}^{(m,0,j)}+\frac{1}{s^{m-1}}\sum_n
    \symb{g}_{4m-1}^{(m,1,n)}K^{m+1}q_{-4m}^{(m,1,n)}+\cdots\right) \\ =
   \frac{1}{s^{m+2}}\symb{p}_{m+2}+\frac{1}{s^{m+1}}\symb{p}_{m+1}+\symb{p}_{0}.
\end{multline}
Here we let $\symb{g}^{(m,k,\cdot)}_{\cdot}$ denote the $k$:th
element in $\symb{h}_1$.

The first of these terms
$\symb{p}_{m+2}=\symb{g}_4^{(1,0,0)}K\symb{d}K\sum_k
\symb{g}_{4m}^{(m,0,k)}K^mq_{-4m}^{(m,0,j)}q_{-4}$. The matrices in
the product all have an outer product structure explicitly given
above and from the observations that
\begin{equation}\label{tild}
K\symb{d}K = K\diagT(d_1,d_2)K=\diagT(d_2,d_1) =: \symb{\tilde{d}},
\end{equation}
and $uv^T=0$, $vu^T=0$ we find that $\symb{p}_{m+2}=0$.

To show that $s^{-m-1}\symb{p}_{m+1}+\symb{p}_0\in\OO_{-m-1}$
consider
\begin{multline}\label{pmo}
\symb{p}_{m+1}=\sum_n
  \symb{g}_{4}^{(1,0,0)}K
\symb{d}K
  \symb{g}_{4m-1}^{(m,1,n)}K^{m+1}q_{-4m}^{(m,1,n)}q_{-4}^{(1,0,0)} \\ +
  \sum_{k,j}
    \symb{g}_{3}^{(1,1,k)}\symb{d}K
    \symb{g}_{4m}^{(m,0,j)}K^m q_{-4m}^{(m,0,j)}q_{-4}^{(1,1,k)} \\ =
\sum_n
  \symb{g}_{4}^{(1,0,0)}
\symb{\tilde{d}}
  \symb{g}_{4m-1}^{(m,1,n)}K^{m+1}q_{-4m}^{(m,1,n)}q_{-4}^{(1,0,0)} +
  \sum_{k,j}
    \symb{g}_{3}^{(1,1,k)}\symb{d}
    \tilde{\symb{g}}_{4m}^{(m,0,j)}K^{m+1} q_{-4m}^{(m,0,j)}q_{-4}^{(1,1,k)} \\
    = \sum_n M_{1n} + \sum_{j,k} M_{2jk},
\end{multline}
where we have once again have used the notation~\eqref{tild}. Upon
multiplying block-diagonal matrices with other block diagonal
matrices all these with elements which are homogeneous polynomials yield
that $\symb{g}_{3}^{(1,1,k)}\symb{\tilde{d}}
\symb{g}_{4m}^{(m,0,j)}$, $\symb{g}_{4}^{(1,0,0)}\symb{d}
\tilde{\symb{g}}_{4m-1}^{(m,1,n)}\in \hP_{4m+4}$. Furthermore,
$q_{-4m}^{(m,\cdot,j)}q_{-4}^{(1,\cdot,k)}\in \hq_{-4m-4}$. This
suffice for $s^{-m-1}\symb{p}_{m+1}$ to be a leading term of an
element in $\OO_{-m-1}$. Similar matrix algebra for a typical term
in $\symb{p}_0$ for a given $s^j$-order we find that each such term
fits into a $\OO_{-m-1}$ element. The remaining issue of the
$\symb{d}$-containing terms is the outer product structure of
$\symb{p}_{m+1}$. Similarly to the $\symb{c}$-terms it is clear that
$M_{1n}$ has the appropriate outer product structure. The term
$M_{2jk}$ is a bit more subtle, and we need to use the
outer-product structure of each of the three matrices. There are two
types of terms in $\symb{g}_{3}^{(1,1,j)}=\symb{p}_3+\symb{p}_4$,
where $\symb{p}_3:=\diag_2(p_1I,p_2I)$ with
$p_1,p_2\in\hp_3(\lambda,\xi';x)$, and
$\symb{p_4}:=\diag_2(p_{11}^Tv+u^Tp_{12},p_{21}^Tu+v^Tp_{22})$ where
$p_{jk}$ are (1,2)-vectors with each element in
$\hp_2(\lambda,\xi';x)$. The $\symb{p}_3$-term in $M_{2jk}$ yields
\begin{multline}
\begin{pmatrix}p_1I & 0 \\ 0 & p_2I\end{pmatrix}
\begin{pmatrix}u^Td_{11}+d_{12}^Tu & 0 \\ 0 & v^Td_{12}+d_{22}^Tv\end{pmatrix}
\begin{pmatrix}v^Tw_1 & 0 \\ 0 & u^Tw_2\end{pmatrix} \\ =
\begin{pmatrix} u^Td_{11}v^Tw_1p_1 & 0 \\ 0 & v^Td_{12}u^Tw_2p_2 \end{pmatrix}
\end{multline}
and it has the desired outer product structure. To obtain this
result we have repeatedly used that $uv^T=0,vu^T=0$. Similarly for
the $\symb{p}_4$ term we have
\begin{multline}
\begin{pmatrix}p_{11}^Tv+u^Tp_{12} & 0 \\ 0 & p_{21}u+v^Tp_{22}\end{pmatrix}
\begin{pmatrix}u^Td_{11}+d_{12}^Tu & 0 \\ 0 & v^Td_{12}+d_{22}^Tv\end{pmatrix}
\begin{pmatrix}v^Tw_1 & 0 \\ 0 & u^Tw_2\end{pmatrix} \\ =
\begin{pmatrix} u^Tp_{12}u^Td_{11}v^Tw_1 & 0 \\ 0 & v^Tp_{22}v^Td_{12}u^Tw_2 \end{pmatrix}
\end{multline}
which once again has the appropriate outer product structure. We
have hence shown that the terms containing $\symb{d}$ are elements
in $\OO_{-m-1}$ with the desired outer product structure.

The last kind of terms are these which contain $\symb{f}$. These
terms are of the form
\begin{multline}
  \ac^{-1}\frac{1}{s}\symb{f}K\symb{h_2} =
  \left(\frac{1}{s}\symb{g}_{4}^{(1,0,0)}Kq_{-4}^{(1,0,0)}+s^0\sum_k
    \symb{g}_{3}^{(1,1,k)}q_{-4}^{(1,1,k)}+\cdots\right)\frac{1}{s}\symb{f}K
  \\ \left(\frac{1}{s^{m-1}}\sum_j
    \symb{g}_{4m-4}^{(m-1,0,j)}K^{m-1} q_{-4m+4}^{(m-1,0,j)}+\frac{1}{s^{m-2}}\sum_n
    \symb{g}_{4m-5}^{(m-1,1,n)}K^{m+2}q_{-4m+4}^{(m-1,1,n)}+\cdots\right) \\ =
   \frac{1}{s^{m+1}}\symb{q}_{m+1}+\symb{q}_{0}.
\end{multline}
The leading order term $\symb{q}_{m+1}$ is explicitly
\begin{equation}\label{ft}
  \symb{q}_{m+1}=\sum_j
    \symb{g}_{4}^{(1,0,0)}K\symb{f}K
    \symb{g}_{4m-4}^{(m-1,0,j)}K^{m-1} q_{-4m+4}^{(m-1,0,j)}q_{-4}^{(1,0,0)}
\end{equation}
Clearly $q_0:=q_{-4m+4}^{(m-1,0,j)}q_{-4}^{(1,0,0)}\in \hq_{-4m}$ and
$\symb{g}_0:=\symb{g}_{4}^{(1,0,0)}\tilde{\symb{f}}
\symb{g}_{4m-4}^{(m-1,0,j)}\in\hP_{4m}$. Observe however that
$q_0/\det\ac\in \hq_{-4m-4}$ and that $\det\ac =
p_0(\lambda,\xi';x)+s^2p_1(\lambda,\xi';x)+s^4p_2(\lambda,\xi';x)$
where $p_0\in \hp_4$, $p_1\in\hp_2$ and $p_2\in\hp_0$.  We note that
$\symb{g}_0p_0\in \hP_{4m+4}$ and that terms of the form
$s^{2j-m-1}\symb{g}_0p_j$, $j=1,2$ fit nicely as lower order terms in
$\OO_{-m-1}$. Similarly we can consider the terms of $\symb{q}_0$ by
explicitly calculating the typical $s^j$-order terms and see that each
such term fits into $\OO_{-m-1}$ to finally draw the conclusion that
$\ac^{-1}\frac{1}{s}\symb{f}K\symb{h_2}\in \OO_{-m-1}$ and
consequently that $\rn\in\OO_{-m-1}$. The outer product structure of
the $\symb{f}$-terms follows directly from \eqref{ft} and the fact
that $\symb{g}_4^{(1,0,0)}$ has the appropriate outer product form.

We can now by a recursion argument draw the conclusion that the
lemma is valid for all $m\geq 1$.
\end{proof}
To show that the operator corresponding to $(\symb{b}_{;-m})$ maps
$\mSob{n-1}\rightarrow \mSob{n}$, it suffices to show that
$(\symb{b}_{;-m})_{jk}\in \set{S}_{1,0}^1$, the space of symbols
first defined by H\"ormander. This means that we need to show that
\begin{equation}\label{H}
|\partial_{\xi'}^\eta\partial_x^\beta (\symb{b}_{;-m})_{jk}|\leq
C_{\eta,\beta}(1+|\xi'|^2)^{1-|\eta|}
\end{equation}
for $\eta,\beta\in \mathbb{N}^2$. From the property of $\OO_{-m}$ we
know that $\partial_x^\beta \symb{r}_{;-m}\in W_{-m}$ for
$(s,\lambda)\in Q_1$.

To show \eqref{H} for $\eta=(0,0)$ we recall that when $(s,\lambda)\in
Q_1$ is $\det\ac$~\eqref{Eeq:ep8} elliptic. This imply that a typical
term of $\symb{r}_{-m}$ can be bounded as
\begin{equation}\label{Lint}
\left|\int_\RR \intd{\lambda_I} \frac{\lambda_I^n}{(\det\ac)^m}\right|\leq C
(|s|^2+|\xi'|^2)^{\frac{n-4m+1}{2}},
\end{equation}
for all $4m-n\geq 1$. For the case $4m-n=1$ the result follows in the
sense of a principal integral. The symbol $\symb{b}$ can be written as
an expansion of terms which are homogeneous in $s,\xi'$, each
constructed by integrating the corresponding $\symb{r}_{-m}$-term.
The $\symb{r}_{-m}$ terms are of the form~\eqref{OO}, and for all
$k\geq 1$ in \eqref{OO} it follows from each matrix element's
homogeneity order and~\eqref{Lint} that the resulting integral is
bounded for all values of $\xi'$. The leading orders $k=0$ have the
outer product structure, and we can therefore apply~\eqref{Lint} to
find that the $\lambda$-integral of these terms grows at most linearly
in $\xi'$ for large values of $\xi'$. Applying the derivative
$\partial_{\xi'}^\eta$ on an element in $\symb{r}_{;-m}$ we find that
this reduces the growth in $\xi'$ with $|\eta|$-order for the
corresponding $\symb{b}_{;-m+1}$ symbol, since the integrand is a sum
of rational functions; it is a sum of polynomials in $s,\xi',\lambda$
over powers of $\det\ac$ both with smooth coefficients. The partial
derivative is hence a bounded or a more regular function in $\xi'$,
and due to the ellipticity of $\det\ac$ and $\RE{s}\geq S_R$ there
exists a $\lambda$-integrable $\xi'$-independent function so that we
can apply the dominated convergence theorem to show that the
interchange of $\lambda$-integral and $\partial_{\xi'}^\eta$ is
allowed. It is clear that the partial derivative exists everywhere
because that the elements of $\symb{r}_{;-m}$ are rational
functions. We thus find that $(\symb{b}_{;-m+1})_{jk}\in S^1_{1,0}$
and the corresponding operator maps $\mSob{n}$ to $\mSob{n-1}$ as
desired. This result is valid for any $m$, and hence we have the same
result for $\oB$. \qed

\subsection{Proof of Lemma~\ref{lem:vv}}

To show this result on iterated integrals is a standard application of
Fubini's theorem for positive integrands, see e.g.~\cite{Lieb+Loss}.
It states that {\it a non-negative measurable function} on the usual
Lebesgue-measure over the $\lambda,\xi'$-domain has its iterated
integrals equal and finite if one of them is finite. We will apply
this to the iterated integrals $V_1$ and $V_2$ with integrand:
\begin{equation}
g(\lambda,\xi';s,x):=\lexp{\iu \xi'\cdot
x'}(w_{;-1}(\lambda,\xi',s;x))_{ij} (\OP{F}u)(\xi),\ i,j=1,2
\end{equation}
where
\begin{equation}
w_{-1}(\lambda,\xi',s;x):=\big(\aS - I(\RL-\iu \IL )\big)^{-1}2(\aS -
I\RL)\big(\aS - I(\RL+\iu \IL )\big)^{-1}.
\end{equation}
We split the integrand into positive real parts
$g=g_{R+}-g_{R-}+\iu(g_{I+}-g_{I-})$, where $g_k>0$ for each
$k\in\{R+,R-,I+,I-\}$. We consider the iterated integral over each of
these positive functions separately. In order to show that $g_k$ for
any $k\in\{R+,R-,I+,I-\}$ is measurable, it is enough to note that
$w_{;-1}$ is continuous in both $\lambda$ and $\xi'$ and so is
$\lexp{\iu \xi'\cdot x'}$, their respective restriction, \eg the
real and non-negative (or any of the other combination) is also
measurable since they are piecewise continuous.  From the
assumptions of the lemma we find that $u\in \mLs$ and it is hence
$\xi'$-measurable and consequently by trivially extending $u$ to be a
function on the product space, it is measurable in both $\lambda$
and $\xi'$ jointly. We then utilize that products of measurable
functions are measurable. Consequently we know that the parts of the
integrals $V_1$ and $V_2$ corresponding \eg the real positive
part exist and are equal. To finish the lemma we note that
the operator corresponding to $\symb{b}_{;-m}$ maps $\setSobolev{1}$
to $\set{L}^2$. This was shown in the previous section. Hence $V_1=V_2\in
\set{L}^2$.\qed

\subsection{Proof of Proposition \ref{prop2:split2}, parts~3 and 4}

That $\oB$ and $\Ac$ commutes in the sense that %(\cf \eqref{Eeq:smooth})
\begin{equation}
            \qB{1}\Ac = \Ac \qB{3} \; ,
\end{equation}
on the set $\mSob{3}$ follows directly form the fact that the
resolvent commutes with $\Ac$ due to the Fubini-Tonelli theorem and
that $\Ac$ commutes in a weak sense with the integral over $\IL$ that
is used to define $\oB$. (See \cite{Jonsson+deHoop01} Proposition~2, Part 3.)

To show that $\sB^2=I$ one can introduce two projectors defined from
the lambda-integral over the resolvent, $\oB$ can be shown to be the
difference of these two projectors, and the sum of the projectors
equals the identity.  The squaring of the operator is hence the
identity operator. The key to this proof is to show that the two
operators $\mathcal{P}_{\pm}:=\frac{1}{2}(I \pm \sB)$, are
projectors, it is done by utilizing the (first) resolvent equation.
A proof of the projector properties is detailed in
\cite{Jonsson+deHoop01} Proposition~2, part~3 and 4 for the acoustic
case. The electromagnetic case follows analogously, and since is
somewhat lengthy, we will not repeat it here. Once this is known we
note that $\mathcal{P}_{+}+\mathcal{P}_{-}=I$, and
$\mathcal{P}_{-}\mathcal{P}_{+}=0$ etc., on an core set.
Consequently, on operator level $\sB^2 =
(\mathcal{P}_{+}-\mathcal{P}_{-})^2=
\mathcal{P}_{+}+\mathcal{P}_{-}=I$.\qed

\subsection{Proof of Proposition \ref{prop2:split2}, part 5}

Let $\gamma$ ba a scalar, find all $(\gamma,\sL)$ with non-zero
$\sL$ such that
\begin{equation}
    \sB \spmL = \gamma \spmL \; , \label{Eeq:be0}
\end{equation}
where $\spmL$ is a `vector' of $2\times 2$-block matrices of scalar
operators. To solve this eigenvalue-like problem, we use that $\sB$
is an involution, that is
\begin{align}
    \sB_{11}\sB_{12} + \sB_{12} \sB_{22} & = 0 \; , \label{Eeq:be3} \\
    \sB_{11}^2 + \sB_{12}\sB_{21} &= I \; . \label{Eeq:be4}
\end{align}
Writing \eqref{Eeq:be0} explicitly with $2\times 2$ blocks gives
\begin{align}
    \sB_{11} \spmL_{1} + \sB_{12} \spmL_{2} & = \gamma \spmL_{1} \; ,\\
    \sB_{21} \spmL_{1} + \sB_{22} \spmL_{2} & = \gamma \spmL_{2} \; .
\end{align}
and collecting similar terms yields
\begin{align}
    \sB_{12} \spmL_{2} & = (\gamma I - \sB_{11}) \spmL_{1} \; , \label{Eeq:be1}\\
    \sB_{21} \spmL_{1} & = (\gamma I -\sB_{22}) \spmL_{2} \; .  \label{Eeq:be2}
\end{align}
Let $(\sB_{11} + \gamma I)$ act on \eqref{Eeq:be1} and use
\eqref{Eeq:be3}. Then
\begin{equation}
\begin{split}
    (\gamma^2 I - \sB_{11}^{2}) \spmL_{1} &= (\sB_{11} + \gamma
    I)\sB_{12}  \spmL_{2}
    \\ & = \sB_{12} (\gamma  I -\sB_{22})\spmL_{2} \; ,
\end{split}
\label{Eeq:be5}
\end{equation}
and analogously let $\sB_{12}$ act on \eqref{Eeq:be2} and use
\eqref{Eeq:be4}, then
\begin{equation}
\begin{split}
    \sB_{12}(\gamma I - \sB_{22}) \spmL_{2} &= \sB_{12}\sB_{21}\spmL_{1} \\
    & = (I-\sB_{11}^2)\spmL_{1} \; .
\end{split}
    \label{Eeq:be6}
\end{equation}
Substituting \eqref{Eeq:be5} into \eqref{Eeq:be6} gives after
simplification
\begin{equation}
    (I- \gamma^2 I) \spmL_{1} = 0 \; .
\end{equation}
Thus $\gamma = \pm 1$ since $\spmL$ is assumed to be non-zero. The
corresponding eigenvectors are obtained by solving the following
linear system
\begin{align}
    (\pm I - \sB_{11}) \spmL_{1} &= \sB_{12} \spmL_{2}  \; ,\label{Eeq:be7}\\
    (\pm I - \sB_{22}) \spmL_{2} &= \sB_{21} \spmL_{1} \; . \label{Eeq:be7b}
\end{align}
Rewrite \eqref{Eeq:be4} into
\begin{equation}
    (\pm I - \sB_{11})(\pm I + \sB_{11}) = \sB_{12} \sB_{21} \; .
\end{equation}
Comparison with \eqref{Eeq:be7} gives that the generalized eigenvectors
have the form
\begin{equation}
    \spmL = \begin{pmatrix}  \pm I + \sB_{11} \\ \sB_{21} \end{pmatrix}
    \spmN
    \label{Eeq:be8}
\end{equation}
for arbitrary normalization operators $\spmN$.
The condition $(\sB)^2=I$ imposes, in addition to \eqref{Eeq:be3} and
\eqref{Eeq:be4} the conditions
\begin{align}
    \sB_{21}\sB_{11} + \sB_{22}\sB_{21} &= 0 \label{eq:bB1}\; , \\
    \sB_{21}\sB_{12} + \sB_{22}^{2} &= I \; . \label{eq:bB2}
\end{align}
Hence with use of \eqref{eq:bB1} we find that $\spmL$ is also a solution
to \eqref{Eeq:be7b}. An alternative form of $\spmL$ is obtained if we
use \eqref{eq:bB2} and \eqref{Eeq:be7b}
\begin{equation}
    (\spmL)' = \begin{pmatrix}  \sB_{12} \\ \pm I + \sB_{22} \end{pmatrix}
    (\spmN)' \; ,
    \label{Eeq:be9}
\end{equation}
which is related to \eqref{Eeq:be8} by the normalization $(\spmN)' =
(\pm I - \sB_{22})(\spmN)''$ and $\spmN = \sB_{12} (\spmN)''$, hence
$\spmL$ and $(\spmL)'$ differ only by a normalization.
\endproof

\subsection{Proof of Proposition \ref{prop2:split2}, part 6}

That $\sB$ is one-to-one follows directly from the fact that $\sB$
is an involution. Indeed for any $F \in \setC$ we have
\begin{equation}
    \sB F = 0 \Rightarrow F = \sB 0 = 0 \; .
\end{equation}
Hence the null space of $\sB$ contains only the element $0$ and thus $\sB$ is
one-to-one. That the null space is trivial implies a condition on
$\sB_{21}$, to see this consider
$F=(F_{1},0)$ and the equation
\begin{equation}
    \sB F = \begin{pmatrix} \sB_{11} F_{1} \\ \sB_{21}
    F_{1}\end{pmatrix} = 0 \; .
\end{equation}
then, since $\sB$ is one-to-one, $\sB_{21}$ is either trivial or
one-to-one, and since the
symbol of $\sB_{21}$ is non-trivial, it follows that $\sB_{21}$ is
non-trivial and hence one-to-one, and thus invertible on its range.
\endproof

\subsection{Proof of Proposition \ref{prop2:split2}, part 7}

We now extend $\spmL$ to a larger domain.  This is done by using
$(\qB{q})_{11}$ and $(\qB{q})_{21}$ in the place of $\sB_{11}$ and
$\sB_{21}$ in \eqref{Eeq:be8}, the generalized eigenvector, together
with the extension of $\spmN$ to a bounded invertible operator
$\pmN:\setSobolev{q}(\mathbb{R}^2,\mathbb{C}^2) \rightarrow
\setSobolev{q}(\mathbb{R}^2,\mathbb{C}^2)$ we obtain a generalization
of $\spmL$ to $\qpmL{q}: \setSobolev{q}(\mathbb{R}^2,\mathbb{C}^2)
\rightarrow \mSob{q-1}$, where the domain of $\qpmL{q}$ follows
from the domain of $\qB{q}$.
\endproof

\section{Directional decomposition}
\label{sec:5}

We have above collected enough information to proceed and answer the
initial question about the existence of $\{\oL,\oV \}$, \ie does the
decomposition of $\Ac$ exists. Most of the proof are done for the
set $\setC$, but in the end we extend the results to the general
case.

With the definition of the splitting matrix in Proposition
\ref{prop2:split2}, in particular the commutation between the splitting
matrix and the electromagnetic system's matrix (see Proposition
\ref{prop2:split2}, part 3), we obtain the
decomposition by the following proposition.
\begin{prop} \label{prop2:dec}
    The equation
    \begin{displaymath}
        \Ac \qL{3} = \qL{1} \OP{V} \; ,
    \end{displaymath}
    has a solution where the columns of $\qL{q}$ are the generalized
    eigenvectors $\qpmL{q}$ to $\qB{q}$ for $q=1,3$ and where
    $\OP{V}$ is a block diagonal matrix with the elements
    $\wpm : \setSobolev{3}(\mathbb{R}^2,\mathbb{C}^2) \rightarrow \setSobolev{1}(\mathbb{R}^2,\mathbb{C}^2)$,
    representing a generalization of the vertical wave number and
    \begin{equation}
        \wpm = (\qN{1})^{-1} ((\qB{1})_{21})^{-1}\left(
        \Ac_{21} (\pm I- (\qB{3})_{11}) + \Ac_{22} (\qB{3})_{21} \right)
        \qN{3} \; ,
    \end{equation}
    where $\qN{q} = \qpN{q} = \qmN{q}$, $q=1,3$.
\end{prop}
\begin{remark}
    If one considers the equation $\sA \spmL = \spmL \swpm$ with the
    particular normalization $\spmN = \sB_{21}^{-1}$ and let
    \begin{equation}
        \spmDtN = (\pm I + \sB_{11})\sB_{21}^{-1} \; ,
    \end{equation}
    then, upon eliminating $\swpm$, one finds that establishing
    the decomposition is
    equivalent to solving the equation
    \begin{equation}
        \spmDtN \sA_{21} \spmDtN + \spmDtN \sA_{22} - \sA_{11} \spmDtN
        - \sA_{12}  = 0 \; , \label{eq:rem31}
    \end{equation}
    \ie an algebraic Riccati operator equation. As $\spmL$ solves the
    decomposition problem we have the fact that $\spmDtN$ solves the
    associated algebraic Riccati operator equation. The map
    $\spmDtN$ is denoted the impedance mapping. Note that one can obtain
    a corresponding admittance mapping, $(\spmDtN)^{-1}$, that solves the
    algebraic Riccati
    operator equation that is obtained by operating with
    $(\spmDtN)^{-1}$ on both sides of \eqref{eq:rem31}.
\end{remark}

\subsection{Proof of Proposition \ref{prop2:dec}}

To show that $\qpmL{q}$ decomposes $\Ac$, we begin with the proof on
$\setC$. By Proposition \ref{prop2:split2}, part 3 and 5 we have
\begin{equation}
    \sB \sA \spmL = \sA \sB \spmL = \pm \sA \spmL \; .
    \label{Eeq:de1}
\end{equation}
Let $\OP{H}^{\pm}:= \sA \spmL$, then by \eqref{Eeq:de1}
\begin{equation}
    \sB \OP{H}^{\pm} = \pm \OP{H}^{\pm} \; ,
\end{equation}
but from Proposition \ref{prop2:split2}, part 5 we know that for
some arbitrary normalization operator, $\OP{M}^{\pm}$, we have
\begin{equation}
    \OP{H}^{\pm} = \begin{pmatrix} \pm I + \sB_{11} \\ \sB_{21}
    \end{pmatrix} \OP{M}^{\pm} = \spmL \swpm \; ,
\end{equation}
by the choice of normalization operator $\OP{M}^{\pm} = \spmN \swpm$,
for some particular $\swpm$.
From the definition of $\OP{H}^{\pm}$, we get
\begin{equation}
    \sA \spmL = \OP{H}^{\pm} = \spmL \swpm \; .
    \label{Eeq:de2}
\end{equation}
Hence $\spmL$ are generalized eigenvectors of $\sA$ with generalized
eigenvectors $\swpm$ of $2\times 2$ matrices of scalar operators.
To obtain an explicit expression for $\swpm$, consider \eqref{Eeq:de2}
explicitly
\begin{align}
    (\sA_{11}(\pm I + \sB_{11}) + \sA_{12} \sB_{21})\spmN &= (\pm I +
    \sB_{11}) \spmN\swpm \; ,
    \label{Eeq:de3} \\
    (\sA_{21}(\pm I + \sB_{11}) + \sA_{22} \sB_{21})\spmN & =
    \sB_{21} \spmN \swpm \; .
    \label{Eeq:de4}
\end{align}
From \eqref{Eeq:de4} we obtain that the range of the left
and right hand sides have to agree, thus the left hand side is within the
range of $\sB_{21}$ and hence the inverse is defined on this range and
thus
\begin{equation}
    \swpm = (\spmN)^{-1} (\sB_{21})^{-1} (\sA_{21}(\pm I + \sB_{11})
    + \sA_{22} \sB_{21})\spmN \; ,
    \label{Eeq:de5}
\end{equation}
is well defined and it also is the generalized eigenvalue to $\Ac$.
To see that \eqref{Eeq:de3} gives the same result, we use two of the
equations implied by the commutation of $\sA$ and $\sB$, \textit{viz.}
\begin{align}
    \sA_{11} \sB_{11} + \sA_{12} \sB_{21} &= \sB_{11}\sA_{11} + \sB_{12}
    \sA_{21} \; ,
    \label{Eeq:de6} \\
    \sA_{21} \sB_{11} + \sA_{22} \sB_{21} &= \sB_{21}\sA_{11} + \sB_{22}
    \sA_{21} \label{Eeq:de7} \; .
\end{align}
We rewrite \eqref{Eeq:de3} and apply \eqref{Eeq:de6} to obtain
\begin{equation}
\begin{split}
    (\pm \sA_{11} + \sA_{11}\sB_{11} + \sA_{12} \sB_{21})\spmN & =
    ((\pm I + \sB_{11})\sA_{11} + \sB_{12}\sA_{21}) \spmN \\ & =
    (\pm I +
    \sB_{11}) \spmN\swpm \; .
\end{split}
    \label{Eeq:de8}
\end{equation}
Applying $\pm I - \sB_{11}$ to both sides and using \eqref{Eeq:be3} and
\eqref{Eeq:be4} gives
\begin{align}
    ((I - \sB_{11}^2)\sA_{11} + (\pm I - \sB_{11})\sB_{12}\sA_{21})
    \spmN &=
    (I -\sB_{11}^2) \spmN\swpm \Leftrightarrow \\
    \sB_{12}(\pm \sA_{21} + \sB_{21}\sA_{11} + \sB_{22}\sA_{21}) \spmN & =
    \sB_{12}\sB_{21} \spmN\swpm \; ,
\end{align}
and using \eqref{Eeq:de7} gives
\begin{equation}
    \sB_{12}(\sA_{21}(\pm I + \sB_{11}) + \sA_{22} \sB_{21}) \spmN =
    \sB_{12}\sB_{21} \spmN\swpm \; ,
    \label{Eeq:de10}
\end{equation}
and hence an equation for $\swpm$ that is equivalent to \eqref{Eeq:de4}. Thus we
have shown that the two expressions \eqref{Eeq:de3} and \eqref{Eeq:de4}
are equivalent and that \eqref{Eeq:de5} is the solution to both.

Before we extend \eqref{Eeq:de2} to the
general one, we introduce the matrix operators
\begin{equation}
    \sV=\begin{pmatrix} \swp & 0 \\ 0 & \swm \end{pmatrix}
\ \ \mbox{and} \ \
    \sL = \begin{pmatrix}  I + \sB_{11} & -I + \sB_{11} \\ \sB_{21} &
    \sB_{21}  \end{pmatrix} \sN \; ,
\end{equation}
where we have made the choice $\sN = \spN = \smN$.
With the introduced notation, \eqref{Eeq:de2} becomes
\begin{equation}
    \sA \sL = \sL \sV \; ,
    \label{Eeq:de13}
\end{equation}
and furthermore, we may rewrite this equation into
\begin{equation}
    \Ac \sL = \qL{1} \sV \; ,
    \label{Eeq:de14}
\end{equation}
where by the definition of $\qB{q}$ we know that $\qL{q}$ is well
defined, the choice of $\qL{1}$ follows since $\qL{1}: \mSob{1}
\rightarrow \mSob{0}$.
To extend the domain to a larger set, let
$\{ F_{n} \}_{n=1}^{\infty} \subset \setC$ be a Cauchy sequence.
Then for fixed $\eta>0$ and large enough $n$ we have
\begin{equation}
    \mLnorm{\Ac \sL F_{n} - \Ac \qL{3} F_{n}} \leq \eta \; ,
    \label{Eeq:de15}
\end{equation}
as long as $\lim_{n\rightarrow \infty} F_{n} = F_{0} \in
\domain{\qL{3}} = \mSob{3}$.
Let
\begin{equation}
    \Ac \qL{3} F_{0} = G \; .
    \label{Eeq:de16}
\end{equation}
Subtracting \eqref{Eeq:de16} from \eqref{Eeq:de14} and using
\eqref{Eeq:de15} we obtain
that for large enough $n$
\begin{equation}
    \eta \geq \mLnorm{\Ac \sL F_{n} - \Ac \qL{3} F_{0}} =
    \mLnorm{\qL{1}\sV F_{n}-G} \; ,
\end{equation}
thus the limit of the right hand side exists, that is there exists an
extension of $\qL{1} \sV$.

Due to the requirement of equal domains of the extensions of $\qL{1} \sV$
and $\Ac\qL{3}$ we find that
\begin{equation}
    G = \qL{1} \qV{3} F_{0}\; ,
\end{equation}
where the elements of $\qV{3}$, $\wpm$ have the form
\begin{equation}
    \wpm = (\qN{1})^{-1} ((\qB{1})_{21})^{-1} (\Ac_{21}(\pm I +
    (\qB{3})_{11}) + \Ac_{22}(\qB{3})_{21})\qN{3} \;
\end{equation}
and $\wpm:
\setSobolev{3}(\mathbb{R}^2,\mathbb{C}^2) \rightarrow \setSobolev{1}(\mathbb{R}^2,\mathbb{C}^2)$. Thus on
$\mSob{3}$ we have obtained
\begin{equation}
    \Ac \qL{3} = \qL{1} \qV{3} \; .
\end{equation}
\endproof

\section{Discussion of the result}

By applying functional analysis to the problem of
decomposition of the wave field for the electromagnetic system's
matrix we have extended the wave-splitting procedure to an
anisotropic media whose properties vary with all three spatial
coordinates. The result extends beyond the up/down symmetric case.
The analysis of the spectrum shows that a strip around the imaginary
axis is in the resolvent set. We define a resolvent integral whose
contour lies in this strip. Due to the explicit form of the systems
matrix, we do not have a full spectral resolution of the operator.
Still, the resolvent integral over a path in the resolvent strip is
shown to be well defined by applying the elliptic theory of
pseudodifferential operators with parameters.

Using this resolvent integral we define a splitting matrix. This
matrix has the feature that one can construct its generalized
eigenvectors of operators corresponding to the (generalized)
eigenvalues $\pm 1$. We have above shown that the splitting matrix
commutes with the electromagnetic system's matrix. One consequence
of this `commutation' of the operators is that the generalized
eigenvectors of the splitting matrix also are generalized
eigenvectors to the electromagnetic system's matrix. The
corresponding generalized eigenvalue to the system's matrix (a $2
\times 2$ matrix operator) is the key ingredient in the definition
of one-way equations for the electromagnetic case. This `eigenvalue'
is the electromagnetic generalization of the vertical wave number
obtained in the linear acoustic case. One of the features of this
procedure of decomposition is that we have constructed the
composition operator without having to invert any of the elements of
the splitting matrix: the construction relies on the fact that
the splitting matrix is an involution than in the corresponding
acoustic case. However, this result can also be carried over to the
acoustic case. The removal of the inverse of an element in the
splitting matrix from the splitting process, is not complete. It
remains in the one-way equation obtained after the splitting, even
though we have been able to remove it from the composition operator.
The generalized eigenvectors to the splitting matrix is used to
generate the composition matrix that decomposes the electromagnetic
system's matrix.

The traditional approach to the decomposition of the system's matrix
gives an algebraic Riccati operator equation. The splitting matrix
construction of the decomposition gives us a family of solutions to
this operator equation in terms of the elements of an integral over
the resolvent of the electromagnetic system's matrix.

Once we have obtained the wave decomposition, we can proceed and use
the one-way representation to study direct problems or by
applying the generalized Bremmer coupling series to study direct and
inverse scattering problems.
%~\cite{MdeHoop9606, vanStralen97,
%Gustafsson00}.

\appendix
\fixNumberingInAppendix
\section{Derivation of $\Ac$}
\label{app:calcA}

Given the normalized Maxwell equations \eqref{Eeq:mr4a}
in a medium with the constitutive relations
\eqref{Eeq:mr3}, we have \Eq \eqref{Eeq:mr4b} with vertical
components \eqref{Eeq:mr5} and transverse coordinates
\eqref{Eeq:mr6}. We replace the explicit appearance of $s H_{3}$ and $s E_{3}$
with \eqref{Eeq:mr5} and obtain
\begin{equation}
\begin{split}
        s \nu_{\alpha \beta} H_{\beta} - \mu_{\alpha 3}\mu_{33}^{-1}
        (\nabla \times
        E)_{3}  + (\nabla \times E)_{\alpha} & =  K_{\alpha}^{\rm e} -\mu_{\alpha
        3}\mu_{33}^{-1} K_{3}^{\rm e}  \; , \\
        -s \varepsilon_{\alpha \beta} E_{\beta} -  \epsilon_{\alpha 3}
        \epsilon_{33}^{-1} (\nabla \times
        H)_{3} + (\nabla \times H)_{\alpha} & =  J_{\alpha}^{\rm e} - \epsilon_{\alpha 3}
        \epsilon_{33}^{-1} J_{3}^{\rm e} \; .
\end{split}
    \label{Eeq:mr7c}
\end{equation}
where
\begin{equation}
        \varepsilon_{\alpha\beta} = \epsilon_{\alpha\beta} -
        \epsilon_{\alpha 3}\epsilon_{33}^{-1}\epsilon_{3\beta} \; , \quad \
        \quad
        \nu_{\alpha\beta}  =  \mu_{\alpha\beta} - \mu_{\alpha 3} \mu_{33}^{-1}
        \mu_{3 \beta} \; .
    \label{Eeq:mr7b}
\end{equation}
The transverse components of $\nabla \times E$ and $\nabla \times H$ are
explicitly
\begin{align}
\begin{split}
        (\nabla \times E)_{1}  & =  \partial_{2}E_{3} -
        \partial_{3}E_{2}\; ,  \\
        (\nabla \times E)_{2}  & =  \partial_{3}E_{1} - \partial_{1}E_{3}
        \; ,
\end{split} &&
\begin{split}
        (\nabla \times H)_{1}  & =  \partial_{2}H_{3} - \partial_{3}H_{2}
        \; , \\
        (\nabla \times H)_{2}  & =  \partial_{3}H_{1} - \partial_{1}H_{3}
        \; .
\end{split}
    \label{Eeq:mr7d}
\end{align}
We replace the vertical components $(E_3,H_3)$ in \eqref{Eeq:mr7d} with the identity \eqref{Eeq:mr5}
to find
\begin{align*}
        (\nabla \times E)_{1}  & =  -\partial_{3} E_{2}
        - \partial_{2}\epsilon_{33}^{-1}\epsilon_{3\alpha} E_{\alpha} +
        s^{-1}\partial_{2}\epsilon_{33}^{-1}(\partial_{1}H_{2} - \partial_{2}H_{1})
        - s^{-1}\partial_{2}\epsilon_{33}^{-1}J_{3}^{\rm e} \; , \\
        (\nabla \times E)_{2} & =  \partial_{3}E_{1} + \partial_{1}\epsilon_{33}^{-1}\epsilon_{3\alpha} E_{\alpha} -
        s^{-1}\partial_{1}\epsilon_{33}^{-1}(\partial_{1}H_{2} - \partial_{2}H_{1})
        + s^{-1}\partial_{1}\epsilon_{33}^{-1}J_{3}^{\rm e} \; ,\\
        (\nabla \times H)_{1} & =   - \partial_{3} H_{2}
        - \partial_{2}
        \mu_{33}^{-1} \mu_{3\beta} H_{\beta} - s^{-1}\partial_{2}\mu_{33}^{-1} (
        \partial_{1}E_{2} - \partial_{2}E_{1}
        ) + s^{-1}\partial_{2}\mu_{33}^{-1} K_{3}^{\rm e} \; ,\\
        (\nabla \times H)_{2} & =  \partial_{3}H_{1} +
        \partial_{1}
        \mu_{33}^{-1} \mu_{3\beta} H_{\beta} + s^{-1}\partial_{1}\mu_{33}^{-1} (
        \partial_{1}E_{2} - \partial_{2}E_{1}) -
        s^{-1}\partial_{1}\mu_{33}^{-1} K_{3}^{\rm e} \; .
%   \label{Eeq:mr7}
\end{align*}
These transverse components of the curl is substituted back into \eqref{Eeq:mr7c}.
Collecting similar terms gives
\begin{multline}
        s \nu_{2 \beta} H_{\beta}  -  \mu_{2 3}\mu_{33}^{-1}
        (\partial_{1} E_{2} - \partial_{2}E_{1}
        ) + \partial_{3}E_{1} + \partial_{1}\epsilon_{33}^{-1}\epsilon_{3\alpha}
        E_{\alpha} \\ -
        s^{-1}\partial_{1}\epsilon_{33}^{-1}(\partial_{1}H_{2} - \partial_{2}H_{1})
        =  K_{2}^{\rm e} -\mu_{23}\mu_{33}^{-1} K_{3}^{\rm e} -
        s^{-1}\partial_{1}\epsilon_{33}^{-1}J_{3}^{\rm e} \; ,
\end{multline}
\begin{multline}
        s \nu_{1 \beta} H_{\beta} - \mu_{1 3}\mu_{33}^{-1}
        (\partial_{1}E_{2} - \partial_{2}E_{1})
        -\partial_{3} E_{2}
        - \partial_{2}\epsilon_{33}^{-1}\epsilon_{3\alpha} E_{\alpha} \\ +
        s^{-1}\partial_{2}\epsilon_{33}^{-1}(\partial_{1}H_{2} - \partial_{2}H_{1})
        =  K_{1}^{\rm e} -\mu_{1 3}\mu_{33}^{-1} K_{3}^{\rm e} +
        s^{-1}\partial_{2}\epsilon_{33}^{-1}J_{3}^{\rm e} \; ,
\end{multline}
\begin{multline}
        s \varepsilon_{1 \beta} E_{\beta} +  \epsilon_{13}
        \epsilon_{33}^{-1} (\partial_{1} H_{2} - \partial_{2} H_{1}
        ) + \partial_{3} H_{2} + \partial_{2}
        \mu_{33}^{-1} \mu_{3\beta} H_{\beta} \\ +
        s^{-1}\partial_{2}\mu_{33}^{-1} (
        \partial_{1}E_{2} - \partial_{2}E_{1}
        )   =  -J_{1}^{\rm e} + \epsilon_{13}
        \epsilon_{33}^{-1} J_{3}^{\rm e} +s^{-1}\partial_{2}\mu_{33}^{-1}
        K_{3}^{\rm e} \; ,
\end{multline}
\begin{multline}
        -s \varepsilon_{2 \beta} E_{\beta} -  \epsilon_{23}
        \epsilon_{33}^{-1} (\partial_{1} H_{2} - \partial_{2} H_{1}
        ) + \partial_{3}H_{1} +
        \partial_{1}
        \mu_{33}^{-1} \mu_{3\beta} H_{\beta} \\ +
        s^{-1}\partial_{1}\mu_{33}^{-1} (
        \partial_{1}E_{2} - \partial_{2}E_{1})  = J_{2}^{\rm e} - \epsilon_{23}
        \epsilon_{33}^{-1} J_{3}^{\rm e} + s^{-1}\partial_{1}\mu_{33}^{-1}
        K_{3}^{\rm e} \; .
\end{multline}
Separating the derivatives in the vertical direction from the
remaining terms gives
\begin{equation}
    (I \partial_{3} + \Ac ) F = N \; ,
    \label{Eeq:mr9}
\end{equation}
in which the elements of the electromagnetic field matrix, $F$, are
given by
\begin{equation}
    F_{1}:= E_{1} \; , F_{2} := -E_{2} \ \ \mbox{and} \ \ F_{3} := H_{2}
    \; , F_{4}:=H_{1} \; .
    \label{Eeq:mr10}
\end{equation}
We write the electromagnetic system's matrix, $\Ac$, as a matrix of
2x2 block matrices
\begin{equation}
    \Ac:=\begin{pmatrix}
       \Ac_{11} & \Ac_{12} \\        \Ac_{21} & \Ac_{22}
    \end{pmatrix}
\end{equation}
where each block-matrix is given by
\begin{equation}\label{Eeq:mrAc}
\begin{split}
    \Ac_{11} & :=  \mu_{33}^{-1}\left( \begin{array}{rr}
            \mu_{23}\partial_{2} & \mu_{23} \partial_{1} \\
            \mu_{13}\partial_{2} & \mu_{13}\partial_{1}
        \end{array} \right)  +
        \left( \begin{array}{rr}
            \partial_{1}\epsilon_{31} & -\partial_{1}\epsilon_{32} \\
            -\partial_{2}\epsilon_{31} & \partial_{2}\epsilon_{32}
        \end{array} \right) \epsilon_{33}^{-1} \; , \\
    \Ac_{12} & :=  s \left(\begin{array}{rr}
        \nu_{22} & \nu_{21} \\  \nu_{12} & \nu_{11} \end{array}\right) -
        s^{-1} \left(\begin{array}{rr}
             \partial_{1}\epsilon_{33}^{-1} \partial_{1} &
            -\partial_{1}\epsilon_{33}^{-1} \partial_{2} \\
            -\partial_{2}\epsilon_{33}^{-1} \partial_{1} &
             \partial_{2}\epsilon_{33}^{-1} \partial_{2}
        \end{array}\right)  \; , \\
    \Ac_{21} & :=  s\left(\begin{array}{rr}
            \varepsilon_{11} & -\varepsilon_{12} \\
            -\varepsilon_{21} & \varepsilon_{22}
        \end{array}\right) - s^{-1}\left(\begin{array}{rr}
            \partial_{2}\mu_{33}^{-1}\partial_{2} &
            \partial_{2}\mu_{33}^{-1}\partial_{1} \\
            \partial_{1}\mu_{33}^{-1}\partial_{2} &
            \partial_{1}\mu_{33}^{-1}\partial_{1}
        \end{array}\right)  \; , \\
    \Ac_{22} & :=  \left(\begin{array}{rr}
            \partial_{2}\mu_{32} & \partial_{2}\mu_{31} \\
            \partial_{1}\mu_{32} & \partial_{1}\mu_{31}
        \end{array}\right) \mu_{33}^{-1} +
        \epsilon_{33}^{-1}\left(\begin{array}{rr}
            \epsilon_{13}\partial_{1} & -\epsilon_{13}\partial_{2} \\
            -\epsilon_{23}\partial_{1} & \epsilon_{23}\partial_{2}
        \end{array}\right) \; ,
\end{split}
\end{equation}
and the elements of the source terms
\begin{equation}
\begin{split}
    N_{1} & :=   K_{2}^{\rm e} -\mu_{23}\mu_{33}^{-1} K_{3}^{\rm e} -
        s^{-1}\partial_{1}\epsilon_{33}^{-1}J_{3}^{\rm e} \; , \\
    N_{2} & :=  K_{1}^{\rm e} -\mu_{1 3}\mu_{33}^{-1} K_{3}^{\rm e} +
        s^{-1}\partial_{2}\epsilon_{33}^{-1}J_{3}^{\rm e} \; , \\
    N_{3} & :=  -J_{1}^{\rm e} + \epsilon_{13}
        \epsilon_{33}^{-1} J_{3}^{\rm e} +s^{-1}\partial_{2}\mu_{33}^{-1}
        K_{3}^{\rm e} \; , \\
    N_{4} & :=  J_{2}^{\rm e} - \epsilon_{23}
        \epsilon_{33}^{-1} J_{3}^{\rm e} +
        s^{-1}\partial_{1}\mu_{33}^{-1} K_{3}^{\rm e} \; .
\end{split}
    \label{Eeq:mr14}
\end{equation}
Note that the adjoint of the 2x2-block matrices with respect to the
$\mLs$-inner product satisfy the following relations
\begin{equation}\label{ssym}
    \Ac_{11}^{*} = - \Ac_{22} \; , \ (\Ac_{12}(s))^{*} =
    \Ac_{12}(\bar{s}) \; , \ (\Ac_{21}(s))^{*} =
    \Ac_{21}(\bar{s}) \; ,
\end{equation}
for self-adjoint $\epsilon,\mu$.

\section{The isotropic homogeneous case}
\label{app:iso}

% Eeq:iso

We consider the normalized Maxwell equations \eqref{Eeq:mr4a} for
isotropic homogeneous
media, \ie we assume the constitutive relations
\begin{equation}
\begin{split}
        B(x,s)  &=  \mu H(x,s) \; , \\
        D(x,s)  &=  \epsilon E(x,s) \; ,
\end{split}
    \label{Eeq:iso1}
\end{equation}
where $\mu$, the permeability, and $\epsilon$, the permittivity, are
both real valued scalars and independent of space and time-Laplace
parameter $s$. The electromagnetic wave field satisfy Maxwell
equations
\begin{equation}
\begin{split}
    \mu s H  + \nabla \times E & =  K^{\rm e} \; ,
    \\
    - \epsilon s E + \nabla \times H & =  J^{\rm e} \; .
\end{split}
    \label{Eeq:iso2}
\end{equation}
The electromagnetic system's matrix, derived in Appendix
\ref{app:calcA}, \cf \eqref{Eeq:mrAc} reduce in the homogeneous
isotropic case to
\begin{equation}
\begin{split}
    \Ac_{11} & =  0 \; , \\
    \Ac_{12} & =  s \mu I -
        s^{-1} \epsilon^{-1} \left(\begin{array}{rr}
             \partial_{1} \partial_{1} &
             \partial_{1} \partial_{2} \\
             \partial_{2} \partial_{1} &
             \partial_{2} \partial_{2}
        \end{array}\right)  \; , \\
    \Ac_{21} & =  s \epsilon I - s^{-1} \mu^{-1} \left(\begin{array}{rr}
            \partial_{2}\partial_{2} &
            -\partial_{2}\partial_{1} \\
            -\partial_{1}\partial_{2} &
            \partial_{1}\partial_{1}
        \end{array}\right)  \; , \\
    \Ac_{22} & =  0 \; ,
\end{split}
\end{equation}
where $I$ is the 2x2 identity matrix. Applying the fourier transform
in transverse space, and using that the coefficients are constant,
gives the spectrum as the set of $\lambda(\xi',s)$ such that $\det(
\aS(\xi',s)-\lambda) = 0$ or
\begin{equation}
    (s^{2} \epsilon\mu + |\xi'|^{2}-\lambda^{2})^{2}=0 \ \ \Rightarrow
    \label{Eeq:iso9}
    \lambda = \pm \sqrt{s^{2}\epsilon\mu + |\xi'|^{2}} \; .
\end{equation}
Note that $\RE{s}>0$ imply $|\RL|>0$. Hence for $\RE{s}>0$ the
spectrum separates into two parts. The inverse of
$\ac(\xi',s,\lambda) := \aS(\xi',s)-\lambda I$, which we denote with
$\symb{r}_{;-1}(\xi',s,\lambda)$ is
\begin{equation}
    \symb{r}_{;-1}(\xi',s,\lambda) =\ac(\xi',s,-\lambda)(s^{2} \epsilon\mu +
    |\xi'|^{2}-\lambda^{2})^{-1} \; .
\end{equation}
The splitting matrix defined as, \cf \eqref{Eeq:p1}
\begin{equation}
    \psb = \frac{1}{\pi \iu} \int_{\RL = 0} \ac^{-1} \; ,
\end{equation}
has two parts, one is proportional to
\begin{equation}
    \frac{1}{\pi \iu}\int_{\RL = 0} \frac{\intd{\lambda}}{(s^{2} \epsilon\mu +
    |\xi'|^{2}-\lambda^{2})} = \frac{1}{\sqrt{s^{2}\epsilon\mu + |\xi'|^{2}}}
    \label{eq:apb1}
\end{equation}
and the second part is
\begin{equation}
    \frac{1}{\pi \iu}\int_{\RL = 0} \frac{\lambda \intd{\lambda}}{(s^{2} \epsilon\mu +
    |\xi'|^{2}-\lambda^{2})} = 0 \; ,
    \label{eq:apb2}
\end{equation}
as a principal value. With the observation that $(\aS_{12}(\xi',s))^{-1} =
\aS_{21}(\xi',s)/(s^{2}\epsilon\mu + |\xi'|^{2})$ and
\eqref{eq:apb1}--\eqref{eq:apb2} we obtain
\begin{equation}
    \psb(\xi',s) = \begin{pmatrix} 0 & \symb{z} \\
    \symb{z}^{-1} & 0 \end{pmatrix} \; ,
    \label{Eeq:iso10}
\end{equation}
where
\begin{equation}
    \symb{z} := \frac{1}{\sqrt{s^{2}\epsilon\mu + |\xi'|^{2}}}
    \aS_{12}(\xi',s) \; .
    \label{Eeq:c97}
\end{equation}
Note that $\psb$ has the expected property: $\psb^{2}=I$ \cf
Proposition \ref{prop2:split2}, part 4.

As the material is homogeneous we have
\begin{equation}
    \oB = \OP{F}^{-1} \psb \; .
\end{equation}
The symbol corresponding to the generalized eigenvector, $\pmL$, Proposition
\ref{prop2:split2}, part 5, is
\begin{equation}
    \symb{l}^{\pm} = \begin{pmatrix} \pm I \\ \symb{z}^{-1}
    \end{pmatrix} \symb{n}^{\pm} \; ,
    \label{Eeq:c98}
\end{equation}
for some normalization $\symb{n}^{\pm}$. The generalized eigenvalue
of $\Ac$ corresponding to $\pmL$ has the symbol representation \cf
Proposition \ref{prop2:dec}
\begin{align}
    \symb{s}^{\pm} & = \pm \symb{n}^{-1} \symb{z} \aS_{21} \symb{n} =
    \pm \symb{n}^{-1} \frac{1}{\sqrt{s^{2}\epsilon\mu + |\xi'|^{2}}}
    \aS_{12} \aS_{21} \symb{n} \nonumber \\ &
    = \pm \sqrt{s^{2}\epsilon\mu +
    |\xi'|^{2}} I \; .
    \label{Eeq:c99}
\end{align}
Thus the generalized eigenvalue problem reduces to an eigenvalue
problem for the isotropic case, \ie $\symb{s}^{\pm}$ reduces to
diagonal matrices. If we consider the wave-splitting problem by
earlier developed techniques see e.g.
\cite{Gustafsson2000,Rikte2001} we find
\begin{equation}
    \aS \begin{pmatrix} \pm I \\ \symb{z}^{-1} \end{pmatrix} =
    \begin{pmatrix} \pm I \\ \symb{z}^{-1} \end{pmatrix} \symb{s}^{\pm} \; .
\end{equation}
Hence the general procedure described in this paper agrees with the
earlier wave-splitting methods available for the homogeneous (and
layered homogeneous cases.
  % isotropa fallet!
\section{Two tools for the proof of Proposition \ref{prop2:split2}}
\label{app:det}

\subsection{The determinant of $\ac$}
The determinant of the symbol of $\Ac-\lambda I$, is
\begin{multline}
    \det\ac =
    \left(\lambda^2 - 2\iu \rpe_{33}^{-1} \RE{\rpe_{3\alpha}}
    \xi_{\alpha} \lambda -
    \rpe_{\alpha\beta}\xi_{\alpha}\xi_{\beta} \right)
    \left(\lambda^2 - 2\iu\rpm_{33}^{-1}\RE{\rpm_{3\gamma}}\xi_{\gamma} \lambda
    \right. \\ \left. \phantom{\lambda^2}-
    \rpm_{\gamma\delta}\xi_{\gamma}\xi_{\delta} \right)
     + s^4 \det \rE \det \rM +
     s^2 \left[
        \lambda^2 ( \rE_{12}\rM_{12} - \rE_{22}\rM_{11} -\rE_{11} \rM_{22}
        +\rE_{21} \rM_{21})  \right. \\ \left.
        + \rE_{\alpha\beta}\xi_{\alpha}\xi_{\beta}
        \rpm_{33}^{-1} \det \rpm +
        \rM_{\alpha\beta}\xi_{\alpha}\xi_{\beta} \rpe_{33}^{-1} \det \rpe
        \right. \\ \left. -
        2 (\rpe_{33}\rpm_{33})^{-1} \RE{(\rpe_{3\cdot} \times \rE_{\alpha
        \cdot})_{3}(\rpm_{3\cdot} \times \rM_{\beta
        \cdot})_{3}\xi_{\alpha}\xi_{\beta}} \right. \\ \left.
        - 2 \iu \lambda \left(\RE{((\rpm_{\cdot 3}
        \times \rM_{\cdot :})_{3}\times \rE_{:\alpha})_{3}\xi_{\alpha}}  +
        \RE{((\rpe_{\cdot 3}
        \times \rE_{\cdot :})_{3}\times \rM_{:\alpha})_{3}\xi_{\alpha}} \right)
     \right] \; ,
     \label{det-ac-em}
\end{multline}
where
\begin{equation}
    (\rpe_{3\cdot} \times \rE_{\alpha \cdot})_{3} =
        \rpe_{31}\rE_{\alpha 2} - \rpe_{32}\rE_{\alpha 1}
\end{equation}
and
\begin{equation}
    ((\rpm_{\cdot 3}
        \times \rM_{\cdot :})_{3}\times \rE_{:\alpha})_{3} =
        \rpm_{1 3} \rM_{2 1} \rE_{2\alpha} -\rpm_{2 3} \rM_{1 1} \rE_{2 \alpha}
        - \rpm_{1 3} \rM_{2 2} \rE_{1\alpha} + \rpm_{2 3} \rM_{1 2}
        \rE_{1\alpha} \; .
\end{equation}
The similar terms are defined analogously. Notice that
\eqref{det-ac-em} for $\RL=0$ this is a second order polynomial in
$s^2$ with real coefficients.

\subsection{Partial result for the symbol of the splitting matrix
in the anisotropic case}

To obtain the symbol of the splitting matrix, integration of the
type $I_{m,n}$ \cf Section \ref{sec:Bpsdo} Eq. \eqref{Eeq:ei1},
is to be evaluated. In the polyhomogeneous expansion it is clear that
the case $4m=n+1$ exists only for $m=1, n=3$ due to the
homogeneous decreasing degree of the polyhomogeneous expansion of
$\symb{r}$. This is the principal integral and it is calculated in
Proposition \ref{prop2:split2}, part 1.
For the remaining terms, $4m>n+1$, we use homogeneity of $\det \ac$ to obtain,
\begin{equation}
    \lim_{|\lambda|\rightarrow \infty} \left|
    \frac{\lambda^{n+1}}{\left(\det \ac\right)^m}\right| \leq
    \lim_{|\lambda|\rightarrow \infty} 2|\lambda|^{-1}=0 \; ,
    \label{Eeq:ei3}
\end{equation}
that allows us to use the residue theorem.
Thus
\begin{equation}
    \frac{1}{2\pi} \int_{-\infty}^{\infty} \intd{\IL}
    \frac{\lambda^n}{(\det \ac)^m} =
    \label{Eeq:ei4}
    - \Res{\frac{\lambda^n}{(\det
    \ac)^m}}{\lambda_{1}^+} - \Res{\frac{\lambda^n}{(\det
    \ac)^m}}{\lambda_{2}^+} \; ,
    \nonumber
\end{equation}
where the roots of the fourth order polynomial $\det \ac$ are denoted
by $\lambda^\pm_{1}$ and $\lambda^\pm_{2}$.
In the evaluation of the integral we have to consider the case when
$\lambda_{1}^+ = \lambda_{2}^+$.

To find the residue at $\lambda_{1}^+$ we first consider the case that
$\lambda_{1}^+ \neq \lambda_{2}^+$ and choose $\lambda$ such that
\begin{equation}
    \left| \frac{\lambda-\lambda_{1}^{+}}{\lambda_{1}^+ - \lambda_{2}^{\pm}}
    \right| < 1 \ \ \mbox{and} \ \
    \left| \frac{\lambda-\lambda_{1}^{+}}{\lambda_{1}^+ - \lambda_{1}^{-}}
    \right|
    < 1 \; .
    \label{Eeq:ei6c}
\end{equation}
Use the identity
\begin{equation}
    \frac{(m-1)!}{(1-y)^m} = \frac{\od^{m-1}(1-y)^{-1}}{\od y^{m-1}} =
    \sum_{j=0}^{\infty} \frac{(m+j-1)!}{j!}y^j \; ,
    \label{Eeq:ei5}
\end{equation}
valid for $|y|<1$ and $m\in\{1,2,3,\ldots\}$, together with the
binomial theorem to rewrite the integrand of $I_{m,n}$ into the
following form
\begin{align}
    \frac{\lambda^n}{(\det
    \ac
    )^m} &= \sum_{p=0}^n \bin{n}{p}
    \frac{(\lambda-\lambda_{1}^+)^p
    (\lambda_{1}^+)^{n-p}}{(\lambda-\lambda_{1}^+)^m
    (\lambda-\lambda_{2}^+)^m (\lambda-\lambda_{1}^-)^m
    (\lambda-\lambda_{2}^-)^m}
%   \label{Eeq:ei6a}
    \nonumber \\
    &= \sum_{p=0}^n \bin{n}{p}
    \frac{
    (\lambda_{1}^+)^{n-p}}{(\lambda-\lambda_{1}^+)^m}
    \frac{1}{((m-1)!)^3}
    \sum_{j_{1},j_{2},j_{3}=0}^\infty
    \frac{(-1)^{j_{1}+j_{2}+j_{3}}
    }{j_{1}!j_{2}!j_{3}!
    }
%   \nonumber
    \label{Eeq:ei6b}
    \\ &\ \times
    \frac{ (m+j_{1}-1)!(m+j_{2}-1)!(m+j_{3}-1)!
    }{
    (\lambda_{1}^+-\lambda_{2}^+)^{j_{1}+m}
    (\lambda_{1}^+-\lambda_{1}^-)^{j_{2}+m}
    (\lambda_{1}^+-\lambda_{2}^-)^{j_{3}+m}}
    (\lambda-\lambda_{1}^+)^{p+j_{1}+j_{2}+j_{3}} \; .
    \nonumber
\end{align}
The residue at $\lambda_{1}^+$ is the coefficient of the sum such that
$j_{1}+j_{2}+j_{3} = m-p-1$.
Thus
\begin{multline*}
%   \lefteqn{
    \Res{\frac{\lambda^n}{(\det \ac)^m}}{\lambda_{1}^+} =
%   }
%   \label{Eeq:ei8} \\
%   &&
    \sum_{p=0}^n \bin{n}{p}
    \frac{
    (\lambda_{1}^+)^{n-p}}{((m-1)!)^3}
    \sum_{\substack{ j_{1}+j_{2}+j_{3}=m-p-1 \\\
    j_{i}\geq 0}} \frac{(-1)^{j_{1}+j_{2}+j_{3}}}{j_{1}!j_{2}!j_{3}!}
    \times
     \\
%   &&
        \frac{ (m+j_{1}-1)!(m+j_{2}-1)!(m+j_{3}-1)!}{
    (\lambda_{1}^+-\lambda_{2}^+)^{j_{1}+m}
    (\lambda_{1}^+-\lambda_{1}^-)^{j_{2}+m}
    (\lambda_{1}^+-\lambda_{2}^-)^{j_{3}+m}} \; .
\end{multline*}
An analogous result is obtained for the root
$\lambda_{2}^+$. Observe that the term
$(\lambda_{1}^{+}-\lambda_{2}^{+})^{-j_{1}-m}$, is not bounded, but
from Proposition \ref{prop2:split2}, part 1 we know that the integral is
bounded, and hence this can be removed by eliminating common factors in
the sum of the two residues, similarly to \eqref{Eeq:pp13} and
\eqref{Eeq:pp14}.

For the case of $\lambda_{1}^+ =
\lambda_{2}^+$ we obtain that the two residue collapse to one and
becomes
\begin{multline}
    \Res{\frac{\lambda^n}{(\det \ac)^m}}{\lambda_{1}^+} =
    \sum_{p=0}^n \bin{n}{p}
    \frac{
    (\lambda_{1}^+)^{n-p}}{((m-1)!)^2} \\ \times
    \label{Eeq:ei9}
    \sum_{\substack{j_{1}+j_{2}=2m-p-1 \\\ j_{i}\geq 0}}
    \frac{(-1)^{j_{1}+j_{2}} (m+j_{1}-1)!(m+j_{2}-1)!}{j_{1}!j_{2}!
    (\lambda_{1}^+-\lambda_{1}^-)^{j_{1}+m}
    (\lambda_{1}^+-\lambda_{2}^-)^{j_{2}+m}} \; .
\end{multline}

Thus given the roots of the polynomial $\det \ac = 0$,
we obtain the integral for each $m,n$. Upon substituting the integral
in the asymptotic series for the parametrix we obtain the symbol.

%\bibliography{EMdec_ref}

%\bibliographystyle{acm}
\newcommand{\SortNoop}[2]{#2}

\end{document}